\documentclass[preprint,aps,preprintnumbers,nofootinbib,superscriptaddress,showpacs]{revtex4}
\usepackage{amsmath,amssymb,amsbsy}
\usepackage{graphicx}
\usepackage{epsfig}
\usepackage{color}

\newcommand{\Delstar}{\ensuremath{\Delta^{\raise0.18ex\hbox{${\scriptstyle *}$}}}}
\def\gtwid{{\,\raise.35ex\hbox{$>$\kern-.75em\lower1ex\hbox{$\sim$}}\,}}
\def\ltwid{{\,\raise.35ex\hbox{$<$\kern-.75em\lower1ex\hbox{$\sim$}}\,}}
\def\leftvec{{\raise1.5ex\hbox{$\leftarrow$}\kern-1.00em}}
\def\rightvec{{\raise1.5ex\hbox{$\rightarrow$}\kern-1.00em}}
\def\half{{\scriptstyle \raise.2ex\hbox{${1\over2}$}}}
\def\threehalves{{\scriptstyle \raise.15ex\hbox{${3\over2}$}}}
\def\third{{\scriptstyle \raise.15ex\hbox{${1\over3}$}}}
\def\third{{\scriptstyle \raise.15ex\hbox{${1\over3}$}}}
\def\twothirds{{\scriptstyle \raise.15ex\hbox{${2\over3}$}}}
\def\fourth{{\scriptstyle \raise.15ex\hbox{${1\over4}$}}}

\newcommand*{\Tr}{\operatorname{Tr}}

\newcommand*{\bea}{\begin{eqnarray}}
\newcommand*{\eea}{\end{eqnarray}}
\newcommand*{\be}{\begin{equation}}
\newcommand*{\ee}{\end{equation}}

\newcommand*{\CPT}{\raise0.4ex\hbox{$\chi$}PT}
\newcommand*{\chpt}{\raise0.4ex\hbox{$\chi$}PT}
\newcommand*{\schpt}{S\raise0.4ex\hbox{$\chi$}PT}

\def\eqref#1{{(\ref{#1})}}

\def\CO{{\cal O}}

\def\bar{\overline}
\def\hat{\widehat}
\def\tilde{\widetilde}

\def\Tr{\textrm{Tr}}

\def\etal{{\it et al.}}

\def\bea{\begin{eqnarray}}
\def\eea{\end{eqnarray}}

\def\spose#1{\hbox to 0pt{#1\hss}}
\def\ltapprox{\mathrel{\spose{\lower 3pt\hbox{$\mathchar"218$}}
 \raise 2.0pt\hbox{$\mathchar"13C$}}}
\def\gtapprox{\mathrel{\spose{\lower 3pt\hbox{$\mathchar"218$}}
 \raise 2.0pt\hbox{$\mathchar"13E$}}}
\def\inapprox{\mathrel{\spose{\lower 3pt\hbox{$\mathchar"218$}}
 \raise 2.0pt\hbox{$\mathchar"232$}}}

\begin{document}

\preprint{FERMILAB-PUB-09-267-T}

\vphantom{}

\title{The neutral kaon mixing parameter $B_K$ \\from unquenched mixed-action lattice QCD\\ 
}

\author{C.\ Aubin}
\email[]{caaubin@wm.edu}
\affiliation{Department of Physics, College of William and Mary, Williamsburg, VA 23187}

\author{Jack Laiho}
\email[]{jlaiho@fnal.gov}
\affiliation{Physics Department, Washington University, St. Louis, MO 63130}

\author{Ruth S. Van de Water}
\email[]{ruthv@bnl.gov}
\affiliation{Physics Department, Brookhaven National Laboratory, Upton, NY 11973}

\date{\today}

\begin{abstract}
We calculate the neutral kaon mixing parameter $B_K$ in unquenched lattice QCD using asqtad-improved staggered sea quarks and domain-wall valence quarks.  We use the ``2+1" flavor gauge configurations generated by the MILC Collaboration, and simulate with multiple valence and sea quark masses at two lattice spacings of $a \approx 0.12$ fm and $a \approx 0.09$ fm.  We match the lattice determination of $B_K$ to the continuum value using the nonperturbative method of Rome-Southampton, and extrapolate $B_K$ to the continuum and physical quark masses using mixed action chiral perturbation theory.  The ``mixed-action" method enables us to control all sources of systematic uncertainty and therefore to precisely determine $B_K$; we find a value of $B_K^{\overline{\textrm{MS}}, \textrm{NDR}}(2 {\rm GeV}) = 0.527(6)(21)$, where the first error is statistical and the second is systematic. 

\end{abstract}

\maketitle

\section{Introduction}
\label{sec:Intro}

The kaon B-parameter ($B_K$), which parametrizes the hadronic part of $CP$-violating neutral kaon mixing, plays an important role in flavor-physics phenomenology.  When combined with an experimental measurement of  indirect $CP$-violation in the kaon sector, $\epsilon_K$, $B_K$ constrains the apex of the  Cabibbo-Kobayashi-Maskawa (CKM) unitarity triangle.  Because $\epsilon_K$ is known to sub-percent accuracy, this constraint is limited by the theoretical uncertainties in several quantities, including $B_K$.  Physics beyond the standard model generically predicts additional quark flavor-changing interactions and $CP$-violating phases.   These will manifest themselves as apparent inconsistencies between measurements that are predicted to be identical within the framework of the standard model.  Thus precise experimental measurements of quark-flavor changing weak-interaction processes are sensitive probes of new physics, provided that the corresponding theoretical calculations are also sufficiently precise.  In this work we calculate $B_K$ using lattice QCD with all sources of systematic uncertainty under control.  This result is needed to interpret the experimental measurement of $\epsilon_K$ as a constraint on the CKM unitarity triangle, and hence to constrain physics beyond the standard model.  

Because an accurate determination of $B_K$ is essential for flavor-physics phenomenology, many lattice QCD calculations of $B_K$ have been done over the past decade, each improving upon the previous one.  The benchmark calculation by the JLQCD Collaboration contains a thorough study of the quark mass and lattice spacing dependence~\cite{Aoki:1997nr}.  Because it does not include the effect of sea quark loops, however, the final result for $B_K$ has a quenching uncertainty which is difficult to estimate.  The HPQCD Collaboration eliminates this source of error in $B_K$ by using dynamical staggered fermions 
at a single lattice spacing~\cite{Gamiz:2006sq}.  The additional species of staggered quarks, referred to as ``tastes", however, complicate the lattice-to-continuum operator matching procedure, and lead to a $\sim 20\%$ systematic error in $B_K$ due to neglected higher-order operators and mixing with operators specific to staggered fermions that break flavor symmetry.  The RBC and UKQCD Collaborations' calculation of $B_K$ contains the effects of three flavors of dynamical domain-wall fermions and employs nonperturbative operator renormalization~\cite{Antonio:2007pb}.
Although their result for $B_K$ has a $\sim 6\%$ total uncertainty, it relies on a single lattice spacing and an estimate of the size of discretization errors based on the earlier quenched calculation in Ref.~\cite{AliKhan:2001wr}.

Our mixed-action lattice QCD calculation combines domain-wall valence quarks and staggered sea quarks, following the method of the LHP Collaboration~\cite{Renner:2004ck}.   We use the ``2+1" flavor asqtad-improved staggered lattices generated by the MILC Collaboration, which include the effects of three light dynamical quarks~\cite{Bernard:2001av}.   These configurations are publicly available with a large range of quark masses, lattice spacings, and volumes and allow for good control over the systematic error from chiral and continuum extrapolation~\cite{Aubin:2004fs}.  We generate domain-wall valence quark propagators using the Chroma lattice QCD software package~\cite{Edwards:2004sx}.  The approximate chiral symmetry of domain-wall quarks simplifies both the extrapolations to physical quark masses and zero lattice spacing and the lattice-to-continuum operator matching.  Because the mixed-action $\Delta S = 2$ lattice operator used to calculate the $B_K$ matrix element is composed of domain-wall valence quarks, which do not carry the taste quantum number, it only mixes with other operators of wrong chirality (due to small residual chiral symmetry breaking), not incorrect tastes.  This makes the relevant mixed action chiral perturbation theory (MA$\chi$PT) more continuum-like than in the purely staggered case.  There is only one new parameter in the one-loop MA$\chi$PT expression for $B_K$ with respect to the purely domain-wall case, and it can easily be determined from the staggered pseudoscalar meson mass spectrum~\cite{Aubin:2004fs, Aubin:2006hg}.  Use of domain-wall valence quarks in the $\Delta S = 2$ lattice operator also makes the 
nonperturbative operator matching via the Rome-Southampton method~\cite{Martinelli:1994ty} as simple as in the purely domain-wall case.  Thus the mixed action method combines the advantages of staggered and domain-wall fermions without suffering from their primary disadvantages and is well-suited to the calculation of $B_K$.

The MILC gauge configurations make use of the rooting procedure to remove the additional staggered quark species (tastes) from the calculation of the fermion determinant.  Although the ``fourth-root trick" has not been proven correct, both theoretical arguments~\cite{Shamir:2004zc,Shamir:2006nj,Bernard:2006zw,Bernard:2007ma} and numerical simulations~\cite{Prelovsek:2005rf,Bernard:2007qf,Aubin:2008wk} support the validity of the rooting procedure.   Most of this evidence is summarized in reviews by D\"urr, Sharpe, Kronfeld, and Golterman~\cite{Durr:2005ax,SharpePlenary,Kronfeld:2007ek,Golterman:2008gt}.  Given the wealth of evidence substantiating the fourth-root trick, we work under the plausible assumption that the continuum limit of the rooted staggered theory is QCD.

Our calculation of $B_K$ relies upon the ability to correctly extrapolate to the physical quark masses and 
zero lattice spacing using MA$\chi$PT~\cite{Bar:2005tu}, which describes the pseudo-Goldstone boson sector of the mixed-action lattice theory.  In Ref.~\cite{Aubin:2008wk} we have therefore performed a strong check of the ability of MA\chpt\ to accurately describe the quark-mass and lattice-spacing dependence of the isovector scalar correlator.  The $a_0$ correlator is particularly sensitive to unitarity-violating discretization effects in the mixed-action theory because it receives contributions from flavor-neutral two-meson intermediate states.  At next-to-leading order (NLO), the size and shape of 
these ``bubble" contributions to the scalar correlator are completely predicted within MA$\chi$PT~\cite{Prelovsek:2005rf}, given knowledge of a few low-energy constants that are easily determined in fits to pseudoscalar meson mass data.   We find that, for all valence-sea mass combinations on both the coarse and fine lattices, the MA\chpt\ prediction is in good quantitative agreement with the numerical lattice data, despite the numerically large discretization effects due to the staggered sea sector.  Thus we conclude that MA\chpt\ describes the dominant unitarity-violating effects in mixed-action lattice simulations.  For the case of most weak-matrix elements, including $B_K$, NLO MA$\chi$PT predicts that unitarity-violating discretization effects in 1-loop chiral logarithms are below a percent on the coarse and fine MILC lattices~\cite{Aubin:2006hg}.  This fact, in conjunction with our successful analysis of the scalar correlator, substantiates the claim that we can use MA\chpt\ to remove these effects from $B_K$ and to precisely determine its value in the continuum.

We have also performed a more general check of our ability to control systematic errors in our mixed-action numerical simulations by calculating the light pseudoscalar meson decay constants, $f_\pi$ and $f_K$, and their ratio~\cite{Aubin:2008ie}.  We use the same gauge configurations and domain-wall valence quark masses as in the calculation of $B_K$ presented in this work.  We determine both $f_\pi$ and $f_K$ with $\sim 3\%$ accuracy, and their ratio with $\sim 2\%$ accuracy.  Given the value of $|V_{ud}|$ from superallowed $\beta$-decay, our result for $f_\pi$ is consistent with experiment.  Similarly, given the $|V_{us}|$ determination from semileptonic kaon decays using non-lattice theory, our result for $f_K$ is consistent with experiment.  Our result for the ratio $f_K/f_\pi$, which is independent of the CKM matrix elements, is consistent with other more precise lattice determinations~\cite{Bernard:2007ps, Follana:2007uv, Lellouch:2009fg}.  Although our decay constant calculation does not check the Rome-Southampton nonperturbative renormalization (NPR) procedure, it does test the remaining ingredients in the calculation of $B_K$, especially the chiral and continuum extrapolation using MA$\chi$PT.  Therefore the successful calculation of the well-known quantities $f_\pi$ and $f_K$ bolsters confidence in the calculation of the weak matrix element $B_K$ presented in this work.

\bigskip

This paper is organized as follows.  In Sec.~\ref{sec:Lat_calc}, we describe the details of our numerical mixed-action lattice simulation; we present the actions and parameters used and describe how the relevant 2-point and 3-point correlators are analyzed.  Next, in Sec.~\ref{sec:Renorm}, we describe two independent calculations of the renormalization factor $Z_{B_K}$ needed to match the lattice matrix element to the continuum. We compute $Z_{B_K}$ using the nonperturbative Rome-Southampton approach; this is used in our preferred determination of $B_K$.  We also compute $Z_{B_K}$ to 1-loop in mean-field improved lattice perturbation theory to provide a cross-check and aid in estimating the systematic error
associated with the matching.  In Sec.~\ref{sec:ChPT}, we describe the extrapolation of $B_K$ to the physical quark masses and the continuum using NLO MA$\chi$PT supplemented by higher-order analytic terms to allow an interpolation about the strange quark mass.  Next, in Sec.~\ref{sec:Error}, we present the systematic error budget for $B_K$, describing each individual uncertainty in a separate subsection for clarity.  Finally, in Sec.~\ref{sec:Conc}, we compare our results to previous unquenched lattice determinations and to the preferred values from the unitarity triangle analyses.  We conclude by discussing the prospects for improvement in our mixed-action lattice calculation and for its phenomenological impact on the search for new physics.

\section{Lattice calculation}
\label{sec:Lat_calc}

In this section we describe the details of our numerical mixed-action lattice calculation.  We first present the valence and sea quark lattice actions and input parameters (such as quark masses and lattice spacings) in Sec.~\ref{sec:Params}. Next, in Sec.~\ref{sec:3pt}, we present the 2-point and 3-point correlation functions needed to determine the unrenormalized lattice value of $B_K$.  We give the valence quark propagator source wavefunctions and boundary conditions.  We also describe the method used to extract $B_K$ from a ratio of 3-point and 2-point functions, and show example correlated plateau fits with jackknife errors.

\subsection{Actions and input parameters}
\label{sec:Params}

We use the unquenched lattices generated by the MILC Collaboration for our numerical lattice calculation of $B_K$, which include the effects of three dynamical flavors of asqtad-improved staggered fermions~\cite{Susskind:1976jm}.  Because the MILC configurations are available at several light quark masses and lattice spacings~\cite{Bernard:2001av,Aubin:2004wf}, they allow us to have good control over the both the chiral extrapolation in the sea sector and the continuum extrapolation.  We calculate $B_K$ on both the ``coarse" ($a\approx 0.12$ fm) and ``fine" ($a\approx 0.09$ fm) MILC ensembles, which have physical volumes ranging from approximately (2.5 -- 3 fm)${}^3$.    For each ensemble, the masses of the up and down sea quarks are degenerate;  our lightest dynamical mass is approximately a tenth of the physical strange quark.  For most of our ensembles, the mass of the dynamical strange quark is close to its physical value.  At each lattice spacing, however, we have data on one ensemble with an unphysically light strange sea quark in order to better constrain the strange sea quark mass dependence and aid in the chiral extrapolation.  The left-hand side of Table~\ref{tab:BK_data} shows the parameters of the MILC gauge configurations used to calculate $B_K$.  

\begin{table}
\caption{Parameters of the MILC improved staggered gauge configurations and domain-wall valence quark propagators used to calculate the unrenormalized lattice value of $B_K$.  Columns one and two list the approximate lattice spacings and lattice volumes (in lattice spacing units).  Columns three and four show the nominal up/down ($m_l$) and strange quark ($m_h$) masses in the sea, along with the corresponding pseudoscalar taste pion mass.  Columns five and six list our partially quenched valence quark masses ($m_x$), along with our lightest available domain-wall pion mass.  Column seven shows the number of configurations analyzed on each ensemble.}
\vspace{3mm}

\label{tab:BK_data}
\begin{tabular}{llccccr}
\hline\hline
\multicolumn{2}{c}{\qquad} & \multicolumn{2}{c}{sea sector} & \multicolumn{2}{c}{valence sector} \\[-0.5mm]
$a$(fm) & $\left(\frac{L}{a}\right)^3 \times \frac{T}{a}$ & \ $a m_l / am_h$ &  \ $a m_\pi$ &  \ $am_x$ &  \ $a m_\pi$ & $N_{\rm conf.}$  \\[0.5mm]  \hline

0.09 & $40^3 \times 96$ & \ 0.0031/0.031 &   \ 0.10538(06) & \ 0.004, 0.0186, 0.046 &  \ 0.0999(12) & 150 \\
0.09 & $28^3 \times 96$ & \ 0.0062/0.0186 &  \ 0.14619(14) & \ 0.0062, 0.0124, 0.0186, 0.046 &  \ 0.1212(17) & 160 \\
0.09 & $28^3 \times 96$  & \ 0.0062/0.031 &  \ 0.14789(18) & \ 0.0062, 0.0124, 0.0186, 0.046 &  \ 0.1222(12) & 210 \\
0.09 & $28^3 \times 96$ & \ 0.0124/0.031 &  \ 0.20635(18) & \ 0.0062, 0.0124, 0.0186, 0.046 &  \ 0.1216(11) & 198 \\
\hline

0.12 & $24^3 \times 64$ & \ 0.005/0.05 &  \ 0.15971(20) & \ 0.007, 0.02, 0.03, 0.05, 0.065 &  \ 0.1718(11) & 216 \\
0.12 & $20^3 \times 64$ & \ 0.007/0.05 &  \ 0.18891(20) & \ 0.01, 0.02, 0.03, 0.04, 0.05, 0.065 &  \ 0.1968(08) & 268 \\
0.12 & $20^3 \times 64$ & \ 0.01/0.03 &  \ 0.22357(19) & \ 0.01, 0.02, 0.03, 0.05, 0.065 &  \ 0.1946(18) & 160 \\
0.12 & $20^3 \times 64$ & \ 0.01/0.05 &  \ 0.22447(17) & \ 0.01, 0.02, 0.03, 0.05, 0.065 &  \ 0.1989(08) & 220 \\
0.12 & $20^3 \times 64$ & \ 0.02/0.05 &  \ 0.31125(16) & \ 0.01, 0.03, 0.05, 0.065 &  \ 0.1949(13) & 117 \\
\hline\hline

\end{tabular}\end{table}

We construct the 2-point and 3-point correlation functions needed to determine $B_K$ using domain-wall valence quark propagators~\cite{Kaplan:1992bt,Shamir:1993zy}.  The approximate chiral symmetry of domain-wall fermions simplifies both the nonperturbative determination of the renormalization coefficient, $Z_{B_K}$, and the extrapolation of $B_K$ to physical quark masses and the continuum;  these advantages will be discussed in greater detail in Secs.~\ref{sec:Renorm} and~\ref{sec:ChPT}, respectively.  We compute the domain-wall propagators using the Chroma software system for lattice QCD~\cite{Edwards:2004sx}.  We use the same input parameters as the LHP Collaboration~\cite{Renner:2004ck};  this allows us to check simple quantities such as the pion masses and the residual quark mass.  We first smear the MILC lattices using the standard hypercubic blocking (HYP) parameters given in Ref.~\cite{Hasenfratz:2001hp} in order to reduce the size of explicit chiral symmetry breaking and proximity to the Aoki phase~\cite{Aoki:1983qi}.
On both the coarse and fine ensembles we simulate with a domain-wall height of $M_5$=1.7 and a fifth dimension of length $L_s$=16.  For each sea quark ensemble, we calculate $B_K$ at several valence quark masses; this allows us both to extrapolate the numerical lattice data to the physical up/down quark mass and to interpolate to the physical strange quark mass.  Our lightest valence quark mass is chosen to be as light as possible while keeping finite-volume effects under control.  Specifically, we restrict the quantity $m_\pi L \gtapprox 3.5$ to keep 1-loop MA$\chi$PT finite volume effects for $B_K$ below $1 \%$.  Thus the mass of our lightest domain-wall pion is $\sim 280$ MeV on the 2.5 fm ensembles, and $\sim 240$ MeV on the 3.5 fm ensemble.  The fifth column of Table~\ref{tab:BK_data} shows the bare domain-wall masses used to calculate $B_K$.      
  
In most mixed staggered sea, domain-wall valence lattice simulations, the bare domain-wall quark mass is tuned so that the mass of the domain-wall pion is equal to the mass of the lightest staggered pion in the sea sector~\cite{Renner:2004ck}.  
Although this procedure does not eliminate unitarity-violating discretization effects in the mixed-action theory at nonzero lattice spacing, tuning the domain-wall pion to one of the staggered pion masses allows one to approach full QCD as the continuum limit is taken numerically, even for quantities for which mixed-action chiral perturbation theory expressions do not exist or are not applicable.
Fortunately, for the case of $B_K$, we can use MA$\chi$PT~\cite{Bar:2005tu,Aubin:2006hg} to properly account for and remove these discretization errors in fits to quantities evaluated at multiple lattice spacings and valence and sea quark masses.  Thus we do not make any attempt to tune the bare domain-wall quark masses in our lattice calculation.

In order to convert dimensionful quantities calculated in our mixed-action lattice simulations into physical units, we need the value of the lattice spacing, $a$, which we determine by computing a known physical quantity that can be directly compared to experiment.  Although all of the coarse (or fine) MILC lattices have approximately the same lattice spacing, slight variations exist due to the choice of simulation parameters in the gauge action.  We account for these differences by converting all of our data from lattice spacing units into $r_1$ units before performing any chiral fits.  Because $r_1$ is related to the force between static quarks, $r_1^2 F(r_1) = 1.0$~\cite{Sommer:1993ce},  this method has the advantage that the ratio $r_1/a$ can be determined precisely from a fit to the static quark potential~\cite{Bernard:2000gd,Aubin:2004wf}.  The absolute scale, $r_1$, can then be determined in several ways.  In this work we use the scale $r_1 = 0.3108(15)(^{+26}_{-79})$ fm to convert our simulation results into physical units.  This value is obtained by combining the recent MILC determination of $r_1 f_\pi$ with the experimentally measured value of $f_\pi$~\cite{Bernard:2007ps}.  We use an alternative determination of $r_1$ from the $\Upsilon$ splitting \cite{Gray:2005ad, Bernard:2005ei}, $r_1=0.318(7)$ fm, in order to estimate the systematic error due to the scale uncertainty.

\subsection{Three-point correlation functions}
\label{sec:3pt}

$B_K$ parametrizes the nonperturbative QCD contribution to CP-violating neutral kaon mixing.  Kaon mixing occurs via electroweak box diagrams.  Integrating out the heavy intermediate $W$-bosons to isolate the hadronic contribution leads to the following $\Delta S = 2$ operator in the effective Hamiltonian:
\begin{eqnarray}
    \CO_K^{\Delta S = 2} &= & [\bar{s} \gamma_\mu (1 - \gamma_5)d][\bar{s}  \gamma_\mu (1 - \gamma_5)d] ,
\label{eq:OK}
\end{eqnarray}
where we omit the color indices for simplicity.  In order to ensure that the value of $B_K$ is close to unity, $B_K$ is defined as a ratio: 
\begin{equation}
    B_K  \equiv  \frac{\langle\bar{K}^0 | [\bar{s} \gamma_\mu (1 - \gamma_5)d][\bar{s}  \gamma_\mu (1 - \gamma_5)d]  | K^0\rangle}{\frac{8}{3} \langle \bar{K}^0 |\bar{s} \gamma_\mu (1 - \gamma_5) d |0\rangle \langle 0|\bar{s}\gamma_\mu(1 - \gamma_5) d|K^0\rangle}  ,
\label{eq:BKdef} 
\end{equation}
where the numerator is the desired $\Delta S = 2$ matrix element, and the denominator is the same matrix element as in the numerator evaluated in the vacuum saturation approximation.  Because the matrix element in the denominator is related to the kaon decay constant, Eq.~(\ref{eq:BKdef}) is often simplified as
\begin{equation}
    B_K = \frac{\langle\bar{K}^0 | \CO_K^{\Delta S = 2}| | K^0\rangle}{\frac{8}{3} m_K^2 f_K^2} .
\end{equation}

\bigskip

In this work we calculate $B_K$ numerically from the following ratio of lattice correlation functions:
 \begin{eqnarray}
 	B_K^\text{lat.} &=& \frac{L^3}{\frac{8}{3}} \frac{\langle \psi_P(t_{src}+T)\, \CO^{\Delta S = 2}_K (t)\ \psi_P(t_{src})^{\dagger} \rangle
}{ \langle \phi_A(t)\, \psi_P(t_{src})^{\dagger} \rangle  \langle \phi_A(t)\, \psi_P(t_{src}+T)^{\dagger} \rangle} ,
\label{eq:BK_corr}
 \end{eqnarray}
where $T$ is the temporal extent of the lattice and we include the superscript ``lat." to emphasize that the quantity in Eq.~(\ref{eq:BK_corr}) needs to be renormalized in order to recover $B_K$ in a continuum regularization scheme.  We fix the locations of the source and sink kaons in the numerator 3-point function at $t_{src}$ and  $t_{src} + T$, respectively, and vary the position of the four-quark operator, $\CO^{\Delta S = 2}_K$, over all time slices $t$ in between.  We use wall sources for our kaons throughout the calculation, but use local sinks for both the four-quark operator in the 3-point function and the axial-current operator in the 2-point functions:
\begin{eqnarray}
	\psi_P (t) &=& \sum_{\vec{x},\vec{y}} \bar{s} (\vec{x}, t) \, \gamma_5 \, d(\vec{y}, t), \\
	\phi_A (t) &=& \sum_{\vec{x}} \bar{s} (\vec{x}, t) \gamma_5 \gamma_\mu d(\vec{x}, t) .
\end{eqnarray}
The volume factor $L^3$ in the numerator of Eq.~(\ref{eq:BK_corr}) accounts for the differing normalizations of the wall sources and point sources used in the determination of $B_K^\text{lat.}$.

For each domain-wall valence quark mass on a given MILC configuration, we compute two Coulomb gauge-fixed wall-source propagators starting from the same lattice timeslice, $t_{src}$:  one with periodic and another with antiperiodic boundary conditions in the temporal direction.  The spatial boundary conditions are always periodic.  The Coulomb gauge-fixed wall-source is used to reduce contamination from excited states.  We then take symmetric and antisymmetric linear combinations in order to produce forward- and backward-moving propagators beginning at $t_{src}$.  We use these symmetrized propagators in the interpolating operators $\psi$ and $\phi$ in order to effectively double the number of lattice timeslices.   This ensures that finite-size effects due to pions circling the lattice in the temporal direction are negligible.  Using the same time slice for the source of the forward- and backward-moving propagators 
also allows us to save a factor of two in computing time.\footnote{This method was suggested to us by N. Christ~\cite{Christ:2006}.}

In order to make the best use of our computing resources, we generate domain-wall quark propagators on every fourth recorded MILC gauge configuration (typically every 20th or 24th trajectory) in order to reduce autocorrelation errors.  
Our earliest runs have propagators with $t_{src} = 0$, which we chose for simplicity.  In order to take advantage of the large temporal extent of the MILC lattices and further reduce autocorrelations, however, our later runs use a randomly chosen $t_{src}$.  Although the two data sets are expected to have somewhat different autocorrelation times, there is nothing \emph{a priori} wrong with combining them in an ensemble average.

\begin{figure}
\begin{center}
\includegraphics[width=4.5in]{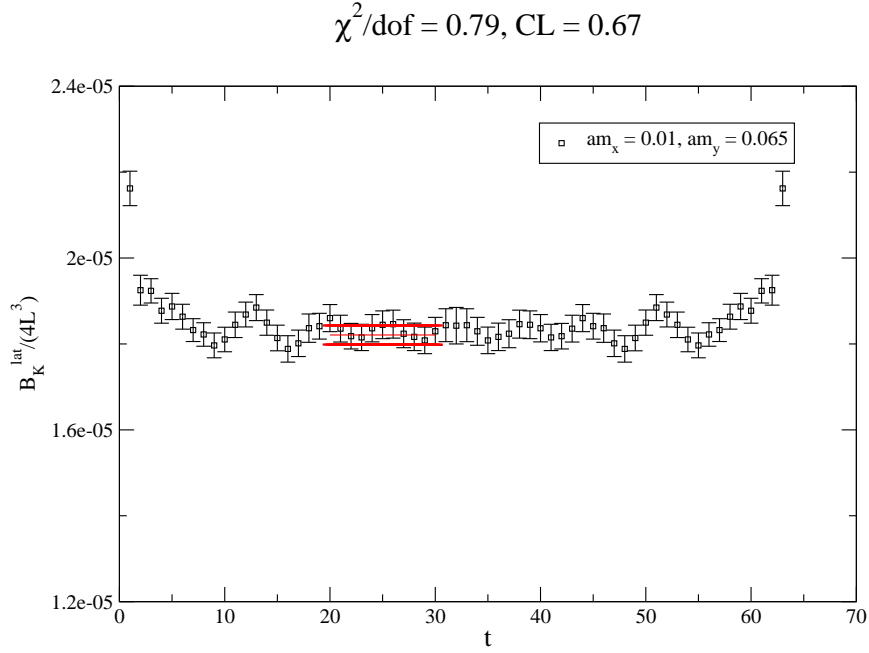}
\caption{Plateau fit to $B_K^{lat}/(4L^3)$ on the coarse $am_l/am_h = 0.007/0.05$ ensemble.  The legend shows the non-degenerate pair of quark masses making up the kaon in the three-point correlation function.  The correlated $\chi^2/{\rm dof}$ and confidence level of the fit are given in the title.}
\label{fig:plat_007}
\end{center}
\end{figure}

\begin{figure}
\begin{center}
\includegraphics[width=4.5in]{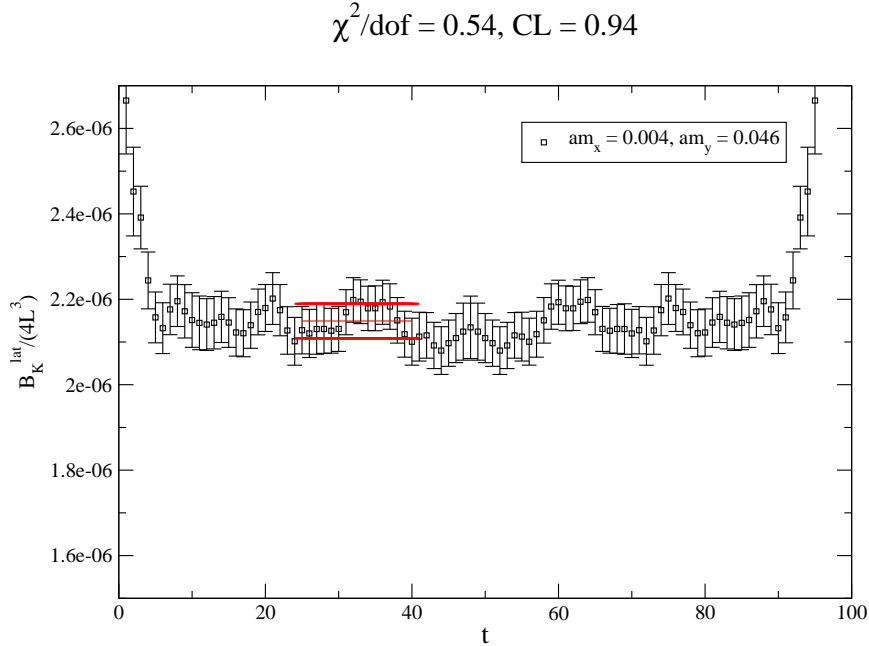}
\caption{Same as Figure~\ref{fig:plat_007} but on the fine $am_l/am_h = 0.0031/0.031$ ensemble.}
\label{fig:plat_0031}
\end{center}
\end{figure}

Figure~\ref{fig:plat_007} shows a representative plateau fit on a coarse ensemble for $B_K^{lat}/(4L^3)$ with a non-degenerate kaon made up of a light quark with mass around 
$m_s/6$ and a heavier quark with mass close to $m_s$.  Figure~\ref{fig:plat_0031} shows a similar plateau fit on a fine ensemble where the heavier quark mass is again close to $m_s$, and the light-quark mass is around $m_s/10$.  The confidence levels of the fits are computed using the full correlation matrix in the minimization of $\chi^2$ in order to assess the quality of the plateaus.  The statistical errors in the fit are determined by performing a separate fit to each single-elimination jackknife sample; the correlation matrix is remade for each jackknife fit.  Excellent fits to a constant are found, and the confidence levels of the fits in Fig.~\ref{fig:plat_007} (${\rm CL}=0.71$) and Fig.~\ref{fig:plat_0031} (${\rm CL}=0.94$) are typical of our numerical data.  
Although our plateau region appears to be quite long by inspection, a correlated fit requires a fit to a smaller span of the time extent so that the correlation matrix does not become too large to resolve with our current statistics ($\sim150$-$270$ configurations per ensemble).  Thus, we limit our plateau fits to $\sim10$-$15$ time slices.  In practice, this is not much of a limitation, since we fold our data in the time direction.  Typical statistical errors on the raw $B_K^{lat}$ lattice data are at the sub-percent level, with $1$-$2\%$ errors on the points with the lightest quark masses.

Autocorrelation errors were studied on the two longest runs on the coarse ($am_l/am_h = 0.007/0.05$) and fine ($am_l/am_h = 0.0031/0.031$) ensembles.  These errors are investigated by blocking the data before performing the single elimination jackknife estimate of the statistical error.  However, there is also a correction to the statistical error coming from the fact that the correlation matrix is not known perfectly, but is determined approximately from the data set for a given fit.  It has been shown in Ref.~\cite{Toussaint:2008ke}
that a jackknife fit with an estimate of the covariance matrix remade with each jackknife sample leads to a slight overestimate of the variance.  This correction to the statistical error is at the $\sim 5-10\%$ level for our data set, and tends to cancel the expected (and difficult to resolve) slight increase in the statistical errors due to autocorrelations.  Corrections to the statistical errors due to autocorrelations are at most a few percent.  Given the rather small total correction to the statistical error due to the combination of these effects, we do not adjust the errors ``by hand'' as input to later (chiral and continuum extrapolation) fits.  This correction to the statistical error due  is also a small fraction of our final total error, and can be neglected.

\section{Renormalization of the $\Delta S=2$ operator}
\label{sec:Renorm}

In this section we describe the calculation of the renormalization factor, $Z_{B_K}$, which is needed to match the lattice matrix element to the continuum.  We present the result renormalized in the $\bar{\textrm{MS}}$ scheme at 2 GeV.  We determine $Z_{B_K}$ using two independent methods:  lattice perturbation theory and the nonperturbative Rome-Southampton approach.  Although we use the nonperturbatively determined $Z_{B_K}$ to calculate our central value for $B_K$, the lattice perturbation theory calculation provides a valuable crosscheck on the nonperturbative renormalization and an indication of the size of the systematic uncertainty on the renormalization factor.

\subsection{Lattice perturbation theory calculation of $Z_{B_K}$}
\label{sec:LPT}

In this subsection, we use lattice perturbation theory to match our lattice calculation of $B_K$ to the $\overline{\textrm{MS}}$ scheme using naive dimensional regularization (NDR). 
Although naive lattice perturbation theory appears to converge slowly, two main causes of this have been identified in Ref.~\cite{Lepage:1992xa}.  The first is that the bare gauge coupling is a poor expansion parameter, and the second is that tadpole diagrams tend to have large coefficients.  If a renormalized coupling is used and one restricts oneself to quantities for which tadpole diagrams largely cancel, then lattice perturbation theory appears to converge as well as continuum perturbation theory.  The difficulties with large tadpole corrections are present even in chiral fermion formulations, where they are just as serious as in other formulations.  We address this issue here.

   In the case of domain-wall quarks, the domain-wall height $M_5$ is additively renormalized, and large tadpole corrections can appear.  It has been shown that mean-field improvement is then necessary to get correct results from lattice perturbation theory for domain-wall quarks \cite{Aoki:1997xg,Aoki:1998vv,Shamir:DWF10}.  Our calculations do not suffer from large tadpole corrections because we use HYP-smeared domain-wall quarks in our simulations.
This HYP-smearing smoothes the gauge fields, and has the effect of dramatically reducing the tadpole contributions in lattice perturbation theory for our simulation parameters.  Thus, the difference between naive and mean-field improved lattice perturbation theory in our renormalization of $B_K$ is small.  Even so, we adopt the correct mean-field improvement in all results presented here.

\begin{figure}
\begin{center}
\includegraphics[width=4in]{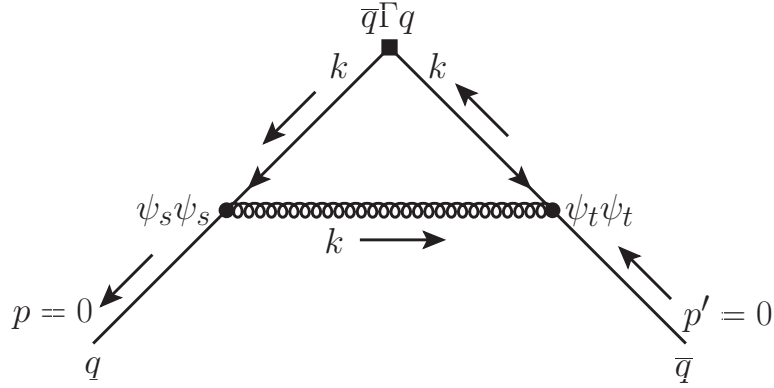}
\caption{Vertex diagram for the correction to bilinear operators in lattice perturbation theory.}
\label{fig:lat_pt}
\end{center}
\end{figure}

The renormalization factor matching the lattice calculation of $B_K$ to the $\overline{\textrm{MS}}$ scheme can be written as \cite{Aoki:2002iq}
\bea\label{eq:ZBKPT} Z_{B_K}(\mu a) = \frac{(1-w_0^2)^{-2}Z_w^{-2}Z_+(\mu a)}{(1-w_0^2)^{-2}Z^{-2}_w Z_A^2(\mu a)} = \frac{Z_+(\mu a)}{Z_A^2(\mu a)},
\eea
where $Z_+$ is the renormalization factor for the operator $O_K^{\Delta S=2}$, $Z_A$ renormalizes the axial current, $w_0=1-M_5$, and $Z_w$ is the quantum correction to the normalization factor $1-w_0^2$.  It is useful to define 
$Z_{B_K}$ in this way, since the tadpole and self-energy corrections cancel.  The renormalization factor contains the running of the operator from the lattice scale $a^{-1}$ to the continuum scale $\mu$.  The relevant Feynman diagram for our particular lattice calculation is shown in Fig.~\ref{fig:lat_pt} and the necessary Feynman rules are given in Appendix \ref{app:Lat_PT}.  In the $\overline{\textrm{MS}}$ scheme with NDR, we have \cite{Aoki:2002iq}
\bea\label{eq:ZBK} Z_{B_K}^{\bar{\textrm{MS}},\textrm{NDR}}(\mu a) = 1+\frac{\alpha_{\overline{\textrm{MS}}}(q^*)}{4\pi}\left[-4 \ln(\mu a) -\frac{11}{3} + 2\ln \pi^2 +\frac{2}{3}(16\pi^2)(I_{S}-I_V) \right],
\eea
with $I_{S,V}$ defined in Eq.~(\ref{eq:Igamma}).

Given the cancellation of tadpoles in Eq.~(\ref{eq:ZBKPT}), the only effect of mean-field improvement in the one-loop renormalization factor $Z_{B_K}$ is to shift the domain-wall height $M_5 \to M_5^{\rm MF}=M_5-4(1-u_0)$, where $u_0$ is the fourth-root of the plaquette.  This shift would be appreciable if not for the HYP smearing of the domain wall quarks, since $u_0\approx0.87$ on the MILC coarse and fine lattices.  However, for our calculation it is appropriate to take the HYP-smeared plaquette in the mean-field improvement factor, and this leads to $u_0^{\rm MF,coarse}=0.984$ and $u_0^{\rm MF,fine}=0.987$ and a significantly smaller shift in $M_5^{\rm MF}$.  The final result for $Z_{B_K}$ decreases by only about one percent (at both coarse and fine lattice spacings) after adopting the mean-field improvement. 

We adopt the Brodsky, Lepage, and Mackenzie (BLM) scheme for setting the scale in the coupling $\alpha_{\overline{\textrm{MS}}}(q^*)$~\cite{Brodsky:1982gc}.\footnote{We actually compute $\alpha_{V}(q^*)$, the strong coupling constant in the $V$ scheme, and exploit the fact that $\alpha_{\overline{\textrm{MS}}}=\alpha_V$ to the order we are working. The scales used to determine the coupling in each scheme are related by $\ln[(aq^*_{V})^{2}] = \ln[(aq^*_{\overline{\textrm{MS}}})^{2}]+5/3$ in the first order BLM prescription where only the first log moment is required.  The $V$ scheme is defined with respect to the heavy-quark potential \cite{Lepage:1992xa, Brodsky:1982gc}.}  The BLM prescription for obtaining $q^*$ is
\begin{equation}\label{eq:q*}
\ln[(aq^*)^{2}] = \frac{\int  d^4k \,  f(k) \ln(k^2)} { \int  d^4k  \, f(k)},
\end{equation}
where $f(k)$ is the one-loop integrand, and the numerator is the first log moment.
Note that throughout this section and in Appendix~\ref{app:Lat_PT}, all momentum integrals run over $-\pi\le k_\mu \le \pi$, with $k_\mu$ in lattice units.
This prescription for computing $q^*$ is well-defined for finite lattice integrals.  In the case of $B_K$, however, where there is an anomalous dimension, the BLM prescription needs to be modified.  We follow the prescription introduced by C. Bernard \emph{et al.} \cite{Bernard:2002pc}, and discussed in detail by DeGrand \cite{DeGrand:2002va}.  A generic integral evaluated in the $\overline{\textrm{MS}}$ scheme will take the following form:
\bea I_{\overline{\textrm{MS}}}&=&16\pi^2\int \frac{d^{2\omega}k}{(2\pi)^{2\omega}}(\mu^2)^{2\omega}\frac{1}{k^2(k^2+\lambda^2)}(A+B\epsilon) \nonumber \\ && = A\left\{\frac{1}{\epsilon}-\gamma_E+\log(4\pi)\right\}+A\log \frac{\mu^2}{\lambda^2} + A +B,
\eea
where $2\omega=4-2\epsilon$ is the dimension of the integral and where the term in curly brackets is discarded to give a finite integral, $I^F_{\overline{\textrm{MS}}}$.  The log moment of the divergent part of the one-loop expression must be handled with care.  The log moments of the finite lattice integrals $I_V$ and $I_S$ are straightforward to evaluate using Eq.~(\ref{eq:q*}), and are denoted $I_V^*$ and $I_S^*$ in Table~\ref{tab:ZBK_LPT}.  However, we need the log moment corresponding to the entire term in brackets in Eq.~(\ref{eq:ZBK}), including the anomalous dimension.  The log moment corresponding to the anomalous dimension [the first three terms in brackets in Eq.~(\ref{eq:ZBK})] can be defined as the log moment of the following finite integral \cite{DeGrand:2002va}
\bea\label{eq:IF} I^F_{\overline{\textrm{MS}}}=AJ_1+BJ_2,
\eea
where the $F$ stands for finite, and
\bea J_1&=&16\pi^2\int \frac{d^4k}{(2\pi)^4}\left[\frac{1-\theta(\pi^2-k^2)}{(k^2)^2}-\frac{1}{(k^2+\mu^2)^2}\right], \\
J_2 &=&16\pi^2\int \frac{d^4k}{(2\pi)^4}\left[\frac{1}{k^2(k^2+\mu^2)}-\frac{1}{(k^2+\mu^2)^2}\right],
\eea
where $\mu$ is the $\overline{\textrm{MS}}$ scale (in lattice units), and $\theta$ is the Heaviside step function.  The values of $A$ and $B$ for $Z_{B_K}$ in the $\overline{\textrm{MS}}$, NDR scheme are $-2$ and $-5/3$, respectively.  The log moment of the one-loop expression for $Z_{B_K}$ can then be used to compute $q^*$ using
\bea  \ln (aq^{*})^2 = \frac{(I^F_{\overline{\textrm{MS}}})^*+\frac{2}{3}(16\pi^2)(I_S^*-I_V^*)}{-4\ln(\mu a)-\frac{11}{3}+2\ln \pi^2 + \frac{2}{3}(16\pi^2)(I_S-I_V)},
\eea
where $(I^F_{\overline{\textrm{MS}}})^*$ signifies that the first log moment is taken in the momentum integrals appearing in Eq.~(\ref{eq:IF}).  The computed values for the integrals $I_V$ and $I_S$, as well as their first log moments, are given in Table~\ref{tab:ZBK_LPT}.  The resulting $q^*$'s and the final values for $Z^{\overline{\textrm{MS}},\textrm{NDR}}_{B_K}$ are also given.  All integrals were evaluated numerically using the Mathematica package \cite{Wolfram:1988}, and results in Ref.~\cite{Aoki:2002iq} using the same action but without HYP-smearing were reproduced.

\begin{table}
\begin{center}
\caption{Computed values of $Z_{B_K}$ in the BLM
prescription.  The first column labels
the approximate lattice
spacing in fm.  The second column is the numerical evaluation of the integral $I_V$, and the third is that of the integral $I_S$.  The fourth and fifth columns are the first moments of $I_V$ and $I_S$, respectively. The sixth column is $aq^*_{BLM}$, and the seventh column is
$Z_{B_K}^{\overline{\textrm{MS}},\textrm{NDR}}(2 \textrm{GeV})$.  Errors from numerical approximation of the integrals are no more than one digit in the last displayed decimal.}
 \label{tab:ZBK_LPT}
\begin{tabular}{ccccccc}
\hline \hline
 $a$ (fm) & $I_V$ & $I_S$ & $I_V^*$ & $I_S^*$ & $aq^*_{BLM}$ \ \ & $Z_{B_K}^{\overline{\textrm{MS}},\textrm{NDR}}(2 \textrm{GeV})$ \\
  \hline
  $0.12$  & 0.0158 \ \ & -0.0161 \ \ & 0.0336 \ \ & -0.0150 \ \ & \ \ 1.56 \ \ & 0.909\\
  $0.09$  & 0.0158 \ \ & -0.0155 & 0.0336 & -0.0154 & 1.42 & 0.955\\
  \hline\hline
\end{tabular}
\end{center}
\end{table}

\subsection{Nonperturbative determination of $Z_{B_K}$} 
\label{sec:NPR}

\subsubsection{Rome-Southampton method}
\label{sec:RomeSH}

We compute the renormalization coefficient for $B_K$ nonperturbatively in the RI/MOM scheme devised by the Rome-Southampton group~\cite{Martinelli:1994ty}.  In this scheme, the simple renormalization condition is that the renormalized $n$-point functions in Landau gauge are equal to their tree-level values.  Because the RI/MOM scheme is regularization-invariant, it is useful for both perturbative or non-perturbative techniques.  Thus it is well-suited to lattice QCD simulations.  Once $Z_{B_K}$ has been determined nonperturbatively in the RI/MOM scheme, it can easily be converted to the $\bar{\textrm{MS}}$ scheme and run to the scale $\mu$ = 2 GeV using continuum perturbation theory.

The Rome-Southampton nonperturbative renormalization technique has already been successfully applied to lattice QCD calculations of $B_K$ with domain wall valence and sea quarks by the RBC and UKQCD Collaborations in Refs.~\cite{Aoki:2005ga,Aoki:2007xm}.  We can determine the renormalization factor $Z_{B_K}$ in the same simple manner for our mixed-action lattice QCD simulations because the properties of the mixing coefficients are largely determined by the symmetries of the valence sector.  In particular, errors of $O(a)$ are suppressed by $\sim e^{-\alpha L_s}$.  Furthermore, mixings between the desired $B_K$ four-fermion operator and other operators of incorrect chirality are suppressed due to the approximate chiral symmetry, as we show in Appendix~\ref{app:NPRmix}.

Our primary goal is to determine the renormalization coefficient of the four-quark operator given in Eq.~(\ref{eq:OK}):
\begin{equation*}
	 \CO_K^{\Delta S = 2}\equiv \CO_{VV+AA} = [\bar{s} \gamma_\mu (1 - \gamma_5)d][\bar{s}  \gamma_\mu (1 - \gamma_5)d] ,
\end{equation*}
where we now show explicitly the chirality of the operator. Because chiral symmetry is slightly broken in our simulations, however, this operator can mix with other $\Delta S=2$ operators of different chiralities:
\begin{eqnarray}\label{eq:wrongchiral}
	\CO_{VV-AA} & = &
	[\bar{s} \gamma_\mu (1 - 	\gamma_5)d]
	[\bar{s}\gamma_\mu (1 + \gamma_5)d] ,\\
	\CO_{SS\pm PP} & = &
	[\bar{s}  (1 - \gamma_5)d]
	[\bar{s} (1 \mp \gamma_5)d] ,\\
	\CO_{TT} & = &
	[\bar{s} \sigma_{\mu\nu} (1 - \gamma_5)d]
	[\bar{s} \sigma_{\mu\nu} (1 - \gamma_5)d] .
\end{eqnarray}
Thus the renormalized $B_K$ operator in principle receives contributions from all of the above operators, and is given in terms of bare lattice operators by
\begin{equation}
	\CO_K^\textrm{ren} = \sum_i Z_{VV+AA,i}\, \CO^{0}_i\, ,
\end{equation}
where $i\in\{VV+AA,VV-AA,SS+PP,SS-PP,TT\}$.  The theoretical arguments of Ref.~\cite{Aoki:2007xm} suggest that the wrong chirality mixing coefficients should be of $\CO[(a m_{\rm res})^2]$, and our data is consistent with this expectation within statistical errors. Thus, despite the fact that chiral perturbation theory predicts the corresponding $B$ parameters of these operators to diverge in the chiral limit~\cite{Becirevic:2004fw,Aoki:2005ga}, their contributions to $B_K$ are negligible.

In order to calculate $Z_{B_K}$ via the Rome-Southampton approach~\cite{Martinelli:1994ty}, we first compute the $5\times5$ matrix
\begin{equation}\label{eq:mixing}
	M_{ij} =  \hat P_j [\Gamma_i^{\rm latt}] \, ,
\end{equation}
where $\Gamma_i^{\rm latt}$ is the amputated four-point Green's function in momentum space, $i,j\in\{VV+AA,VV-AA,SS+PP,SS-PP,TT\}$, and the projector $\hat P_j$ selects out the component with chirality $j$.  The details of this procedure are described thoroughly in Appendix~B of Ref.~\cite{Aoki:2005ga}.
We also compute the tree-level value of this matrix by setting all of the momentum-space propagators in the amputated Green's functions equal to the identity:
\begin{equation}\label{eq:mixing_tree}
	M^{\rm tree}_{ij} =  \hat P_j [\Gamma_i^{\rm tree}] \, .
\end{equation}
We then impose the RI/MOM renormalization condition,
\begin{equation}\label{eq:RIMOM_cond}
	\frac{Z_{ij}}{Z_q^2} M_{jk} = M^{\rm tree}_{ik}\, ,
\end{equation}
where $Z_q$ is the quark wavefunction renormalization factor, in order to obtain the quantity
\begin{equation}
	\frac{Z_{ij}}{Z_q^2} = M^{\rm tree}_{ik} M_{jk}^{-1}\, .
\end{equation}
The renormalization coefficients for the various four-fermion operators are then given by 
\begin{equation}
	\frac{Z_{ij}}{Z_A^2} = \frac{Z_{ij}}{Z_q^2} \left(\frac{Z_q}{Z_A}\right)^2 \, ,
\label{eq:Zij}
\end{equation}
where $Z_A$ is the renormalization factor for the axial current.
For example, the dominant contribution to the $B_K$ lattice operator renormalization comes from the diagonal mixing coefficient:
\begin{equation}\label{eq:ZBK_def}
	Z_{B_K} 
	\equiv \frac{Z_{VV+AA,VV+AA}}{Z_A^2}
	= \frac{Z_{VV+AA,VV+AA}}{Z_q^2}
	\left(\frac{Z_q}{Z_A}\right)^2 \, .
\end{equation}

In order to determine the four-fermion operator mixing coefficients using Eq.~(\ref{eq:Zij}), we also need the ratio $Z_q/Z_A$.  Fortunately, the renormalization factors for the quark bilinears can also be calculated in a simple manner using the Rome-Southampton method.  The renormalization coefficients relate the bare and renormalized quark bilinear operators in the following manner:
\begin{equation}\label{eq:ren_con}
	[\bar u \Gamma d]_{\rm ren} = Z_{\Gamma} [\bar u \Gamma d]_0\ .
\end{equation}
In order to determine $Z_\Gamma$, we first compute the bare Green's functions between off-shell quarks in momentum space.  We then amputate the Green's function and separately project out the components with each chirality to obtain the bare vertex amplitudes:
\begin{eqnarray}\label{eq:Lamba_bilins}
	\Lambda_S & = & \frac{1}{12} \Tr[G^{\rm amp}_{1} 1]\ ,\\
	\Lambda_P & = & \frac{1}{12} \Tr[G^{\rm amp}_{\gamma_5} 
	\gamma_5]\ ,\\
	\Lambda_V & = & \frac{1}{48} \Tr\left[\sum_\mu G^{\rm amp}_{\gamma_\mu} \gamma_\mu\right]\ ,\\
	\Lambda_A & = & \frac{1}{48} \Tr\left[\sum_\mu 
	G^{\rm amp}_{\gamma_\mu\gamma_5}\gamma_5 \gamma_\mu \right]\ ,\\
	\Lambda_T & = & \frac{1}{72} \Tr\left[\sum_{\mu<\nu} 
	G^{\rm amp}_{\sigma_{\mu\nu}} \sigma_{\nu\mu}\right]\ .
\end{eqnarray}
Finally, we impose the RI/MOM renormalization condition
\begin{equation}\label{eq:bil_ren_cond}
	\Lambda_{i,\rm ren} = \frac{Z_i}{Z_q} \Lambda_i = 1\, .
\end{equation}
The renormalization coefficients for the quark bilinears are then given by 
\begin{equation}
	\frac{Z_i}{Z_q} = \frac{1}{\Lambda_i }\, .
\end{equation}

In the RI/MOM prescription, the four-fermion operator renormalization coefficients are given as functions of the momenta in the amputated Green's functions used to determine $Z_{ij}/Z_q^2$, which are chosen to be identical for all four quarks in our computation of $Z_{B_K}$.  We therefore need to extract $Z_{B_K}$ at a sufficiently high momentum that hadronic effects are negligible and the momentum-dependence can be described by perturbation theory.  We cannot use too high a momentum, however, or lattice discretization errors will become large.  Thus use of the Rome-Southampton technique requires the existence of a momentum window in which both hadronic effects and discretization errors can be neglected:
\begin{equation}\label{eq:NPR_window}
	\Lambda_{\rm QCD} \ll p \ll a^{-1}\ .
\end{equation}
In practice, however, we need to work in the region
	$(ap)^2 \gtwid 1$
in order to avoid large violations of chiral symmetry which we observe at low momenta.  Fortunately, discretization effects in the region of interest, $p \approx$ 2 GeV, are generally rather small and can be taken into account by a simple linear fit in $(ap)^2$, as discussed in Ref.~\cite{Blum:2001sr}.  This is the approach that we take in the calculation of $Z_{B_K}$ in Sec.~\ref{sec:ZBK_calc}.

\subsubsection{Chiral symmetry breaking and $\Lambda_A - \Lambda_V$}
\label{sec:LA_LV}

Although the calculation of $Z_{B_K}$ only requires the renormalization factor for the axial current, $Z_A$, 
the vector and axial-vector current renormalization factors should be equal in the chiral limit for sufficiently large momenta due to chiral symmetry:
\begin{equation}\label{eq:WardZ}
	Z_A = Z_V\ ,
\end{equation}
or equivalently,
\begin{equation}\label{eq:WardLambda}
	\Lambda_A = \Lambda_V\ .
\end{equation}
Thus we can take the average of these two quantities in order to reduce the statistical error in $Z_q/Z_A$ using the relationship
\begin{equation}\label{eq:ZqoverZa}
	\Lambda_A = \frac{1}{2}(\Lambda_A + \Lambda_V)\ .
\end{equation}

\begin{figure}
\begin{center}
\includegraphics[width=4in,angle=0]{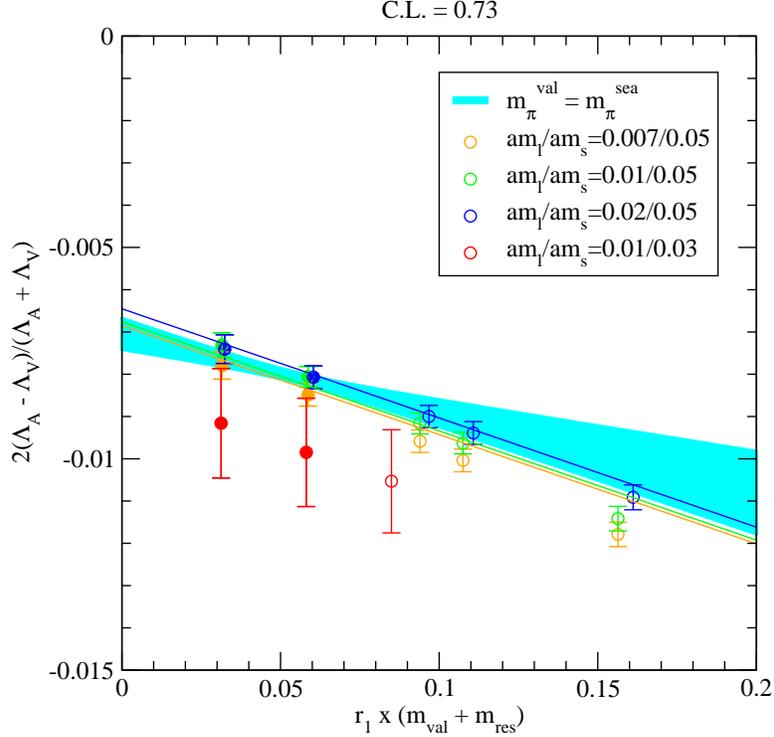}
\caption{Chiral extrapolation of $2(\Lambda_A - \Lambda_V)/(\Lambda_A + \Lambda_V)$ on the coarse lattice at $(ap)^2 = 1.468$ using a linear function in $m_x$ and $m_l$.   Although only the data points with filled symbols were used in the fit, the fit line does a reasonable job of describing the heavier data points that were not included.  The cyan error band shows the extrapolation/interpolation for points where the domain-wall pion mass is tuned to equal the lightest (taste pseudoscalar) staggered pion mass.}
\label{fig:LA_LV_extrap}
\end{center}
\end{figure}

\begin{figure}
\begin{center}
\includegraphics[width=4in,angle=0]{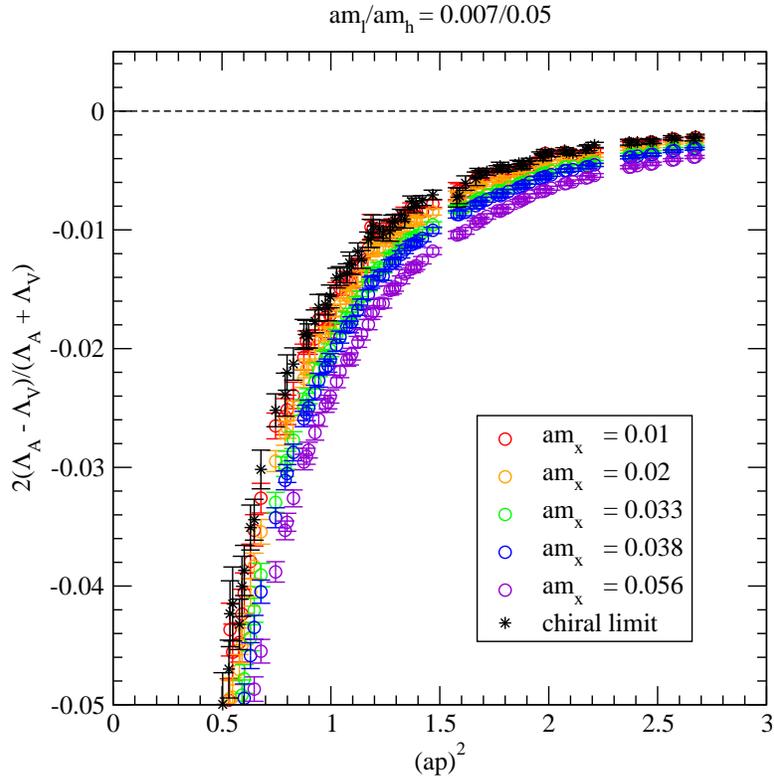}
\caption{The quantity $2(\Lambda_A - \Lambda_V)/(\Lambda_A + \Lambda_V)$ versus $(ap)^2$ on the $am_l/am_h = 0.007/0.05$ coarse ensemble for several valence quark masses and in the chiral limit $a m_x = a m_l = 0$.}
\label{fig:LA_LV_vs_ap2}
\end{center}
\end{figure}

In practice, however, $\Lambda_A\ne \Lambda_V$ in the chiral limit for any value of the momentum in our nonperturbative determination.  Figure~\ref{fig:LA_LV_extrap} shows the extrapolation of the quantity $2(\Lambda_A - \Lambda_V)/(\Lambda_A + \Lambda_V)$ on the coarse lattice to the chiral limit at $p \approx \textrm{2 GeV}$ using a function that is linear in both the valence and sea quark masses.  This quantity provides an indication of the amount of chiral symmetry breaking in the computation.
At nonzero quark mass,  $2(\Lambda_A - \Lambda_V)/(\Lambda_A + \Lambda_V)$ can be as large as $\sim1\%$ in the momentum region $(ap)^2 \gtapprox 1$ that we are using to extract $Z_{B_K}$.   The difference between $\Lambda_A$ and $\Lambda_V$ decreases towards the chiral limit, as is expected, but never becomes consistent with zero.  Figure~\ref{fig:LA_LV_vs_ap2} shows $2(\Lambda_A - \Lambda_V)/(\Lambda_A + \Lambda_V)$ versus $(ap)^2$ on the $am_l / am_h = 0.007/0.05$ coarse ensemble for the five available valence quark masses and in the chiral limit.  Again, it decreases in magnitude as expected at larger momenta, but is never zero.  We observe similar behavior on the fine lattice.  This persistent difference between $\Lambda_A$ and $\Lambda_V$ has also been observed and studied in detail by the RBC and UKQCD collaborations \cite{Blum:2001sr,Christ:2005xh,Aoki:2007xm}, and can be attributed to several sources.  

The first source is explicit chiral symmetry breaking due to the nonzero quark masses used in simulations~\cite{Aoki:2007xm}.  Use of the operator product expansion shows that  $\Lambda_A$ and $\Lambda_V$ can receive contributions proportional to 
\begin{equation}\label{eq:chiral_sym_br}
	\frac{m_q^2}{p^2}\, , \qquad \frac{m_q\langle \bar qq\rangle}{p^4}\, ,
\end{equation}
at lowest order in $1/p^2$~\cite{Blum:2001sr}.  Because these operators are proportional to $m_q$, they explicitly break chiral symmetry and need not contribute equally to $\Lambda_A$ and $\Lambda_V$.  The contribution of the operators in Eq.~(\ref{eq:chiral_sym_br}) can be seen clearly in the data.  As Fig.~\ref{fig:LA_LV_vs_ap2} shows, the difference between $\Lambda_A$ and $\Lambda_V$ increases rapidly as the momentum approaches zero and decreases slowly as the momentum becomes larger than $\approx$ 2 GeV.
The contributions from the operators in Eq.~(\ref{eq:chiral_sym_br}) can be removed by extrapolating to the chiral limit at fixed momentum.   
Figs.~\ref{fig:LA_LV_extrap} and~\ref{fig:LA_LV_vs_ap2} show that, although this procedure does indeed reduce the splitting between $\Lambda_A$ and $\Lambda_V$, it does not eliminate the difference.  Thus there must be additional sources of chiral symmetry breaking, which we now discuss.

The next source is chiral symmetry breaking due to the use of a finite $L_s$.  Theoretical arguments suggest, however, that this would lead to errors that are much smaller, of $\CO((a m_\textrm{res})^2) \sim 10^{-6}$ in our numerical simulations~\cite{Christ:2005xh,Aoki:2007xm}.  This would produce a negligible difference between $\Lambda_A$ and $\Lambda_V$, and cannot account for the size of the difference that we observe in the data.

\begin{figure}
\begin{center}
\includegraphics[width=4in,angle=0]{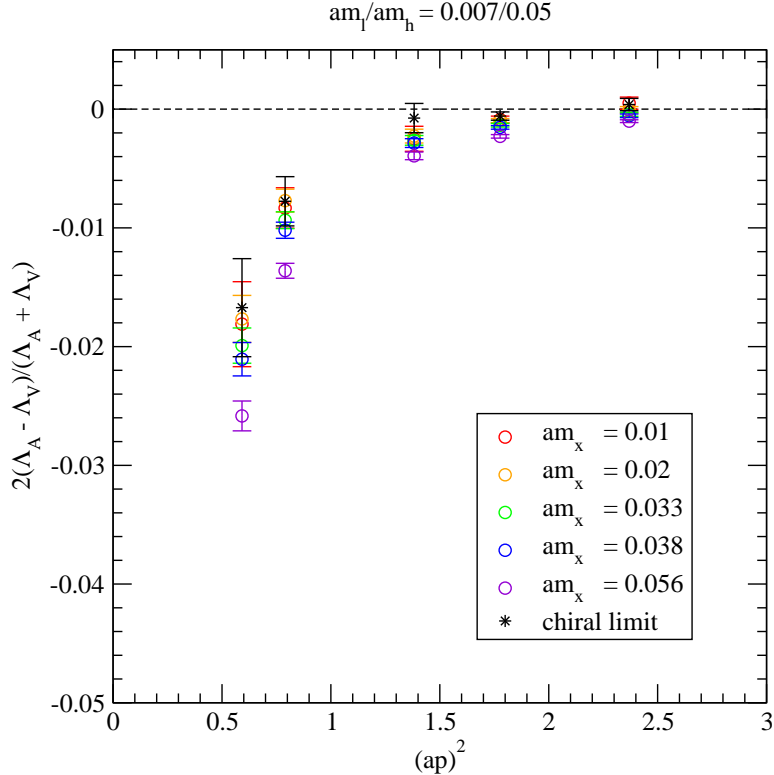}
\caption{The quantity $2(\Lambda_A - \Lambda_V)/(\Lambda_A + \Lambda_V)$ versus $(ap)^2$ computed using non-exceptional momenta on the $am_l/am_h = 0.007/0.05$ coarse ensemble for several valence quark masses and in the chiral limit $a m_x = a m_l = 0$.}
\label{fig:LA_LV_vs_ap2_nonex}
\end{center}
\end{figure}

A more significant source of chiral symmetry breaking that does not vanish in the chiral limit is the choice of kinematics used to compute both $Z_{ij}/Z_q^2$ and $\Lambda_i$.  As in the standard RI/MOM prescription, we are using ``exceptional momenta" configurations in which there is no momentum transferred between the initial and final states, or more precisely
\be
	p_i = p_f \equiv p \ ,
\ee
where $p_{i,f}$ are the momenta of the initial and final states.
It was shown in Ref.~\cite{Aoki:2007xm} that this, unfortunately, leads to contributions to $\Lambda_A - \Lambda_V$ of the form 
\begin{equation}\label{eq:qqqqoverp2}
	\frac{\langle \bar q q \bar q q\rangle}{p^2}\, .
\end{equation}
Because this operator is not proportional to the quark mass, it does not vanish in the chiral limit at fixed momentum.  This contribution can be removed by performing the nonperturbative renormalization calculation at non-exceptional kinematics, in which the sum of any subset of external momenta is nonzero. In this case we have
\be
	p_i^2 = p_f^2 = (p_i - p_f)^2 \equiv p^2\ ,
\ee
but $p_i \ne p_f$.
In order to check that this is indeed the source of the difference between $\Lambda_A$ and $\Lambda_V$ in our data, we have also computed $2(\Lambda_A - \Lambda_V)/(\Lambda_A + \Lambda_V)$ at non-exceptional kinematics;  the results are shown in Fig.~\ref{fig:LA_LV_vs_ap2_nonex}.  Although the statistical errors are not as small, the results in the chiral limit are consistent with zero for sufficiently large values of $(ap)^2$.

For the calculation of $Z_{B_K}$ in this work, we use exceptional kinematics, despite the resulting chiral symmetry breaking.  This is because the continuum perturbation theory needed to convert the result from the RI/MOM scheme to the $\bar{\textrm{MS}}$ scheme has not yet been calculated for non-exceptional kinematics.\footnote{The expressions needed to convert the quark bilinears from the RI/MOM scheme to the $\bar{\textrm{MS}}$ scheme have recently been computed to one-loop order for non-exceptional kinematics by Sturm \etal\ in Ref.~\cite{Sturm:2009kb}.} We therefore include the difference between $\Lambda_A$ and $\Lambda_V$ as a source of systematic uncertainty in  $Z_{B_K}$.

\subsubsection{Nonperturbative renormalization factor calculation}
\label{sec:ZBK_calc}

We now present the nonperturbative determination of $Z_{B_K}$, which we compute from the quantity
\begin{equation}
	\frac{Z_{VV+AA,VV+AA}}{Z_q^2}\left(\frac{\Lambda_A + \Lambda_V}{2}\right)^2
\end{equation}
using the method of Rome-Southampton.  Table~\ref{tab:NPR-confs} shows the parameters used in generating the NPR lattice data set.  We have several valence and sea quark mass combinations on both the coarse and fine lattices in order to allow an extrapolation of $Z_{B_K}$ to the chiral limit.  For those ensembles that are listed as ``blocked'' in the table, we computed $Z_{B_K}$ on every sixth trajectory and blocked the data in groups of four in order to reduce autocorrelations.  On those ensembles for which the data was not blocked, we computed $Z_{B_K}$ only on every 24$^{th}$ trajectory.

Because we must extrapolate $Z_{B_K}$ to the chiral limit in both the valence and sea quark sectors, on the coarse lattice we have generated data on three ensembles at the nominal strange quark mass ($am_h = 0.05$) and on one ensemble with a lighter than physical strange sea quark mass ($am_h = 0.03$).  At our current level of statistics the results for $Z_{B_K}$ on the $am_l/am_h = 0.01/0.05$ and $am_l/am_h = 0.01/0.03$ coarse ensembles are consistent.  Because we do not observe any dependence on the strange sea quark mass in our data, we fit our data assuming only a dependence on the light sea quark mass to determine the central value of $Z_{B_K}$.  We use an alternative fit that includes strange sea quark mass dependence to estimate the systematic error associated with extrapolating the strange sea quark mass to the chiral limit.

\begin{table}
\caption{Lattice QCD data used in the nonperturbative renormalization of $B_K$. For those configurations that were blocked, $Z_{B_K}$ was computed on every 6$^{th}$ configuration and blocked in groups of 4.  For those configurations that were not blocked, $Z_{B_K}$ was computed on every 24$^{th}$ configuration.}
\begin{center}
\begin{tabular}{llcccc}
\hline\hline
$a$(fm) & $\left(\frac{L}{a}\right)^3\times \frac{T}{a}$ & $am_l/am_h$ & $am_x$ & $N_{\rm conf.}$ & blocked?\\
\hline
0.09 & $28^3\times 96$ \quad & $0.0062/0.031$ & $0.0119,0.0171,0.0287,0.04$ & 387 & no\\
0.09 & $28^3\times 96$ & $0.0093/0.031$ & $0.0287$ & 251 & no\\
0.09 & $28^3\times 96$ & $0.0124/0.031$ & $0.0287$ & 381 & no\\
\hline
0.12 & $20^3\times 64$ & $0.007/0.05$ & $0.01,0.02,0.033,0.038,0.056$ & 836 & yes\\
0.12 & $20^3\times 64$ & $0.01/0.05$ & $0.01,0.02,0.033,0.038,0.056$ & 540 & yes\\
0.12 & $20^3\times 64$ & $0.02/0.05$ & $0.01,0.02,0.033,0.038,0.056$ & 484 & yes\\
0.12 & $20^3\times 64$ & $0.01/0.03$ & $0.01,0.02,0.03$ & 81 & no\\
\hline\hline
\end{tabular}
\end{center}
\label{tab:NPR-confs}
\end{table}

We first extrapolate $Z_{B_K}$ using a polynomial function in the valence and sea quark masses:
\begin{equation}\label{eq:fit_funcs}
	f_{n_{\rm sea},n_{\rm val}}(\chi_{\rm val}, \chi_{\rm sea};p^2)
	= Z_{B_K}^\textrm{RI/MOM}(p^2)
	+ \sum_{i=1}^{n_{\rm sea}} C_{i,{\rm sea}}(p^2) \chi_{\rm sea}^i
	+ \sum_{i=1}^{n_{\rm val}} C_{i,{\rm val}}(p^2) \chi_{\rm val}^i\ ,
\end{equation}
where
\begin{eqnarray}\label{eq:chi}
	\chi_{\rm sea} & =&  \frac{2\mu_{\rm stag}}{(4\pi f_\pi)^2}
	(2 m_l)\ , \\
	\chi_{\rm val} & = & \frac{2\mu_{\rm dw}}{(4\pi f_\pi)^2}
	[2 (m_x + m_{\rm res})]\ ,
\end{eqnarray}
are dimensionless ratios $< 1$, and the parameters $\mu_{\rm stag}$ and $\mu_{\rm dw}$ are obtained from tree-level MA$\chi$PT fits to the pseudoscalar meson masses.  In order to determine the preferred fit ansatz, we independently increase $n_{\rm sea}$ and $n_{\rm val}$ until the correlated confidence level of the fit no longer increases significantly.  We find that this occurs when $n_{\rm sea} = 1$ and $n_{\rm val} = 2$.  We use the fits with additional terms to estimate the systematic uncertainty due to the choice of fit function.

\begin{figure}
\begin{center}
\includegraphics[width=4in]{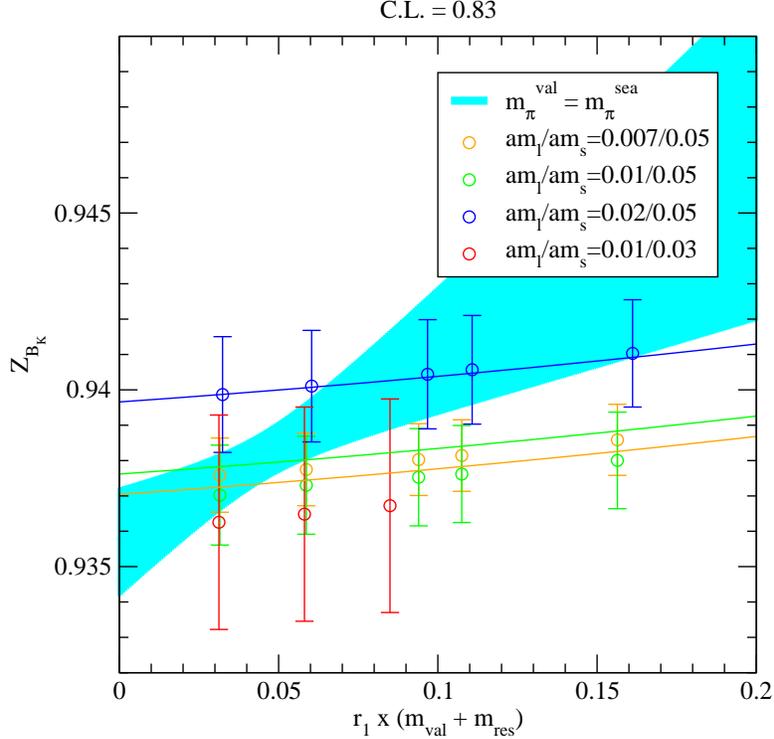}
\caption{Chiral extrapolation of $Z_{B_K}^{\textrm{RI/MOM}}$ on the coarse lattice at $(ap)^2 = 1.468$.  
Note that the fit lines for the $am_l/am_h = 0.01/0.05$ and $am_l/am_h = 0.01/0.03$ ensembles are identical because we have not included any strange sea quark mass dependence in the fit function.  The cyan band shows the extrapolation along the trajectory $m_\pi^{val} = m_\pi^{sea}$.}
\label{fig:coarse_22fit}
\end{center}
\end{figure}

\begin{figure}
\begin{center}
\includegraphics[width=4in]{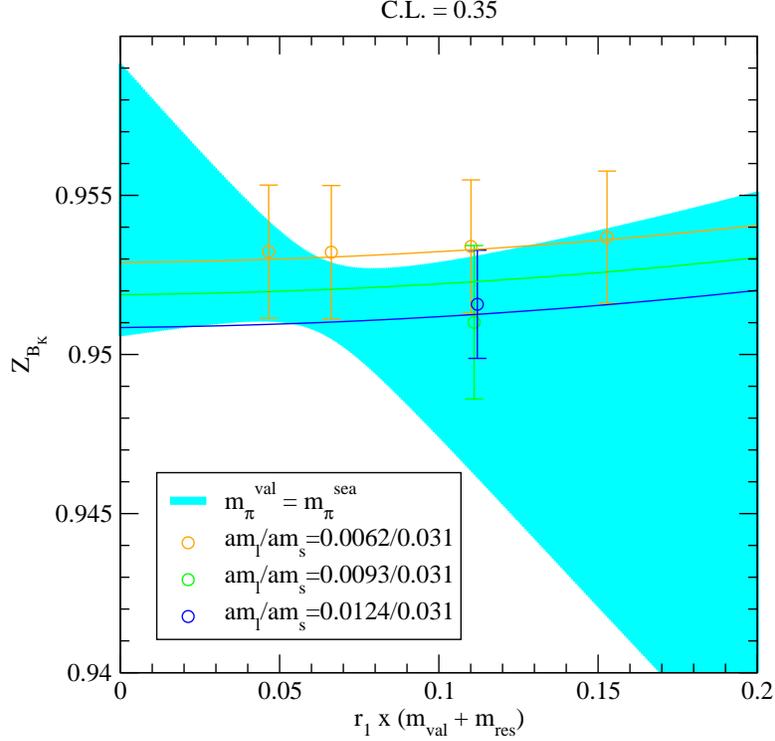}
\caption{Chiral extrapolation of $Z_{B_K}^{\textrm{RI/MOM}}(p^2)$ on the fine lattice at $(ap)^2 = 0.744$.  The cyan band shows the extrapolation along the trajectory $m_\pi^{val} = m_\pi^{sea}$.}
\label{fig:fine_12fit}
\end{center}
\end{figure}

Figures~\ref{fig:coarse_22fit} and \ref{fig:fine_12fit} show the chiral extrapolation of $Z_{B_K}$ on the coarse and fine lattices at $p^2 \approx (2 \textrm{ GeV})^2$ using a fit function linear in the light sea quark mass and quadratic in the valence quark mass.  Although this extrapolation is nominally at ``fixed $(ap)^2$'', this is not quite true.  This is because, although all of the coarse (or fine) MILC ensembles are generated with approximately the same lattice spacing, 
there are slight fluctuations in the lattice spacing from ensemble to ensemble.  Thus data on different ensembles with the same value of $(ap)^2$ do not correspond to precisely the same physical momentum.  We convert our data into $r_1$ units using the value of $r_1/a$ determined on each ensemble before performing the chiral extrapolation, so that everything is in the same units.  Fortunately, the variation in $r_1$ leads to only a slight variation in the momentum-squared, of $\sim 0.1\%$.  Because this is even smaller than the statistical errors in our data points, the resulting systematic error can be safely neglected.

\begin{figure}
\begin{center}
\includegraphics[width=4in]{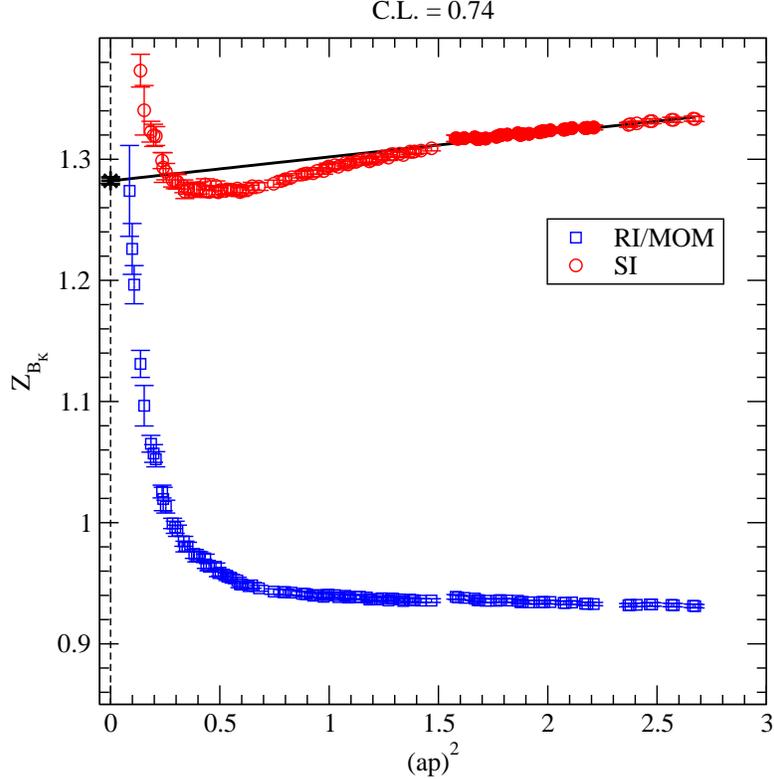}
\caption{Linear-in-$(ap)^2$ extrapolation of  $Z_{B_K}^\textrm{SI}$ on the coarse lattice to remove lattice artifacts using the fit range $1.5 < (ap)^2 < 2.3$. The data points used in the fit are denoted by the filled circles and the true value of $Z_{B_K}^\textrm{SI}$ obtained at $(ap)^2=0$ is denoted by the star.
}
\label{fig:coarse_SI_fit}
\end{center}
\end{figure}

\begin{figure}
\begin{center}
\includegraphics[width=4in]{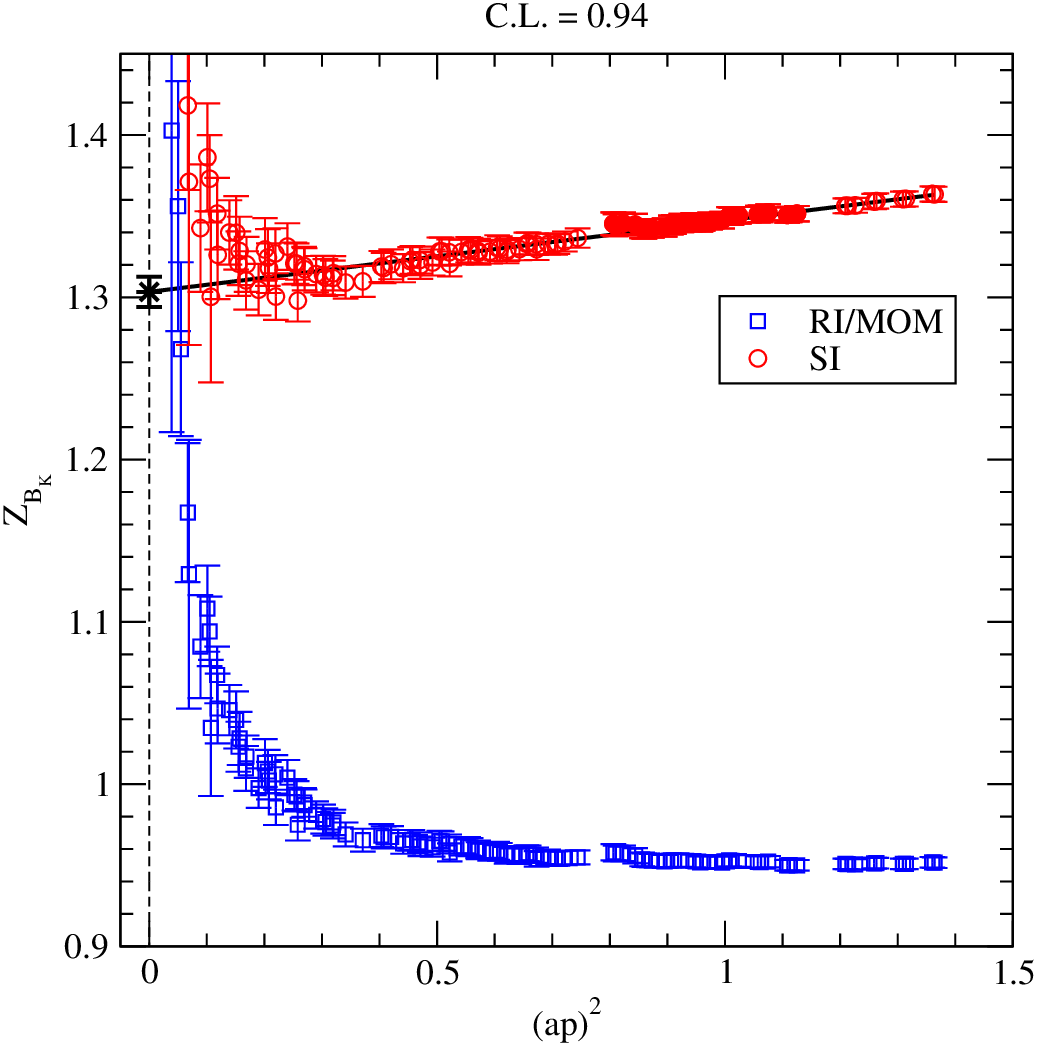}
\caption{Linear-in-$(ap)^2$ extrapolation of  $Z_{B_K}^\textrm{SI}$ on the fine lattice to remove lattice artifacts using the fit range $0.8 < (ap)^2 < 1.2$. The data points used in the fit are denoted by the filled circles and the true value of $Z_{B_K}^\textrm{SI}$ obtained at $(ap)^2=0$ is denoted by the star.}
\label{fig:fine_SI_fit}
\end{center}
\end{figure}

Next we attempt to remove discretization errors in $Z_{B_K}$ due to the fact that we are extracting $Z_{B_K}$ at momenta that are of $\CO(a^{-1})$.  Following the same procedure as in Ref.~\cite{Aoki:2007xm}, we use the continuum 1-loop perturbation theory expressions in Eqs.~(\ref{eq:Z_bk_SI})--(\ref{eq:J-MSbar}) to convert $Z_{B_K}^\textrm{RI/MOM}$ to $Z_{B_K}^\textrm{SI}$.  This is shown in Fig.~\ref{fig:coarse_SI_fit} (Fig.~\ref{fig:fine_SI_fit}) for the coarse (fine) lattice.  Although the quantity $Z_{B_K}^\textrm{SI}$ should be scale-invariant, we observe that $Z_{B_K}^\textrm{SI}$ in fact has an approximately linear dependence upon $(ap)^2$ in the region of interest.  We believe that this scale-dependence is primarily from lattice artifacts that can be removed by performing a linear extrapolation in $(ap)^2$ .  We therefore fit the data to the form 
\begin{equation}
	A + B (ap)^2\, 
\end{equation}
and interpret the intercept $A$ as the true $Z_{B_K}^\textrm{SI}$.  We extrapolate $Z_{B_K}^\textrm{SI}$ to its true value as shown in Figs.~\ref{fig:coarse_SI_fit} and \ref{fig:fine_SI_fit} using the momentum range $2 \textrm{ GeV} \ltapprox p \ltapprox 2.5 \textrm{ GeV}$.  This choice satisfies the criterion $p \gg \Lambda_\textrm{QCD}$ needed to avoid hadronic effects.  Specifically, for the coarse data, we fit within $1.5 < (ap)^2 < 2.3$ and for the fine data we fit within $0.8 < (ap)^2 < 1.2$.  Variations of these fit regions do not alter the final results significantly.  We obtain 
\begin{eqnarray*}\label{eq:ZBK_SI_results}
	Z_{B_K}^{\rm SI, coarse} & = &  1.2822(29)\ , \\
	Z_{B_K}^{\rm SI, fine} & = & 1.3033(93)\ ,
\end{eqnarray*}
where the errors are statistical only.  We note, however, that some of the scale-dependence in $Z_{B_K}^\textrm{SI}$ may in fact be due to the lack of higher-order terms in the matching factor.  We therefore account for this and other errors due to the truncation of perturbation theory in the systematic error budget for $Z_{B_K}$.

Finally, we convert $Z_{B_K}^\textrm{SI}$ to $Z_{B_K}^{\rm \bar{MS}}$ and run it to 2 GeV again using Eqs.~(\ref{eq:Z_bk_SI})--(\ref{eq:J-MSbar}).  We obtain
\begin{eqnarray*}\label{eq:ZBK_MSbar_results}
	Z_{B_K}^{\rm \bar{MS}, coarse} (\textrm{2 GeV}) & = &  0.9339(21)\ , \\
	Z_{B_K}^{\rm \bar{MS}, fine} (\textrm{2 GeV}) & = & 0.9493(68)\ ,
\end{eqnarray*}
where the errors are statistical only.  We estimate the systematic uncertainties in  $Z_{B_K}^{\rm \bar{MS}}(\textrm{2 GeV})$ later in Sec.~\ref{sec:NPR_err}.

\section{Determination of $B_K$}
\label{sec:ChPT}

In this section we describe the extrapolation of $B_K$ to physical quark masses and the continuum.  In Sec.~\ref{sec:MAChPT}, we present the expression for $B_K$ at next-to-leading order (NLO) in MA$\chi$PT and describe those features that are most relevant for the chiral-continuum extrapolation.  We then discuss the details of the chiral-continuum extrapolation procedure in Sec.~\ref{sec:ChPT_extrap}.

\subsection{$B_K$ at NLO in MA$\chi$PT}
\label{sec:MAChPT}

We first review the tree-level mass relations for light pseudoscalar mesons in MA$\chi$PT since they are useful in understanding the leading-order lattice-spacing contributions to mixed-action numerical simulations~\cite{Bar:2005tu}.  In a mixed-action theory one can have mesons composed of two sea quarks, two valence quarks, or one of each.   At tree-level in MA$\chi$PT, discretization effects lead to different additive shifts to the masses of the three types of mesons.   These mass-shifts are the only new parameters as compared to the continuum at this order, and their values have all been determined for our choice of mixed-action simulation parameters.  The tree-level mass-shifts on both the coarse and fine MILC lattices are given in Table~\ref{tab:mass_split}.  

\bigskip

\begin{table}
\caption{Tree-level mass-shifts on the coarse and fine MILC lattices for our choice of mixed-action simulation parameters.  The taste-singlet mass-splitting, $\Delta_I$, is independent of the valence sector~\cite{Aubin:2004fs}, while the residual quark mass, $m_\text{res}$, and mixed meson mass-splitting, $\Delta_\text{mix}$, both depend upon the choice of HYP-smearing~\cite{Aubin:2008wk}.  Errors shown are statistical only.}
\label{tab:mass_split}
\begin{tabular}{lccccc} \\ \hline\hline

\qquad & sea sector & \quad & valence sector & \quad & mixed sector \\[-0.5mm]

$a$(fm) & $r_1^2 a^2 \Delta_I$  &&  $r_1 m_\text{res}$ && $r_1^2 a^2 \Delta_\text{mix}$  \\[0.5mm] \hline

0.12 & 0.537(15) && 0.0044(1) && 0.207(16)  \\
0.09  & 0.206(17) && 0.0016(2) && 0.095(20) \\\hline\hline

\end{tabular}\end{table}

In the sea sector of the mixed action theory, each flavor of staggered quark comes in four species, or ``tastes"; consequently, each flavor of staggered pseudoscalar meson comes in sixteen tastes.  In the continuum, these tastes are identical and are related by an $SU(4)$ symmetry~\cite{Susskind:1976jm}.  At nonzero lattice spacing, however, discretization effects split the degeneracies among the sixteen pseudoscalar meson tastes~\cite{Lee:1999zx}:
\begin{equation}
	m^2_{ss',t} = \mu_\text{stag} (m_s + m_{s'}) + a^2 \Delta_t ,
\label{eq:m2_ss}
\end{equation}
where $s$ and $s'$ are the staggered quark flavors, $\mu_\text{stag}$ is a regularization-dependent continuum low-energy constant, and $\Delta_t$ is the mass-splitting of a pion with taste $t$.  At leading-order in  staggered $\chi$PT (S$\chi$PT), a residual $SO(4)$ taste symmetry ensures that the mass-splittings are identical for pions in the same $SO(4)$-irrep:  $P, V, A, T, I$~\cite{Lee:1999zx}.  An exact $U(1)_A$ symmetry protects the taste pseudoscalar meson from receiving a mass-shift to all-orders in S$\chi$PT, implying that $\Delta_P = 0$.

At NLO in the mixed-action theory, the only nonzero staggered mass-splitting that is relevant is that of the taste-singlet, $\Delta_I$~\cite{Bar:2005tu}.  This is because the domain-wall valence quarks do not carry the taste quantum number; therefore mixed valence-sea four-fermion operators must contain two domain-wall quarks and two taste-singlet staggered quarks in order to be overall taste-invariant.    The mass-splitting $\Delta_I$ has been calculated by the MILC Collaboration on both the coarse and fine MILC lattices~\cite{Aubin:2004fs}, and is given in Table~\ref{tab:mass_split}.  Because the parameter $\Delta_I$ is known, we reduce the number of free parameters in the chiral and continuum extrapolation of $B_K$ by fixing $\Delta_I$ to the values in Table~\ref{tab:mass_split}.  The mass-splitting $\Delta_I$ turns out to be the largest of the taste-splittings, and comparable to the taste-Goldstone pion mass on the coarse MILC lattices, so the taste-singlet sea-sea mesons are quite heavy on the coarse lattices.  Because the mass-splittings arise from discretization effects, however, they become smaller as the lattice spacing decreases.  Specifically, the staggered taste-splittings scale as $\CO(\alpha_s^2 a^2)$ since the Asqtad staggered action is $\CO(a^2)$-improved.  Thus $a^2\Delta_I$ is already more than a factor of two smaller on the fine MILC lattices than on the coarse.

In the valence sector of the mixed-action theory, domain-wall quarks receive an additive contribution to their mass from explicit chiral symmetry breaking~\cite{Kaplan:1992bt,Shamir:1993zy}:
 \begin{equation}
 	m^2_{vv'} = \mu_{\rm dw} (m_v + m_{v'} + 2 m_{\rm res}),
\label{eq:m2_vv}
 \end{equation}
where $v$ and $v'$ are the domain-wall quark flavors and $m_{\rm res}$ is the residual quark mass.  The size of $m_{\rm res}$ parametrizes the amount of chiral symmetry breaking in the valence sector, and is controlled by the length of the fifth dimension.  We have determined the value of $m_{\rm res}$ in our mixed-action simulations in a previous publication (Ref.~\cite{Aubin:2008wk}) and present the results in Table~\ref{tab:mass_split}.  On the coarse MILC lattices, we find that the value of $m_{\rm res}$ in the chiral limit is about 3/4 the physical light quark mass; $m_{\rm res}$ is approximately a factor of three smaller on the fine MILC lattices, i.e. 1/4 the physical light quark mass.  The small values of the residual quark mass indicate that chiral symmetry breaking is under control in our mixed action lattice simulations.  

In order to reduce the number of fit parameters in our chiral and continuum extrapolation of $B_K$, we fix the value of $m_{\rm res}$ in our chiral fits.  We do not, however, use the values of $m_{\rm res}$ given in Table~\ref{tab:mass_split}, which are found by taking the chiral limit ($m_l = m_h = m_x = 0$) in both the valence and sea sectors.  Instead, for each lattice data point, we fix $m_{\rm res}$ to the value determined at that particular combination of valence and sea-quark masses.  This effectively includes higher order corrections to $m_{\rm res}$ and improves the confidence levels of our chiral fits to $f_\pi$ and $m_\pi^2$.

Because the mixed-action lattice theory has new four-fermion operators, the chiral effective theory has new low-energy constants due to discretization effects.  It turns out, however, that the mixed-action chiral Lagrangian has only one new constant at lowest order~\cite{Bar:2005tu}.  This coefficient combines with coefficients coming from the taste-symmetry breaking operators in the staggered sector \cite{Chen:2009su} to produce an $\CO(a^2)$ shift to the mixed valence-sea meson mass-squared:
\begin{equation}
	m^2_{vs} = \mu_{\rm dw} ( m_v + m_{\rm res} ) + \mu_{\rm stag} m_s + a^2 \Delta_\textrm{mix},
\label{eq:m2_vs}
\end{equation}
where $v$ is the domain-wall quark flavor, $s$ is the staggered quark flavor,  and $\Delta_\text{mix}$ is the effective mixed valence-sea meson mass-splitting obtained in lattice simulations.\footnote{The explicit expression for the effective mixed meson mass splitting $\Delta_\textrm{mix}$, given as a linear combination of the staggered sea taste-splittings and the new splitting unique to the mixed action theory, is derived in Ref.~\cite{Chen:2009su}.}  We have calculated the value of $\Delta_\text{mix}$ for the parameters of our mixed-action lattice simulations in Ref.~\cite{Aubin:2008wk}, and present the results in Table~\ref{tab:mass_split}.  We find that the size of $\Delta_\text{mix}$ is less than half of the taste-singlet staggered mass-splitting, $\Delta_I$, on both the coarse and fine MILC lattices.  

We do not need to fix the value of $\Delta_\text{mix}$ during the chiral and continuum extrapolation of $B_K$ because it turns out that the parameter $\Delta_\text{mix}$ does not enter the expression for $B_K$ in MA$\chi$PT at NLO, Eq.~(\ref{eq:BK2+1})~\cite{Aubin:2006hg}.  Although the mass-splitting enters the mixed-action expression for $f_K$, it cancels exactly at NLO between the numerator and denominator in the ratio of matrix elements that defines $B_K$, Eq.~(\ref{eq:BKdef}).

Finally, we note that, for the purpose of our chiral and continuum extrapolation of $B_K$, it is useful to express the tree-level meson masses in terms of the bare lattice quark masses given in Table~\ref{tab:BK_data}, not in terms of the renormalized quark masses.  Because the valence and sea quarks are renormalized according to different schemes, we absorb the scheme-dependent quark-mass renormalization factors into separate coefficients of proportionality, $\mu_{\rm dw}$ and $\mu_{\rm stag},$ in the tree-level mass relations, Eqs.~(\ref{eq:m2_ss})--(\ref{eq:m2_vs}).

\bigskip

The NLO $\chi$PT expression for $B_K$ in a mixed-action domain-wall valence, staggered sea theory with 2+1 flavors of dynamical sea quarks is~\cite{Aubin:2006hg}
\bea
\label{eq:BK2+1} 
	\left( \frac{B_K}{B_0}\right)^{\textrm{PQ},2+1}&=&1+\frac{1}{16\pi^2 f_{xy}^2 m^2_{xy}}\left[I_{conn} +  I^{2+1}_{disc} \right]   \nonumber \\ 
	&&  +\, c_1 a^2+ \frac{8}{f_{xy}^2}\left[c_2 m^2_{xy} + c_3\frac{(m^2_X-m^2_Y)^2}{m^2_{xy}}+c_4 (2m^2_{L_P} + m^2_{H_P})\right] \,,
\eea
where $m^2_{X(Y)}$ is the mass-squared of a meson composed of two $x(y)$ valence quarks and $m ^2_{L_P(H_P)}$ is the mass-squared of a taste-pseudoscalar meson composed of two $l(h)$ sea quarks.  The 1-loop chiral logarithms are separated into contributions from quark-level connected and disconnected diagrams.    
The parameter $B_0$ is the tree-level value of $B_K$ obtained in the continuum and $SU(3)$ chiral limits.  The four analytic terms, $c_1$ - $c_4$, are the only additional free parameters in the expression for $B_K$ at NLO.  Although the analytic term proportional to $a^2$ is not present in the continuum, it is present for chiral lattice fermions.  Thus $B_K$ in the mixed-action theory, Eq.~(\ref{eq:BK2+1}), has no more undetermined coefficients than in the purely domain-wall case.

The connected contribution to $B_K$ is 
\bea  
	I_{conn} = 2m^4_{xy}\widetilde{\ell}(m^2_{xy})-\ell(m^2_X)(m^2_X+m^2_{xy})- \ell(m^2_Y)(m^2_Y+m^2_{xy}) . 
\eea
The chiral logarithms,  $\ell$ and $\widetilde{\ell}$, are defined as
\bea 
	\ell(m^2) &=& m^2\left(\ln \frac{m^2}{\Lambda^2_{\chi}} + \delta^{FV}_1(m\textrm{L})\right)\,, \qquad \delta^{FV}_1(m\textrm{L})=\frac{4}{m\textrm{L}}\sum_{\vec{r} \neq 0} \frac{K_1(|\vec{r}|m\textrm{L})}{|\vec{r}|}\,, \\
	\widetilde{\ell}(m^2) &=& -\left(\ln \frac{m^2}{\Lambda^2_{\chi}}+1 \right) + \delta^{FV}_3(m\textrm{L})\,, \qquad
\delta^{FV}_3(m\textrm{L})=2\sum_{\vec{r} \neq 0}K_0(|\vec{r}|m\textrm{L})\,, 
\eea
where the difference between the finite and infinite volume result is given by $\delta_i^{FV}(m\textrm{L})$, and $K_0$ and $K_1$ are modified Bessel functions of imaginary argument.  The disconnected contribution to $B_K$ is
\bea
	I^{2+1}_{disc}=\frac{1}{3}(m^2_X-m^2_Y)^2\frac{\partial}{\partial m^2_X}\frac{\partial}{\partial m^2_Y}\left\{\sum_j
\ell(m^2_j)\left(m^2_{xy}+m^2_j \right)R^{[3,2]}_j(\{M^{[3]}_{XY,I}\};\{\mu^{[2]}_I \}) \right\} ,
\eea
where
\bea 
	R^{[n,k]}_j(\{m\},\{\mu\}) & \equiv & \frac{\prod_{a=1}^k (\mu^2_a-m^2_j)}{\prod^n_{i=1,i\neq j}
(m^2_i-m^2_j)}, \\
	\{M^{[3]}_{XY,I}\} & \equiv & \{m_X, m_Y, m_{\eta_I} \}, \\ 
	\{\mu^{[2]}_I \} & \equiv & \{m_{L_I},m_{H_I} \} . 
\eea
When the up and down sea quark masses are degenerate, the flavor-neutral, taste-singlet mass eigenstates are
\begin{eqnarray} 
	m^2_{\pi^0_I}  &= & m^2_{L_I} , \nonumber \\
	m^2_{\eta_I} & = & \frac{m^2_{L_I}}{3} + \frac{2m^2_{H_I}}{3},
\end{eqnarray}
and the disconnected contribution to $B_K$ simplifies:
\bea\label{eq:sea}
    I_{disc}^{2+1}=\frac{1}{3}  \left(I_X+I_Y+I_\eta\right) ,
\eea
with
\bea 
	I_X &&= \widetilde{\ell}(m^2_X) \frac{(m^2_{xy}+m^2_X)(m^2_{L_I}-m^2_X)(m^2_{H_I}-m^2_X)}{(m^2_{\eta_I}-m^2_X)}  \nonumber \\ 
	&& -\ell(m^2_X)\left[\frac{(m^2_{xy}+m^2_X)(m^2_{L_I}-m^2_X)(m^2_{H_I}-m^2_X)}{(m^2_{\eta_I}-m^2_X)^2} + \frac{2(m^2_{xy}+m^2_X)(m^2_{L_I}-m^2_X)(m^2_{H_I}-m^2_X)}{(m^2_Y-m^2_X)(m^2_{\eta_I}-m^2_X)} \right. \nonumber \\ 
	&& \left. +\frac{(m^2_{L_I}-m^2_X)(m^2_{H_I}-m^2_X)-(m^2_{xy}+m^2_X)(m^2_{H_I}-m^2_X) -(m^2_{xy}+m^2_X)(m^2_{L_I}-m^2_X)}{(m^2_{\eta_I}-m^2_X)} \right],
\eea
\bea I_Y=I_X(X \leftrightarrow Y),\eea
\bea I_\eta =
	\ell(m^2_\eta)\frac{(m^2_X-m^2_Y)^2(m^2_{xy}+m^2_{\eta_I})(m^2_{L_I}-m^2_{\eta_I})(m^2_{H_I}-m^2_{\eta_I})} {(m^2_X-m^2_{\eta_I})^2(m^2_Y-m^2_{\eta_I})^2}. 
\eea
All of the sea quark dependence in the chiral logarithms appears in the disconnected terms, the sum of which vanishes for degenerate valence quark masses.  The contribution $I_\eta$ vanishes identically when $m_X = m_Y$.  In the limit that $m_X \to m_Y$, $I_X \to -I_Y$.  Thus the sum $I_X + I_Y + I_\eta \to 0$.

\subsection{Chiral and continuum extrapolation of $B_K$}
\label{sec:ChPT_extrap}

We use the $SU(3)$ MA$\chi$PT formula of Eq.~(\ref{eq:BK2+1}) in the extrapolation of our numerical lattice data to the continuum and to physical quark masses.  The choice of $SU(3)$ $\chi$PT is appropriate given the parameters of our numerical simulations because our light pion masses range from 240-500 MeV and are not much lighter than the physical kaon, which is integrated out in $SU(2)$ $\chi$PT. Furthermore, the largest of the taste-splittings on the coarse lattices is not much smaller than the kaon mass 
[$a^2\Delta_I\approx (460 \textrm{MeV})^2$], though on the fine lattices it is a factor of 2.7 times smaller [$a^2\Delta_I \approx (280 \textrm{MeV})^2$].

There have been several studies showing that NLO $SU(3)$ mixed-action and staggered chiral perturbation theory accurately describe the lattice discretization effects even at the rather large taste-splittings on the coarse MILC lattices \cite{Beane:2005rj, Bernard:2007qf, Beane:2007xs, Aubin:2008wk}.  For example, the agreement between the NLO mixed-action $SU(3)$ $\chi$PT prediction of the scalar correlator at large times and the lattice data is excellent over the range of masses and lattice spacings used in the present calculation of $B_K$ \cite{Aubin:2008wk}.  The agreement is good to within the statistical accuracy of the scalar correlator data, which is about $10\%$.  This agreement is a highly non-trivial test because the heavy taste-singlet $\eta$ plays the dominant role in modifying the continuum $\chi$PT form of the correlator.  The MILC collaboration also finds that the scalar correlator in the staggered theory is well-described by staggered chiral perturbation theory, and that when the low energy constants in the prediction for the scalar correlator are allowed to vary, they agree with those determined in fits to the light pseudoscalar sector \cite{Bernard:2007qf}.       

The statistical errors on $B_K^{lat}$ are $\sim 0.5\%-2\%$ for most of our data points.  It is now well-established that NLO $\chi$PT does not describe quantities such as pseudoscalar masses, decay constants, or $B_K$ to percent-level accuracy at the physical kaon mass, nor is it expected to based on power counting.  Our data set confirms this picture for $B_K$.  In order to get good fits to our $B_K$ data in the region of interest we must include next-to-next-to-leading order (NNLO) analytic terms.  Fits without these terms give terrible correlated $\chi^2/\textrm{d.o.f.}$'s and miniscule confidence levels.  The two-loop NNLO logarithmic corrections, however, are not known for $B_K$.  These expressions would also have to be modified to account for the staggered sea sector, though, given our experience with the one-loop modifications due to the mixed action, this is likely a small effect.  In the region where the NNLO analytic terms that we have added are important, we expect the NNLO logarithms to vary slowly enough that their effect is well approximated by the analytic terms.  We vary the assumptions we make for the (thus far unknown) NNLO behavior in order to estimate a systematic error for the chiral extrapolation.  When we apply the same approach to our calculation of $f_\pi$, $f_K$, and their ratio \cite{Aubin:2008ie}, we find systematic errors due to the chiral extrapolation of a similar size to other lattice determinations \cite{Lellouch:2009fg} and in excellent agreement with phenomenology.  This good agreement between our lattice calculations and known quantities lends confidence in our methods for quantities such as $B_K$ that are not known from experiment.

Our approach to chiral fits using $SU(3)$ $\chi$PT plus higher-order analytic terms is related to the approach of other groups using $SU(2)$ $\chi$PT with the kaon integrated out \cite{Allton:2008pn} in the following way.  The chiral logarithms due to pion loops are common to $SU(3)$ and $SU(2)$ $\chi$PT, and are the dominant non-analytic contribution when the strange quark is much heavier than the two lightest quarks.  In order to get acceptable fits, we need to introduce polynomial dependence in the valence quark mass at higher order than NLO in the chiral expansion, as discussed above.  This is necessary to describe the data in the region where the $SU(3)$ $\chi$PT is not especially convergent, and higher-order corrections are important.  Nonetheless, our simulations interpolate about the physical strange quark mass, so the higher-order dependence of the $SU(3)$ expansion, with terms involving kaon and eta masses, is expected to be well described.  Towards the physical value of $B_K$, where we extrapolate in the light quark masses, the heavier meson masses in the $SU(3)$ expansion decouple and the $SU(3)$ form becomes that of $SU(2)$.  In this decoupling region, the taste-breaking in the heavier mesons containing the strange quark reduces to terms analytic in $a^2$ and the light quark masses, which are included in the fits.  Thus, we expect our fit function to describe our light-mass data well, and we expect the additional analytic terms to capture higher-order effects in the heavier-mass region where terms beyond NLO are necessary.  Recent work by the MILC collaboration corroborates this approach. They find excellent agreement between $SU(2)$ and $SU(3)$ staggered chiral perturbation theory approaches for $f_\pi$ and light quark masses \cite{Bazavov:2009tw,Bazavov:2009ir}.  The $\pi$-$\pi$ scattering results of NPLQCD also show good agreement for the $I=2$ scattering length $a_{\pi\pi}^{I=2}$ between $SU(2)$ and $SU(3)$ NLO MA$\chi$PT determinations \cite{Beane:2005rj, Beane:2007xs}.

There are six new continuum NNLO analytic terms for $B_K$, as well as NNLO terms that modify the NLO constants $c_1$-$c_4$ by terms proportional to $a^2$.  We include only a subset of the NNLO terms that are needed to obtain good correlated $\chi^2$ values, and include the others in alternative fits for systematic error estimation.  The number of new continuum low energy constants can be constrained using CPS symmetry~\cite{Bernard:1985wf}, chiral symmetry, and the fact that there is only one mass scale in the tree-level diagrams with the external kaons at rest.  The new continuum analytic NNLO contributions to $B_K$ are
\bea\label{eq:NNLO} && d_1m_{xy}^4, \ \ d_2(m_X^2-m_Y^2)^2, \ \ d_3(2m_{L_P}^4+m_{H_P}^4), \ \ d_4(2m_{L_P}^2+m_{H_P}^2)^2, \nonumber \\ && d_5(m_{xy}^2)(2m_{L_P}^2+m_{H_P}^2), \ \ d_6\frac{(m_X^2-m_Y^2)^2}{m_{xy}^2}(2m_{L_P}^2+m_{H_P}^2).
\eea
We also test for higher order analytic terms proportional to $a^2$.  We find an improvement to the fit when including a term of the form $a^2 m_{xy}^2$.  Our systematic error estimate includes the effects of generic NNLO non-analytic terms on the extrapolation and an estimate of the size of NNLO taste-violations not accounted for in the fitting procedure. 

\begin{figure}
\begin{center}
\includegraphics[width=4in]{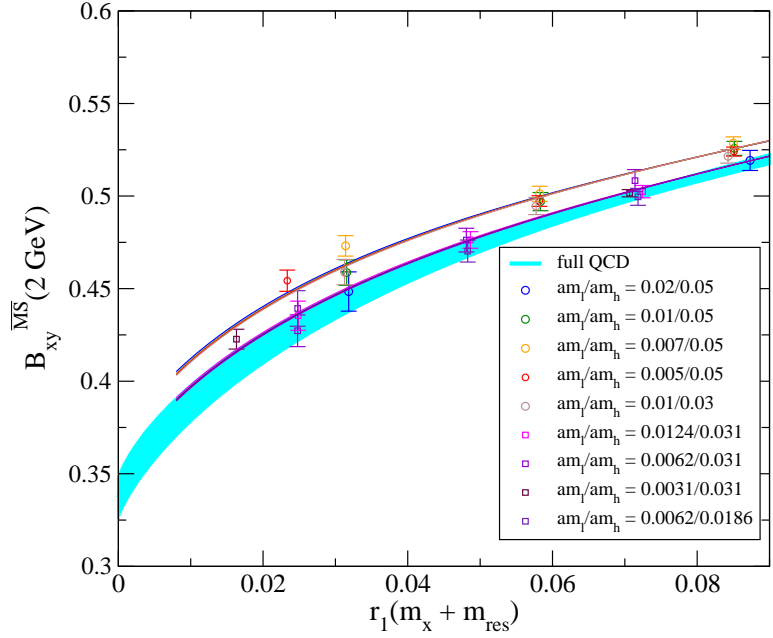}
\caption{$B_K$ versus degenerate light valence quark mass $r_1(m_x + m_{\rm res})$ on different ensembles.  The fit lines are the same partially quenched contours as the data.  Note that all of the coarse fit lines lie on top of each other, and likewise for the fine, indicating very little sea quark dependence.  The band is the degenerate quark mass full QCD curve ($m_x = m_y = m_l = m_h$) in the continuum limit.  The y-intercept of the full QCD curve gives the low-energy constant $B_0$, which
is the value of $B_K$ in the $SU(3)$ chiral limit.}
\label{fig:BK_deg}
\end{center}
\end{figure}

\begin{figure}
\begin{center}
\includegraphics[width=4in]{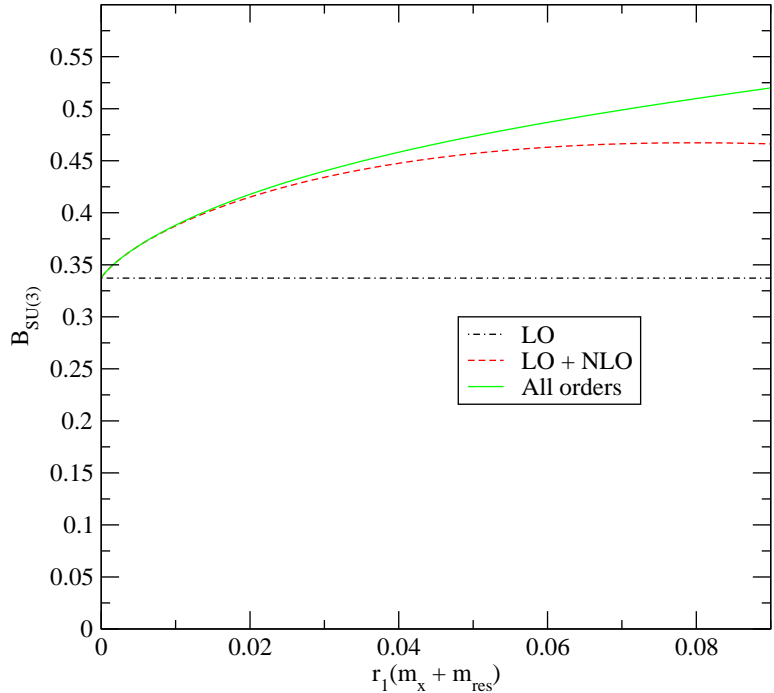}
\caption{Comparison of higher order $\chi$PT corrections for $B_K$.  The rightmost point on the graph corresponds to $\sim m_s/2$.}
\label{fig:BK_NLO}
\end{center}
\end{figure}

\begin{table}
\caption{Tree-level $\mu$-parameters, defined in Eqs.~(\ref{eq:m2_ss})--(\ref{eq:m2_vv}), on the coarse and fine MILC lattices.  The staggered coefficient, $r_1 \mu_{\rm stag}$, is independent of the valence sector~\cite{Aubin:2004fs}, while the domain-wall coefficient, $r_1 \mu_{\rm dw}$, depends upon the choice of HYP-smearing~\cite{Aubin:2008wk}.}
\label{tab:mu_values}
\begin{tabular}{lccccc} \\ \hline\hline

$a$(fm) & $r_1 \mu_{\rm stag}$  &&  $r_1 \mu_{\rm dw}$  \\[0.5mm] \hline

0.12 & 6.234 && 4.13 \\
0.09  & 6.382 && 3.83 \\\hline\hline

\end{tabular}\end{table}

Figure~\ref{fig:BK_deg} shows our preferred fit to the data using NLO partially quenched MA$\chi$PT supplemented by some of the above NNLO analytic terms.  In order to obtain a fit with a good correlated confidence level, we include the NNLO continuum analytic terms proportional to $d_1$, $d_2$, $d_5$ and $d_6$ in Eq.~(\ref{eq:NNLO}) plus an NNLO analytic term containing discretization effects, $d_{a^2 m_{xy}^2}(8/f_{xy}^2)a^2m_{xy}^2$.  We fix the following parameters in the fit: the tree-level (continuum) coefficients $\mu_{\rm dw}$ and $\mu_{\rm stag}$, the decay constant $f_{xy}$, and the taste-splitting $a^2\Delta_I$.  We take for the parameters $\mu_{\rm dw}$ and $\mu_{\rm stag}$ the values obtained from fits to the light pseudo-scalar masses squared to the tree-level forms given in Eqs.~(\ref{eq:m2_ss}) and (\ref{eq:m2_vv}).
This accounts for higher-order chiral corrections and is more accurate than using $\mu$ obtained in the chiral limit (which is found by fitting to the one-loop pseudoscalar mass and decay constant expressions), giving a
better approximation to the pion mass squared at a given light quark mass.  The values for $\mu_{\rm dw}$ and $\mu_{\rm stag}$ are given in Table~\ref{tab:mu_values}.  We take the decay constant $f_{xy}$, which appears as the inverse square in the coefficient of the chiral logarithms, to be the physical $f_K = 156.5$ MeV \cite{Bernard:2007ps} for our preferred fit, though we vary $f_{xy}$ in order to estimate the systematic error.  We use the value for the taste-singlet splitting  $a^2\Delta_I$ obtained by the MILC Collaboration in Ref.~\cite{Aubin:2004fs} and given in Table~\ref{tab:mass_split}.  Given these choices, our preferred combined chiral-continuum extrapolation fit function contains only ten free parameters.

Figure~\ref{fig:BK_deg} shows only the degenerate valence mass points and the corresponding partially quenched fit lines, although the fit includes nondegenerate masses as well.  The heaviest valence kaon masses included in this fit are slightly larger than the physical kaon mass.  We restrict the degenerate valence ``kaon" masses to below 500 MeV, but we allow slightly heavier non-degenerate valence kaons up to masses of around 600 MeV in order to interpolate about the physical strange quark mass.  In the sea sector, we restrict the taste-pseudoscalar pions to be less than 550 MeV on the coarse ensembles and less than 500 MeV on the fine ensembles.   Our lightest degenerate valence ``kaon'' is $\sim 230$~MeV, while our lightest taste-pseudoscalar sea pion is $\sim 240$~MeV.  Given these mass restrictions, the number of data points in our preferred fit is 69, which is more than sufficient to constrain ten parameters.

Although most of our degenerate-mass data points are far from the physical kaon mass, including this data in the fit allows us to constrain the parameters of the $SU(3)$ chiral Lagrangian and to study the convergence of $SU(3)$ $\chi$PT.  The (cyan) band in Fig.~\ref{fig:BK_deg} shows the full QCD curve with statistical errors in the $SU(3)$ ($m_x = m_y = m_l = m_h$) and continuum limits that is obtained from our preferred fit.  In order to examine the convergence of $SU(3)$ $\chi$PT, we plot the separate contributions to the degenerate $SU(3)$ curve through LO, NLO, and ``all orders''  in Fig.~\ref{fig:BK_NLO}.  The right-most part of the $x$-axis corresponds to $\sim m_s/2$ (i.e., a $500$ MeV pion), where we do not expect $\chi$PT to be especially convergent.  Because we are interpolating in the quark mass in this region, we expect the ``all orders'' curve to be accurate, with the NNLO terms approximating the correct higher order behavior.  Closer to the physical pion mass, however, the NLO contributions are the dominant corrections, and the particular choice of NNLO analytic terms has little impact on the fit result in this light quark-mass region as long as the fit has a good correlated confidence level.  

It should be noted that the relative contributions of NLO versus NNLO terms can change significantly, depending on whether one uses bare expansion parameters or the physical masses and decay constants in the one-loop expressions.  Following Bijnens \cite{Bijnens:2009pq}, we prefer to use the physical masses and decay constants because the chiral logs are created by particles propagating with their physical momentum, and (though not relevant for $B_K$) thresholds appear in the right places at each order in perturbation theory.  Using the bare parameters in our fits yields poor confidence levels, and we do not consider them further.  Using the physical parameters in the expansion, we find the results of our fits to be consistent with the expectations from chiral power-counting.  This is consistent with the findings of the JLQCD collaboration that using a resummed physical expansion parameter significantly improves the ability of NLO (or NNLO) $\chi$PT to describe data at heavier masses near the kaon by accelerating the convergence of the chiral expansion \cite{Noaki:2008iy}.  The NPLQCD collaboration also uses the physical parameters  in the $SU(3)$ MA$\chi$PT expansion for their determination of the $\pi$-$\pi$ scattering length, and they find good agreement between their data and the leading order form, which is completely predicted \cite{Beane:2005rj, Beane:2007xs}.  When NPLQCD include the NLO $SU(3)$ corrections, they also find good fits to their lattice data.  Despite these successes of the $SU(3)$ formalism, it would be valuable to continue our $B_K$ study with the complete NNLO formula once it is available.  It should also be noted that the systematic error in the chiral extrapolation to the $SU(3)$ chiral limit is fairly large, since the simulated strange quark masses in the sea are close to the physical value.  Thus, there is a large systematic error in the value of the leading order term $B_{0}$, and the picture in Fig.~\ref{fig:BK_NLO} may change appreciably once we can better constrain the approach to the $SU(3)$ limit.  Given the central value of our current best fit, however, the convergence of the chiral expansion appears reasonable.

\bigskip

\begin{figure}
\begin{center}
\includegraphics[width=4in]{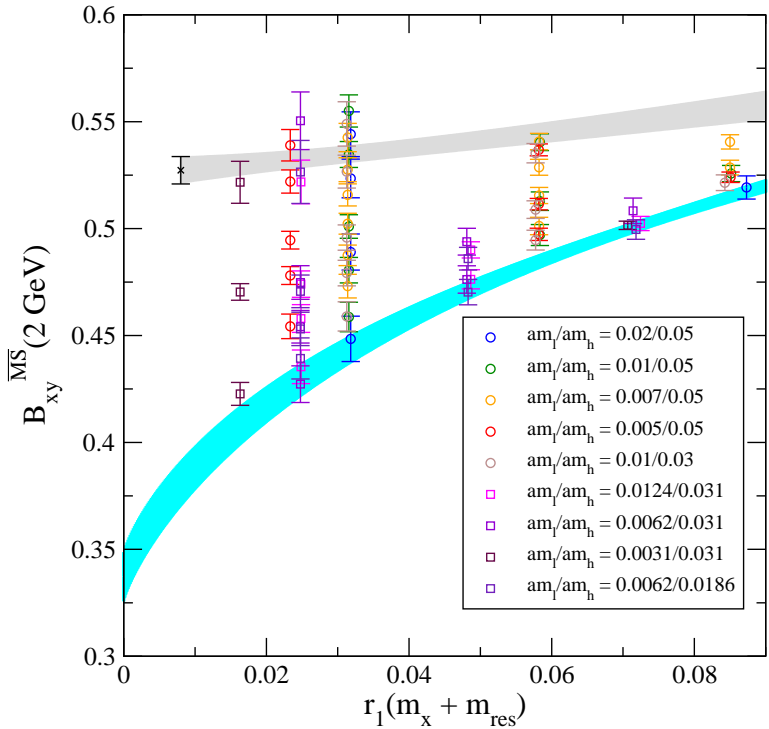}
\caption{$B_K$ versus light valence quark mass $r_1(m_x + m_{\rm res})$.  All data points used in the fit (same fit as in Fig.~\ref{fig:BK_deg}) are shown.  The upper band is the full QCD continuum extrapolated curve with the strange quark fixed to its tuned value.  The lower band is the degenerate quark mass, full-QCD band, as in Fig.~\ref{fig:BK_deg}.  The extrapolated value of $B_K$ at the physical quark masses with statistical errors is given by the ``X" with solid black error bars. }
\label{fig:BK_all}
\end{center}
\end{figure}

We obtain a value of $B_K^{\overline{MS}}(2 \textrm{GeV})= 0.5273(64)$ from our preferred fit when the matching factor is calculated using NPR, where the error is statistical only.  We take this result as our central value.  For comparison, the value for $B_K$ obtained using lattice perturbation theory to compute the matching factor is $B_K^{\overline{MS}}(2 \textrm{GeV})= 0.541(6)$.  The result of our preferred fit is shown in Fig.~\ref{fig:BK_all}.  All data points used in the fit are shown.  The upper band is the full QCD continuum extrapolated curve with the strange quark fixed to its tuned value.  The lower band is the degenerate 
quark mass, full-QCD band, as in Fig.~\ref{fig:BK_deg}.  
The extrapolated value of $B_K$ at the physical quark masses with statistical errors is shown as an ``X" with solid black error bars.

Table~\ref{tab:fit_results} shows the low-energy constants determined in our preferred fit, with statistical errors only.  We do not attempt to estimate a systematic uncertainty in these parameters because this is not necessary for determining $B_K$ at the physical quark masses.  We expect that the systematic uncertainties, however, will be large given the size of the extrapolation to the SU (3) chiral limit.  We present the values only to illustrate a few important points.  Discretization errors in our data are small; this can be seen from the size of the parameters  $c_1$ and $d_{a^2 m^2_{xy}}$.  Further, we do not observe any clear sea-quark mass dependence in our data, as shown by the fact that $c_4$, $d_5$, and $d_6$ are zero within errors.  We estimate the systematic uncertainty in $B_K$ due to the choice of chiral and continuum extrapolation fit function in the following section.

\begin{table}
\caption{Fit parameters obtained in our preferred fit.  Errors shown are statistical only, and do not include the extrapolation uncertainty.  The coefficient $B_0$ in the top panel is the only leading-order low-energy constant, the coefficients in the middle panel are the NLO low-energy constants, and the coefficients in the bottom panel are the NNLO parameters included in the fit.  Definitions of the parameters are given in Eqs.~(\ref{eq:BK2+1}) and~(\ref{eq:NNLO}). }
\label{tab:fit_results}
\begin{tabular}{lr} \\ \hline\hline

$B_0$ & 0.338(12) \\\hline
$c_1$  &  0.057(25) \\
$c_2$  & 0.00531(60)  \\
$c_3$  & 0.00033(15)  \\
$c_4$  & 0.00003(15)  \\\hline
$d_1$  & 0.351(23)  \\
$d_2$  & -0.004(38)  \\
$d_5$  & -0.006(27)  \\
$d_6$  & 0.006(35)  \\
$d_{a^2 m^2_{xy}}$ & -0.00049(29) \\
\hline\hline

\end{tabular}\end{table}

\section{Systematic uncertainties}
\label{sec:Error}

In the following subsections, we examine the uncertainties in our calculation due to the chiral/continuum extrapolation, scale and light quark mass uncertainties, finite volume effects, and uncertainties in the matching factor $Z_{B_K}$.

\subsection{Chiral and continuum extrapolation errors}

We estimate the systematic error in the chiral extrapolation by varying the fit function used to extrapolate the data over a variety of different reasonable choices and taking the spread between them.  By reasonable, we mean theoretically motivated fits that also describe the data with good confidence levels, determined by the correlated $\chi^2$ per degree of freedom.  These fits always involve the known one-loop mixed action chiral logarithms, since including them incorporates the leading non-analytic dependence on the light quark masses.  We consider variations of the fit function by including different types of terms beyond NLO that still give acceptable confidence levels and in order to determine the systematic error in the chiral extrapolation.  We also vary other assumptions, such as the values of parameters used in the prediction for the chiral logarithms, and the lattice spacing dependence of the continuum extrapolation.  Each of these variations is addressed in turn.

We combine the chiral and continuum extrapolations using MA$\chi$PT to control the approach to physical light quark masses and to the continuum.  Combining the data sets on coarse and fine lattices, we have seven different valence quark masses and nine different sea quark mass combinations.  Our valence kaons range from around 600 MeV down to as light as 230 MeV.  These lighter kaons are useful for constraining the low energy constants of the chiral Lagrangian.  As can be seen in Fig.~\ref{fig:BK_NLO}, even at the physical kaon mass near the rightmost part of the plot, the chiral behavior is mostly accounted for by a combination of leading order and NLO terms.  The combinations of sea quark masses include values of the simulated strange sea quark above and below the physical strange quark mass on both the coarse and fine lattices, allowing us to interpolate in the strange sea quark mass.  

The light sea quark masses used in our simulation are as low as $m_s/10$, and this translates into a taste-Goldstone pion (the lightest of the staggered pions) of around 240 MeV.  In the chiral extrapolation of $B_K$, however, the only sea pion mass that appears in the NLO MA$\chi$PT expression is the taste-singlet pion mass (the heaviest of the staggered pions), the lightest of which in our simulations is still a rather heavy 370 MeV.  Fortunately, the sea quark contribution to the NLO chiral logarithms vanishes for degenerate valence quarks, and gives only a small contribution for nondegenerate valence quarks (the region of interest for the physical kaon).  This is in part because the terms that contain the taste-singlet pion are suppressed by a factor of 
$1/N_{\rm sea}$.  The taste-breaking is corrected for explicitly through NLO, and partially at NNLO by including analytic terms proportional to $a^2$ and $a^2 m_{xy}^2$.       

As described in the previous subsection, we take as our central value the result of a fit that includes all of the terms through NLO, the NNLO continuum analytic terms proportional to $d_1$, $d_2$, $d_5$ and $d_6$ in Eq.~(\ref{eq:NNLO}), and an NNLO analytic term containing discretization effects proportional to $a^2m_{xy}^2$.  We estimate the systematic uncertainty due to the chiral extrapolation by including additional NNLO analytic terms and taking the difference between the new value of $B_K$ and that obtained with the preferred fit.  The largest contribution to the systematic error comes from adding terms quadratic in the sea quark masses to the above fit and taking the spread between the two.  Although we analyze ensembles with different light and strange sea quark masses, we do not observe any clear sea-quark mass dependence in our data.  When we include the terms proportional to $d_3$ and $d_4$ in Eq.~(\ref{eq:NNLO}) to our fit the result shifts to 0.5175(72), yielding a difference of $1.9\%$.  We take this as the error due to approximating higher order terms in the chiral expansion.

We also consider NNLO non-analytic terms of the generic form $m^2 \log(m^2)$ to estimate higher order effects.  A term of the form $m_X^4[\log(m_X^2/\Lambda_\chi^2)]$ appears at NNLO [in both $SU(2)$ and $SU(3)$ $\chi$PT], but it is subleading in the chiral expansion and should therefore have a smaller impact than the terms that we are already including as $m_X$ approaches the physical $m_\pi$.  In order to test that the effects of such a term are indeed small, we added this term to our preferred fit leaving its coefficient as a free parameter.  We obtain a small coefficient for such a term, which leads to a slight $0.9\%$ shift upwards in our central value.  This is within our estimate of the error due to approximating higher order terms in the chiral expansion. 

Higher-order taste-breaking is considered as well.  If we set the staggered singlet taste-splitting to zero in the NLO logarithms this amounts to using the continuum-like expression appropriate to a purely domain-wall simulation; if we do this it shifts our chiral extrapolation to $B_K$ at the physical quark masses by only $0.7\%$.  Thus, the discretization effects particular to the use of staggered quarks in the sea sector are small.  The formulas we use do not explicitly remove taste-breaking at NNLO. Note, however, that the higher order analytic term $a^2 m_{xy}^2$ is needed to get good confidence levels in our fits, and is expected to absorb some of the higher-order taste-violating effects, especially those in the kaon sector of $SU(3)$ $\chi$PT in the limit that the kaon can be integrated out.  We estimate the residual effect of the higher-order taste-breaking by varying the splittings used for the sea-mesons in the analytic terms [Eq.~(\ref{eq:NNLO})] over the full range of staggered meson masses.  When we do this, we find that the central value decreases by $1.8\%$, which is within our systematic error due to neglecting higher order terms in the chiral expansion.

We consider other variations to the fit, but they lead to a much smaller shift in the central value.  Although we fix the tree-level (continuum) coefficients $\mu_{\rm dw}$ and $\mu_{\rm stag}$, the decay constant $f_{xy}$, and the taste-splitting $a^2\Delta_I$ in the NLO chiral extrapolation formula, we vary them within their statistical uncertainties in order to estimate their contribution to the error in $B_K$.  The staggered and domain-wall $\mu_{\rm tree}$ parameters are well-determined, and their error is negligible in $B_K$.  In the chiral fit used for our central value we take the decay constant $f_{xy}$, which appears as the inverse square in the coefficient of the chiral logarithms, to be the physical $f_K$.  We vary this coefficient between $f_\pi$ and $f_K$ as an additional way of estimating higher order corrections.  Note that a change in the coefficient of the chiral logarithms will change the other fit parameters, so that this produces a much smaller effect than simply changing the overall coefficient of the chiral logarithms in the final result by a factor of $(f_K/f_\pi)^2 \sim 1.4$.  The change of $f_{xy} $ from $f_K$ to $f_\pi$ leads to a $0.1\%$ shift in the value for $B_K$.  The approximately $10\%$ statistical error in $a^2\Delta_I$ leads to a similarly negligible error in $B_K$.

Although we include terms proportional to $a^2$ in the preferred fit, there is some ambiguity (with only two lattice spacings) in the dominant source of discretization errors, which may be purely $a^2$ corrections, taste-breaking terms proportional to $\alpha_s^2 a^2$, or chiral symmetry breaking terms proportional to $m_{\rm res} a^2$.  It is also possible that the different sources lead to discretization effects of the same size.  We therefore vary the change in the effective $a^2$ between coarse and fine lattice spacings, taking the resulting spread in the value of $B_K$ as part of the systematic error.  If discretization effects decrease as $m_{\rm res} a^2$ or as $\alpha_s^2 a^2$ then they should go down by about a factor of three from our coarse to fine lattices.  If they decrease as $a^2$ they should decrease by about a factor of two.  This difference leads to a $0.3\%$ change in the continuum extrapolated central value.

\bigskip

\begin{figure}
\begin{center}
\includegraphics[width=4in]{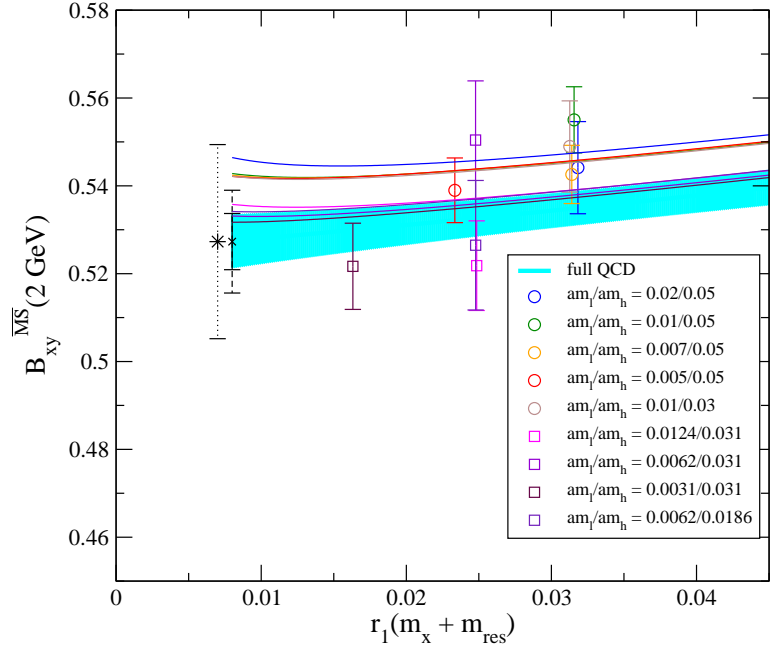}
\caption{Same fit as Fig.~\ref{fig:BK_all}, but with a subset of the data points for illustration.  In this case, the valence masses are nondegenerate, with the heavier mass fixed close to the strange quark mass.  The light-quark mass is the lightest simulated on each ensemble.  The fit curve is the full QCD continuum extrapolated curve with the strange quark fixed to its physical value.  The extrapolated value of $B_K$ is shown, including the statistical error (solid error bar with X) and the systematic error due to the chiral extrapolation, combined with the statistical error in quadrature (dashed error bar).  The dotted error bar (star, slightly offset) shows the total error for $B_K$.}
\label{fig:BK_nondeg}
\end{center}
\end{figure}

In summary, the largest source of chiral-continuum extrapolation error comes from uncertainty in the sea-quark mass dependence.  The parametric uncertainty on quantities used as inputs in the chiral logarithms is negligible, and the residual errors due to the lattice spacing dependence in the continuum extrapolation are small.  Adding the uncertainty due to approximating higher order terms in the chiral expansion and the residual continuum extrapolation error in quadrature, we quote a total systematic error due to the chiral and continuum extrapolation of $1.9\%$.

Figure~\ref{fig:BK_nondeg} illustrates the chiral extrapolation error.  This is the same fit as that shown in Fig.~\ref{fig:BK_all}, but with a subset of the data points.  In this case the valence masses are nondegenerate, with the heavier mass fixed close to the strange quark mass.  The light quark mass is the lightest simulated on each ensemble.  The fit curve is the full QCD continuum extrapolated curve with the strange quark fixed to its tuned value.  We extrapolate the light valence quark mass to the physical $d$ quark mass, while we extrapolate the light sea quark mass to the average of the $u$ and $d$ quark masses.  The band shows the full QCD curve ending at the full QCD $d$ quark mass.  The error bar is centered on the final result, which has a small (not visible) shift due to setting the light sea quark mass equal to the isospin-averaged quark mass.  The extrapolated value of $B_K$ is shown, including the statistical error (solid error bar with X) and the systematic error due to the chiral extrapolation, combined with the statistical error in quadrature (dashed error bar).  The dotted error bar (star, slightly offset) shows the total error for $B_K$ including the matching error.

\subsection{Scale and quark mass uncertainties} 

In order to convert lattice quantities into physical units we use the MILC Collaboration's determination of the scale, $r_1$, where $r_1$ is related to the force between static quarks, $r_1^2F(r_1)=1.0$~\cite{Sommer:1993ce,Bernard:2000gd}.  The ratio $r_1/a$ can be calculated precisely on each ensemble from the static quark potential.  We use the mass-independent prescription for $r_1$ described in Ref.~\cite{Bernard:2007ps} so that all of the mass dependence is explicit in MA$\chi$PT and none is hidden in the scale-fixing scheme. 
In order to fix the absolute lattice scale, one must compute a physical quantity that can be compared directly to experiment; we use the $\Upsilon$ 2S--1S splitting \cite{Gray:2005ad} and the most recent MILC determination of $f_\pi$ \cite{Bernard:2007ps}.  The combination of the $\Upsilon$ mass-splitting and the continuum-extrapolated $r_1$ value at physical quark masses leads to the determination $r_1^{\rm phys}=0.318(7)$ fm~\cite{Bernard:2005ei}.  The use of $f_\pi$  to set the scale yields $r_1^{\rm phys}=0.3108(15)(^{+26}_{-79})$ fm \cite{Bernard:2007ps}.  This difference between the two scale determinations leads to a systematic error in our result for $B_K$.  Since $B_K$ is a dimensionless quantity, the scale enters only through the quark mass determinations.
We determine the light valence quark masses using MA$\chi$PT fits to the pseudoscalar masses and decay constants, as described in Ref.~\cite{Aubin:2008ie}.  The sea-quark masses are taken from the most recent update of the MILC pseudoscalar analysis \cite{Bernard:2007ps}.

MILC finds for the bare staggered quark masses in $r_1$ units \cite{Bernard:2007ps, Bailey:2005ss}
\bea r_1 \hat{m}^{\rm stag} \times 10^3 &=& 3.78(16), \\
r_1 m_s^{\rm stag} \times 10^3 &=& 102(4)\ , 
\eea
where $\hat m \equiv (m_u + m_d)/2$.
Although the masses are in dimensionless $r_1$ units, they are scale and scheme dependent quantities.  The scheme, of course, is the improved staggered lattice action used in the MILC simulations.  The scale is the fine lattice scale of $a^{-1}\sim 2.3$ GeV, but with discretization effects removed by fits to multiple lattice spacings using rooted staggered $\chi$PT.  Both our fits and the MILC fits use as inputs from experiment the averaged meson masses with electromagnetic effects removed as well as possible.  We (and MILC) take for the squared meson masses $m^2_{\hat{\pi}}$ and $m^2_{\hat{K}}$,
\bea
&& m^2_{\hat{\pi}} \equiv m^2_{\pi^0} , \\ \nonumber &&
m^2_{\hat{K}} \equiv \frac{1}{2}(m^2_{K^0}+m^2_{K^+}-(1+\Delta_E)(m^2_{\pi^+}-m^2_{\pi^0})),
\eea
where $\Delta_E \approx 1$ parametrizes corrections to Dashen's theorem.

We find from our mixed action chiral fits to the pseudoscalar sector the values for the bare domain-wall quark masses (also evaluated at the fine lattice scale)
\bea r_1 \hat{m} \times 10^3 &=& 5.87(8)(41) , \\
r_1 m_s \times 10^3 &=& 168(2)(8) ,
\eea
where the first error is statistical and the second is systematic.
Following MILC, we also obtain the masses of the two lightest quarks.  Given $m_s$, we can obtain $m_u$ by extrapolating not to the mass of the $\hat{K}$ but to the mass of the $K^+$ (with EM effects removed).  We take
\bea
(m^2_{K^+})_{\rm QCD} \equiv m^2_{K^+} - (1+\delta_E)(m^2_{\pi^+}-m^2_{\pi^0}),
\eea
where $\delta_E=1$, which corresponds to vanishing EM corrections to the $K^0$ mass.  We then obtain
\bea r_1 m_u \times 10^3 &=& 3.7(22)(7), \\
r_1 m_d \times 10^3 &=& 8.0(27)(7),
\eea
where again the first error is statistical and the second is systematic.  

It is useful to observe for our $B_K$ error analysis that the systematic error for the domain-wall $m_s$ is dominated by the scale error.  However, the error in $m_d$ (needed since a kaon is an $\bar{s}d$ state) is dominated by statistical uncertainty.  Thus we can treat the errors from the $s$ and $d$ quark masses as uncorrelated.  Note that all of the above masses are the bare lattice masses, so no error has been included for the renormalization needed to match to a continuum scheme like $\overline{\textrm{MS}}$.  The bare quark masses are sufficient for the purpose of calculating $B_K$.  The error in $r_1 m_s$ leads to an $0.8\%$ uncertainty in $B_K$, while the error in $r_1 m_d$ leads to an $0.2\%$ error in $B_K$.  The errors in the sea quark masses produce a negligible uncertainty in $B_K$.  Combining these errors in quadrature we obtain an error due to scale and quark mass uncertainties for $B_K$ of $0.8\%$.

\subsection{Finite volume error} 

The finite volume error is estimated using one-loop finite volume MA$\chi$PT \cite{Aubin:2006hg}.  We have simulated at fairly large volumes, such that $m_\pi L \gtapprox 3.5$, and we have corrected our data using the appropriate one-loop MA$\chi$PT expressions, which are never larger than $0.6\%$.  There could still be non-negligible residual finite volume corrections, however, as numerical studies by the MILC collaboration of $f_\pi$ and $m_\pi^2$ show that the one-loop $\chi$PT corrections can be off by as much as $50\%$ for similar simulation parameters using staggered quarks \cite{Bernard:2007ps}.  Even so, given that the largest finite-volume correction to any individual data point in our analysis is $0.6\%$, we expect the residual corrections to be only as much as $0.3\%$.  However, in order to be conservative, we take the entire $0.6\%$ as our total finite volume error.

\subsection{Renormalization factor uncertainty} 
\label{sec:NPR_err}

In this section we estimate the systematic uncertainty in $B_K$ due to the nonperturbative determination of the renormalization factor $Z_{B_K}$.  We consider several sources, discussing each in turn.

\subsubsection{Chiral extrapolation fit ansatz}

In order to remove explicit chiral symmetry breaking contributions to $Z_{B_K}$ from operators such as those in Eq.~(\ref{eq:chiral_sym_br}), we first extrapolate $Z_{B_K}^\textrm{RI/MOM}$ to the chiral limit at fixed values of $(ap)^2$.  Although we choose to use a fit function that is linear in the light sea quark mass and quadratic in the valence quark mass, we can obtain an equally good correlated confidence level using a fit function with even more terms.  We must therefore consider the systematic uncertainty introduced by the choice of chiral extrapolation fit ansatz.  We do so by adding a quadratic term in the light sea quark mass to the fit function and re-doing the chiral extrapolation at each value of $(ap)^2$.  We then re-compute $Z_{B_K}^{\rm \bar{MS}} (\textrm{2 GeV})$ and take the difference between this result and the central value to be the systematic error.  This leads to an uncertainty in $Z_{B_K}^{\rm \bar{MS}, coarse} (\textrm{2 GeV})$ of 0.0063, or $\sim 0.7 \%$, on the coarse lattice and an uncertainty of 0.0111, or $\sim 1.2 \%$, on the fine lattice.  The addition of yet another term cubic in the valence quark mass produces a negligible difference in $Z_{B_K}$.  We take the larger, $1.2 \%$ difference, to be the uncertainty in $Z_{B_K}$ from the choice of chiral extrapolation fit function.

\subsubsection{Strange sea-quark mass dependence}

When we extrapolate $Z_{B_K}^\textrm{RI/MOM}$ to the chiral limit at fixed values of $(ap)^2$, we do not, in fact, take the value of the strange sea quark mass to zero.  This is because, within statistical errors, $Z_{B_K}$ is independent of the strange sea quark mass.  We can explicitly calculate the strange sea-quark mass dependence, however, by taking the chiral limit of $Z_{B_K}^\textrm{RI/MOM}$ at fixed $(ap)^2$ using a function that is linear in the sum of the sea 
quark masses and quadratic in the valence quark mass
on the coarse lattices, so we make the replacement $2m_l\to(2m_l + m_h)$ in Eq.~(\ref{eq:chi}).  We find that this leads to a difference in $Z_{B_K}^{\rm \bar{MS}, coarse} (\textrm{2 GeV})$ of 0.0027, or $\sim 0.3 \%$, and take this to be the uncertainty in $Z_{B_K}$ due to the nonzero strange sea quark mass.  

\subsubsection{$\Lambda_A - \Lambda_V \neq 0$}

The use of exceptional kinematics in our nonperturbative renormalization factor calculation leads to a difference between $\Lambda_A$ and $\Lambda_V$ of $\sim 1\%$ at nonzero quark masses and $p \approx \textrm{2 GeV}$. This is shown in Figs.~\ref{fig:LA_LV_extrap} and~\ref{fig:LA_LV_vs_ap2}.  Because we do not know \emph{a priori} which of the two quantities has less contamination from chiral symmetry breaking, we use the average $(\Lambda_A + \Lambda_V)/2$ to determine the central value for $Z_{B_K}$.  In order to estimate the systematic uncertainty that is introduced by this choice, we also calculate $Z_{B_K}$ using $\Lambda_A$ for the normalization.  This leads to a difference in $Z_{B_K}^{\rm \bar{MS}, coarse} (\textrm{2 GeV})$ of 0.0084, or $\sim 0.9 \%$, on the coarse lattice and a difference of 0.0112, or $\sim 1.2 \%$, on the fine lattice.  We take the larger, $1.2 \%$ difference, to be the uncertainty in $Z_{B_K}$ due to chiral symmetry breaking between $\Lambda_A$ and $\Lambda_V$.

\subsubsection{Mixing with wrong-chirality operators}

The use of exceptional kinematics also leads to mixing between the standard model operator $\CO_K$, which has a $VV+AA$ chiral structure, and other operators of different chiralities that do not contribute to $K^0 -\bar{K}^0$ mixing in the standard model.  Although the size of the mixing coefficents, shown in Figs.~\ref{fig:coarse_VVmAAsea007}--\ref{fig:coarse_TTsea007}, are small, the matrix elements for the wrong-chirality operators diverge in the chiral limit and are much larger than the desired matrix element~\cite{Aoki:2005ga}.  Thus a small mixing coefficient can still potentially lead to a non-negligible error in $Z_{B_K}$.  Fortunately, we can bypass this concern by computing the mixing coefficients at non-exceptional kinematics.  Theoretically, we expect their size to be of $\CO((a m_\textrm{res})^2) \sim 10^{-6}$~\cite{Christ:2005xh,Aoki:2007xm}.  Numerically, we find that all of the mixing coefficients are consistent with zero on both the coarse and fine lattices, as shown in Figs.~\ref{fig:coarse_VVmAAsea007_nonexp}--\ref{fig:fine_TTsea0062_nonexp}.  
Because the contribution to $B_K$ in the $\bar{\textrm{MS}}$ scheme from each wrong-chirality lattice operator is independent of the lattice scheme initially used to obtain the mixing coefficients, we conclude it is safe to neglect them in our calculation of $Z_{B_K}$, despite the fact that we are using exceptional kinematics.  We therefore do not add any systematic uncertainty to $Z_{B_K}$ due to operator-mixing. 

\subsubsection{Perturbative matching and running}

Although we compute $Z_{B_K}$ in the RI/MOM scheme nonperturbatively, we must still convert its value to the SI scheme to remove lattice discretization effects and ultimately to the $\overline{\rm MS}$ scheme using 1-loop continuum perturbation theory.  This introduces uncertainty into $B_K$ due to the omission of higher-order terms.  Because the true truncation error cannot be known without the computation of the next term in the perturbative series, we consider several ways to estimate the uncertainty here.

The first is to multiply the largest individual 1-loop conversion that is used by an additional factor of $\alpha_s^{\bar{\textrm{MS}}}$.  We determine $Z_{B_K}$ from data in the momentum window $2 \textrm{ GeV} \ltapprox p \ltapprox 2.5 \textrm{ GeV}$; thus the largest value of $\alpha_s^{\bar{\textrm{MS}}}(p)$ used is that at 2 GeV.  The largest correction comes from the conversion between the RI/MOM scheme and the SI scheme, and leads to the following estimate of the truncation error:
\begin{eqnarray}
	\alpha_s^{\bar{\textrm{MS}}}(2 \textrm{ GeV}) \times \frac{\alpha_s^{\bar{\textrm{MS}}}(2 \textrm{ GeV})}{4 \pi} J^{(3)}_{RI/MOM} & = & 0.0188 ,
\end{eqnarray}	
or $\sim 2\%$.

The second is to take the size of the entire 1-loop correction from the RI/MOM scheme to the SI scheme to the $\bar{\textrm{MS}}$ scheme:
\begin{eqnarray}
	\frac{\alpha_s^{\bar{\textrm{MS}}}(2 \textrm{ GeV})}{4 \pi} (J^{(3)}_{RI/MOM} - J^{(3)}_{\bar{MS}}) & = &  0.0204 ,
\end{eqnarray}
which also leads to an estimate of $\sim 2\%$.  Because the conversion factors, however, are only known to 1-loop, they are particularly sensitive to the scale at which they are evaluated.  We did not attempt to determine an optimal scale for the process, for example using the BLM prescription~\cite{Brodsky:1982gc}, and must therefore estimate the error due to scale ambiguity.  The standard, although somewhat arbitrary, prescription used in the continuum literature is to take the variation in the quantity when the scale $\mu$ is varied between $2\mu$ and $\mu/2$.  For our case, this leads to the estimates
\begin{eqnarray}
	\frac{\alpha_s^{\bar{\textrm{MS}}}(4 \textrm{ GeV})}{4 \pi} (J^{(3)}_{RI/MOM} - J^{(3)}_{\bar{MS}}) & = &  0.0152 , \\\
	\frac{\alpha_s^{\bar{\textrm{MS}}}(1 \textrm{ GeV})}{4 \pi} (J^{(3)}_{RI/MOM} - J^{(3)}_{\bar{MS}}) & = &  0.0327 .
\end{eqnarray}
Thus, the 1-loop correction can be as large as $\sim 3\%$, if we use a scale of 1 GeV.

The third is to take the difference between $Z_{B_K}$ determined using the nonperturbative Rome-Southampton approach and using lattice perturbation theory.  Each method for computing $Z_{B_K}$  relies on 1-loop perturbation theory, but involves a different series expansion, so one does not know \emph{a priori} which leads to a faster converging series and smaller truncation error.  Thus having two independent calculations of $Z_{B_K}$ provides a valuable independent crosscheck.  To estimate the error, we replace the values of $Z_{B_K}$ determined using nonperturbative renormalization with those from lattice perturbation theory and repeat the extrapolation to the physical quark masses and the continuum.  We obtain  $B_K^{\bar{MS}} {(\rm 2 GeV)} = 0.541(6)$, where the error is statistical only.  We then take the difference between the resulting $B_K$ and our preferred central value:
\begin{eqnarray}
	|B_K^{NPR} - B_K^{LPT}| / B_K^{NPR} = 0.027,
\end{eqnarray}
which is $\sim 3\%$.  This is comparable to, but slightly smaller than, the estimate from the scale ambiguity.  Nevertheless, we think that it is a more reasonable estimate of the uncertainty, given that it comes from two independent perturbative computations.  

Finally, we consider the possibility that the scale dependence observed in $Z_{B_K}^{\rm SI}$ (see Figs.~\ref{fig:coarse_SI_fit} and~\ref{fig:fine_SI_fit}) is not due solely to discretization errors.   Although we made this assumption when obtaining the central value for $B_K$, some of the slope in $Z_{B_K}^{\rm SI}$ versus $(ap)^2$ may in fact be due to the lack of higher-order terms in the matching factor, which is currently known to only 1-loop in perturbation theory.  Our fourth method for estimating the truncation error is therefore to take the difference between $Z_{B_K}^{\rm SI}$ at 2 GeV and that obtained in the limit $p \to 0$.  To estimate the resulting error in $B_K$, we replace the values of $Z_{B_K}$ used in our preferred analysis with those obtained without extrapolating $Z_{B_K}^{\rm SI}$ to zero momentum, and repeat the combined chiral and continuum extrapolation.  We obtain  $B_K^{\bar{MS}} {(\rm 2 GeV)} = 0.542(7)$ (statistical error only), which is remarkably close to the lattice perturbation theory value.  We then take the difference between the resulting $B_K$ and our central value:
\begin{eqnarray}
	|B_K^{p \to 0} - B_K^{p = \rm{2 GeV}}| / B_K^{p \to 0} = 0.028,
\end{eqnarray}
which is $\sim 3\%$, and similar to our previous estimate.  Although taking $Z_{B_K}^{\rm SI}$ directly at 2 GeV provides a sensible alternate method for obtaining the renormalization factor, it still does not eliminate truncation errors from the lack of higher-order perturbative matching.  Nevertheless, because the values of $B_K$ obtained using this alternate nonperturbative determination and from lattice perturbation theory are so similar, we estimate the residual truncation error to be small:
\begin{eqnarray}
	|B_K^{p = \rm{2 GeV}} - B_K^{LPT}| / B_K^{p = \rm{2 GeV}} = 0.002 ,
\end{eqnarray}
which leads to a negligible increase in the systematic error when added in quadrature.  Since this method for estimating the perturbative truncation error leads to a slightly more conservative estimate than the third approach, we take $2.8\%$ to be the uncertainty in $Z_{B_K}$ due to the use of 1-loop perturbation theory.

\subsubsection{Total uncertainty in $Z_{B_K}$}

We summarize the contributions to the ``renormalization factor" uncertainty in $B_K$ in Table~\ref{tab:NPR_err} and add them in quadrature.  The $\sim 2.8\%$ error due to the use of perturbation theory is the largest single contribution to the total error in $B_K$, and can only be reduced by a calculation of the necessary matching factors at 2-loops.

\begin{table}
\caption{Error contributions to $B_K$ from the nonperturbative renormalization procedure.  Each source of uncertainty is discussed in Sec.~\ref{sec:NPR_err}, and is given as a percentage of $B_K$.}
\label{tab:NPR_err}
\begin{tabular}{lr} \\ \hline\hline

uncertainty & $\qquad Z_{B_K}$  \\[0.5mm] \hline

statistics & 0.7\%  \\
chiral extrapolation fit function &  1.2\% \\
strange quark mass dependence & 0.3\% \\
chiral symmetry breaking & 1.2\% \\
perturbation theory & 2.8\% \\
\hline
total & 3.4\% \\
\hline\hline
\end{tabular}\end{table}

\section{Result and conclusions}
\label{sec:Conc}

We obtain the following result for $B_K$ in the $\bar{\textrm{MS}}$ scheme at 2 GeV:
\begin{equation}
	B^{\bar{\textrm{MS}}}_K (2 \textrm{ GeV}) = 0.527(6)(10)(4)(3)(18) ,
\end{equation}
where the errors are from statistics, the chiral-continuum extrapolation, scale and quark mass uncertainties, finite volume errors, and the renormalization factor uncertainty, respectively.  The total error is $\sim 4\%$, and the error budget is presented in Table~\ref{tab:total_err}.  It is often more convenient to use the scale-invariant parameter $\hat{B}_K$ in new physics analyses, for which we find the value
\begin{equation}
	\hat{B}_K = 0.724(8)(29) . 
\end{equation}

\begin{table}
\caption{Total error budget for $B_K$.  Each source of uncertainty is discussed in Sec.~\ref{sec:Error}, and is given as a percentage of $B_K$.}
\label{tab:total_err}
\begin{tabular}{lr} \\ \hline\hline

uncertainty & $\qquad B_K$  \\[0.5mm] \hline

statistics & 1.2\%  \\
chiral \& continuum extrapolation &  1.9\% \\
scale and quark mass uncertainties & 0.8\% \\
finite volume errors  & 0.6\% \\
renormalization factor & 3.4\% \\
\hline
total & 4.2\% \\
\hline\hline
\end{tabular}\end{table}

Our 2+1 flavor lattice QCD calculation of $B_K$ is the first to have all lattice sources of systematic uncertainty under control.  The largest errors in our result for $B_K$ come from the chiral-continuum extrapolation ($1.9\%$) and from the determination of the renormalization factor ($3.4\%$).  The former uncertainty can be improved by the addition of statistics and the use of more lattice spacings.  The MILC collaboration has generated ensembles with a lattice spacing of $a \approx 0.06$ fm which we plan to analyze in  the near future.  The latter uncertainty can be improved in several ways.  The use of Landau gauge-fixed momentum-source propagators will reduce the size of the statistical errors in $Z_{B_K}$~\cite{Wennekers:2008sg,Lytle:2009xm}, and may consequently better constrain the extrapolation to the chiral limit.  The use of non-exceptional kinematics will reduce the contamination from chiral-symmetry breaking~\cite{Sturm:2009kb} and also provide an alternative nonperturbative renormalization scheme with independent truncation errors from the standard RI/MOM scheme~\cite{Aoki:Lat09}.  A calculation of the 2-loop continuum perturbation theory formulae needed to match $Z_{B_K}$ in the RI/MOM scheme to $Z_{B_K}$ in the $\bar{\textrm{MS}}$ scheme would allow for a better estimate of the perturbative truncation error of $Z_{B_K}$ in the RI/MOM scheme.  Nevertheless, our calculation of the matching factor $Z_{B_K}$ in mean-field improved lattice perturbation theory provides a robust alternative to our nonperturbative determination in the RI/MOM scheme since the systematic uncertainties are uncorrelated between the two methods.  In particular, the difference between the two results allows for a more reliable estimate of the matching error than from the RI/MOM scheme alone.  This is important because some errors, such as perturbative truncation errors, are difficult to estimate within a single scheme.  The error in $B_K$ from all sources except the renormalization error is only 2.5\% because the use of domain-wall valence quarks and staggered sea quarks allows us to control the remaining sources of uncertainty quite well.  Thus, if the use of non-exceptional kinematics or 2-loop continuum perturbation theory does reduce the matching error, we can obtain an even more precise determination of $B_K$ without making any further improvements to the lattice calculation.  

Our result is consistent with the determination by the RBC and UKQCD Collaborations using 2+1 flavors of domain-wall fermions, $\hat{B}_K=0.720(13)(37)$  \cite{Antonio:2007pb},
but our result has a smaller total error.  The largest error in the RBC/UKQCD calculation is the 4\% scaling uncertainty due to the use of only a single lattice spacing, which we reduce by using two lattice spacings.  Our result also has smaller statistical errors because of the large number of available staggered gauge configurations.  Our result has comparable matching errors to RBC/UKQCD because the dominant error in both calculations of $Z_{B_K}$ is from the use of precisely the same one-loop continuum perturbation theory results when converting from the RI-MOM scheme to the $\overline{\textrm{MS}}$ scheme.  Our error estimate is slightly more conservative, however, because we take the difference between the renormalization factors determined using NPR and using lattice perturbation theory to be the error due to the omission of higher-order terms.  

Our result is 1.9-$\sigma$ lower than the value currently preferred by the global unitarity triangle analysis, $\hat{B}_K = 0.92 \pm 0.10$~\cite{Laiho:2009eu}, which comes from an update of the work of Lunghi and Soni in Ref.~\cite{Lunghi:2008aa} using the latest determinations of all of the input parameters.  The tension with the standard model is enhanced by the inclusion of the correction factor $\kappa_\epsilon$ derived by Buras and Guadagnoli~\cite{Buras:2008nn,Buras:2009pj}, which raises the location of the $\epsilon_K$ band.  The uncertainty in the standard model constraint on $\hat{B}_K$ is $\sim$11\% .  This is largely due to the error in the CKM matrix element  $|V_{cb}|$, which is known to $\sim 2\%$ accuracy, but enters the constraint from $B_K$ on the unitarity triangle as the fourth power.  Thus the error in $|V_{cb}|$ must be reduced in order to maximize the constraint on new physics from neutral kaon mixing.  Fortunately work on improving the exclusive determination of $|V_{cb}|$ is ongoing by the Fermilab Lattice and MILC collaborations~\cite{Bernard:2008dn}, and work on improving the inclusive determination of $|V_{cb}|$ is in progress by Becher and Lunghi~\cite{Becher:2009}. 

Lattice QCD calculations of the hadronic weak matrix element $B_K$ that incorporate the effects of the dynamical up, down, and strange quarks can now reliably control all sources of uncertainty.   Because our result for $B_K$ is consistent with the determination of the RBC and UKQCD Collaborations, one can safely average the two values (taking correlations between systematic errors into account) for use in future unitarity triangle analyses.  There is already a hint of the presence of new physics in the quark flavor sector as indicated by the tension between the unitarity triangle constraints from $\epsilon_K$ and $\sin(2\beta)$~\cite{Lunghi:2008aa}.  We expect the errors in both lattice QCD calculations of $B_K$ to be reduced in the future, such that indirect $CP$-violation in the kaon system will play a valuable role in the search for new physics.

\section*{Acknowledgments}

We thank Sam Li for allowing the use of his nonperturbative renormalization code, Enrico Lunghi for providing us with updated results from his unitarity triangle analysis, and Christian Sturm for helpful discussions on perturbation theory.  We thank Claude Bernard, Urs Heller, and Yigal Shamir for valuable comments on the manuscript.  We thank Robert Edwards, B\'alint Jo\'o, and Kostas Orginos for
invaluable assistance in using Chroma.  We also thank Saul Cohen,
Huey-Wen Lin and Meifeng Lin for help in compiling and running
CPS.  We thank Jim Simone for help using FermiQCD and QDP++.  We thank Martin Savage for sharing an example of how to
write and compile new code that uses the Chroma libraries, and we
thank MILC for the use of their configurations.  We thank Claude Bernard, Peter Boyle, Norman
Christ, Chris Dawson, and Enno Scholz for useful physics discussions and
suggestions.  Computations for this work were carried out in part on facilities of the USQCD Collaboration, which are funded by the Office of Science of the U.S. Department of Energy, and on facilities of the NSF Teragrid under allocation TG-MCA93S002. This work was supported in part by the United States Department of Energy under Grant Nos. DE-FG02-92ER40699, DE-FG02-04ER41302 (C.A.) and DE-FG02-91ER40628 (J.L.), and by the National Science Foundation under Grant No.~PHY-0555235 (J.L.).  This manuscript has been co-authored by an employee of Brookhaven Science Associates, LLC under Contract No. DE-AC02-98CH10886 with the U.S. Department of Energy. 

\appendix

\section{Feynman rules for lattice perturbation theory}
\label{app:Lat_PT}

In this appendix we present the Feynman rules and the integrals needed to calculate $Z_{B_K}$ to one-loop in lattice perturbation theory with a Symanzik-improved gauge action and HYP-smeared domain wall quarks.

\subsection{Gluon propagator} 

The gluon propagator for the Symanzik-improved gauge action used by the MILC Collaboration is
\bea D_{\mu \nu}(k) = (\hat{k}^2)^{-2}\left[(1-A_{\mu \nu})\hat{k}_\mu \hat{k}_\nu +\delta_{\mu \nu}\sum_\sigma \hat{k}^2_\sigma A_{\nu \sigma} \right] -(1-\alpha)\frac{\hat{k}_\mu \hat{k}_\nu}{(\hat{k}^2)^2},
\eea
where
\bea A_{\mu \nu}(k) &=& A_{\nu \mu}(k)=(1-\delta_{\mu \nu})\Delta(k)^{-1}\left[({\hat{k}}^2)^2 - c_1 \hat{k}^2\left(2\sum_\rho \hat{k}_\rho^4 + \hat{k}^2 \sum_{\rho \neq \mu, \nu}\hat{k}_\rho^2\right) \right. \nonumber \\ &&\left. +c^2_1\left((\sum_{\rho}\hat{k}^4_\rho)^2 + \hat{k}^2\sum_{\rho}\hat{k}^4_\rho \sum_{\tau \neq \mu, \nu}\hat{k}^2_\tau +(\hat{k}^2)^2 \prod_{\rho \neq \mu, \nu} \hat{k}^2_\rho \right) \right], \\
\Delta(k) &=& \left(\hat{k}^2 - c_1\sum_{\rho}\hat{k}^4_\rho\right)\left[ \hat{k}^2 -c_1\left((\hat{k}^2)^2 + \sum_\tau \hat{k}^4_{\tau}\right) \right. \nonumber \\ && \left. + \frac{1}{2}c^2_1\left((\hat{k}^2)^3 +2\sum_\tau \hat{k}^6_\tau - \hat{k}^2\sum_\tau \hat{k}^4_\tau \right) \right] - 4c_1^3 \sum_\rho \hat{k}^4_\rho \prod_{\tau\neq \rho} \hat{k}^2_\tau,
\eea
with $c_1=-\frac{1}{12u_0^2}$, ($u_0$ is the fourth root of the plaquette) and
\bea 
	\hat{k}_\mu = 2 \sin \frac{k_\mu}{2}\ , \hat{k}^2 = 
	\sum_\mu \hat{k}_\mu^2\ .
\eea
Without loss of generality, we adopt the Feynman gauge $\alpha=1$.  The above propagator is that of the tree-level (tadpole) improved gauge action \cite{Weisz:1982zw, Weisz:1983bn}.  The gluon propagator in the improved case is significantly more complicated then that from the Wilson plaquette gauge action, where $A_{\mu\nu}^{\rm plaquette}=1-\delta_{\mu\nu}$.  The action used in the generation of the MILC ensembles is further improved through 1-loop, but this additional improvement introduces corrections of higher order than 1-loop in $Z_{B_K}$, and is not needed here.

\subsection{Domain wall propagator}

For the domain-wall propagator, we make use of the results of Ref.~\cite{Aoki:2002iq}. There are three types of domain-wall quark propagators.  The first connects general flavor indices:
\bea S(p)_{st} &=& \sum_{u=1}^{N}(-i\gamma_\mu \sin p_\mu + W^- +m M^-)_{su} \ G_R(u,t) P_R \nonumber \\ && 
+ \sum_{u=1}^{N}(-i\gamma_\mu \sin p_\mu + W^+ + m M^+)_{su} \ G_L(u,t) P_L,
\eea
where $P_{R,L}=(1\pm \gamma_5)/2$ are projection matrices, $s$, $t$, and $u$ are flavor indices, the mass matrices are
\bea W^+ = \left(\begin{array}{cccc}-W & 1 &  &  \\ & -W & ... &  \\ &  & ... & 1 \\ &  &  & -W\end{array}\right), \ \ W^- = \left(\begin{array}{cccc}-W &  &  &  \\1 & -W &  &  \\ & ... & ... &  \\ &  & 1  & -W\end{array}\right), 
\eea
\bea
M^+=\left(\begin{array}{ccc} &  & \\1 & & \end{array}\right),  \ \ M^- = \left(\begin{array}{ccc} & & 1  \\ & & \end{array}\right),
\eea
and $G_{R,L}$ are
\bea  G_R(s,t) &=& \frac{A}{F}[-(1-m^2)(1-W e^{-\alpha})e^{\alpha(-2N + s + t)} - (1-m^2)(1-W e^\alpha)e^{-\alpha(s+t)} \nonumber \\ && - 2W \sinh(\alpha)(e^{\alpha(-N+s-t)}+e^{\alpha(-N-s+t)})] + A e^{-\alpha|s-t|}, \\
G_L(s,t) &=& \frac{A}{F}[-(1-m^2)(1-W e^{\alpha})e^{\alpha(-2N + s + t -2)} - (1-m^2)(1-W e^{-\alpha})e^{\alpha(-s-t+2)} \nonumber \\ && - 2W \sinh(\alpha)(e^{\alpha(-N+s-t)}+e^{\alpha(-N-s+t)})] + A e^{-\alpha|s-t|}, \\ 
&& \cosh(\alpha) = \frac{1+W^2+\sum_{\mu} \sin^2 p_\mu}{2W}, \\ &&
A=\frac{1}{2W \sinh(\alpha)}, \\ &&
F=1 - e^\alpha W - m^2(1-W e^{-\alpha}), \\ &&
W = 1-M_5 +\sum_\mu (1-\cos p_\mu).
\eea
In these formulas $m$ is the domain-wall quark mass, and $M_5$ is the domain-wall height.  $N$ is the number of sites in the fifth dimension, i.e. the number of generalized flavors.

The second propagator connects the physical quark field $q$ with the fermion field of general flavor index,
\bea \langle q(p) \overline{\psi}(-p,s) \rangle &=& \frac{1}{F}(i\gamma_\mu \sin p_\mu -m(1-We^{-\alpha}))(e^{-\alpha(N-s)}P_R + e^{-\alpha(s-1)}P_L) \nonumber \\ && + \frac{1}{F}[m(i\gamma_\mu \sin p_\mu - m(1-W e^{-\alpha}))-F]e^{-\alpha}(e^{-\alpha(s-1)}P_R+e^{-\alpha(N-s)}P_L), \nonumber \\ &&
\eea
\bea
\langle \psi(p,s)  \overline{q}(-p) \rangle &=& \frac{1}{F}(e^{-\alpha(N-s)}P_L+e^{-\alpha(s-1)}P_R)(i\gamma_\mu \sin p_\mu - m(1-W e^{-\alpha})) \nonumber \\ && +\frac{1}{F}(e^{-\alpha(s-1)}P_L+e^{-\alpha(N-s)}P_R)e^{-\alpha}[m(i\gamma_\mu \sin p_\mu - m(1-W e^{-\alpha}))-F]. \nonumber \\ &&
\eea

The third propagator is that of the physical quark field
\bea S_q(p) \equiv \langle q(p) \overline{q}(-p) \rangle = \frac{-i \gamma_\mu \sin p_\mu + (1-W e^{-\alpha})m}{-(1-e^\alpha W)+m^2(1-We^{-\alpha})},
\eea
which reduces in the continuum limit to 
\bea S_q(p) = \frac{(1-w_0^2)}{i \! \not \! p +(1-w^2_0)m},
\eea
where $w_0=1-M_5$.

The following form of the propagators, where we perform the sum over generalized flavor indices, is useful for evaluating the vertex diagrams needed to renormalize $B_K$ \cite{Aoki:2002iq}, 
\bea S_{Lq}(p) \equiv \sum_{s=1}^{N}L(s)\langle \psi(p,s) \overline{q}(-p)\rangle &=& \left(\frac{e^{-\alpha}}{F(1-w_0 e^{-\alpha})}\right)\left(i m \gamma_\mu \sin p_\mu - (1-W e^\alpha) \right), \quad \\
S_{qR}(p) &\equiv& \sum_{s=1}^{\infty} \langle q(p)\overline{\psi}(-p,s) \rangle R(s) = S_{Lq}(p), \\
S_{qL}(p) \equiv \sum_{s=1}^{N} \langle q(p)\overline{\psi}(-p,s) \rangle L(s) &=& \frac{1}{1-w_0 e^{-\alpha}}\frac{1}{F}\left(i\gamma_\mu \sin p_\mu -m(1-W e^{-\alpha}) \right), \\
S_{Rq}(p) &\equiv& \sum_{s=1}^{\infty} R(s)\langle \psi(p,s)\overline{q}(-p) \rangle = S_{qL}(p)\ ,
\eea
with
\bea L(s)=(w_0^{(N-s)}P_R + w_0^{(s-1)}P_L), \\
R(s)=(w_0^{(s-1)}P_R+w_0^{(N-s)}P_L),
\eea
where in the rightmost expressions we take the limit that the number of lattice sites in the fifth dimension $N$ is infinite.  In principle, this limit should be taken after the momentum integral, but there is no difficulty with taking the limit first.  We use the mean-field improved value $M_5^{\rm MF}=M_5-4(1-u_0)$ throughout the perturbative calculation, as discussed in subsection~\ref{sec:LPT}. 

\subsection{Quark gluon vertices}

The quark gluon interaction vertices are \cite{Aoki:2002iq}
\bea 
	V^a_{1\mu}(k,p)_{st} 
	&=& V^a_{1\mu}(k,p) \delta_{st} = 
	-igT^a \left(\gamma_\mu \overline{V}_{1\mu}(k,p)
	+\tilde{V}_{1\mu}(k,p)\right)\delta_{st}, \\
	V^{ab}_{2\mu\nu}(k,p)_{st} &=& 
	V^{ab}_{2\mu\nu}(k,p) \delta_{st} = 
	\frac{1}{2}g^2\frac{1}{2}\{T^a,T^b\}
	\left(\gamma_\mu \tilde{V}_{1\mu}(k,p)
	+\overline{V}_{1\mu}(k,p)\right)\delta_{\mu\nu}\delta_{st},
\eea
where $g$ is the coupling constant, $T^a$ are the $SU(3)$ generators, and
\bea \overline{V}_{1\mu}(k,p) &=& \cos \frac{1}{2}(-k_\mu + p_\mu), \\
\tilde{V}_{1\mu}(k,p) &=& i \sin \frac{1}{2}(-k_\mu + p_\mu).
\eea

To account for the HYP-smearing of the valence quarks to the order we are working, the vertices must be modified by a form factor $h_{\mu\nu}$.  Since all gluons begin and end on fermion lines, the gluon propagator gets replaced by a more complicated propagator
$D_{\mu\nu} \rightarrow h_{\mu\lambda} D_{\lambda \sigma} h_{\nu\sigma}$.
The form factor is \cite{Lee:2003sk}
\bea h_{\mu\lambda} = \delta_{\mu\lambda}D_\lambda +(1-\delta_{\mu\lambda})G_{\mu\lambda}, \eea
where
\bea\label{eq:HYP1} D_\lambda &=& 1 - d_1\sum_{\nu\neq \lambda} \overline{s}^2_\nu + d_2\sum_{\begin{subarray}{l}\nu < \rho \\ \nu,\rho\neq \lambda \end{subarray}} \overline{s}^2_\nu \overline{s}^2_\rho - d_3 \overline{s}^2_\nu \overline{s}^2_\rho \overline{s}^2_\sigma, \\
G_{\mu\lambda} &=& \overline{s}_\mu\overline{s}_\lambda \tilde{G}_{\mu\lambda}(k), \\
\label{eq:HYP3} \tilde{G}_{\mu\lambda}(k) &=& d_1 - d_2 \frac{\overline{s}^2_\rho+\overline{s}^2_\sigma}{2} + d_3 \frac{\overline{s}^2_\rho\overline{s}^2_\sigma}{3},
\eea
and
\noindent $\overline{s}_\mu = \sin \frac{k_\mu}{2}$.  In Eqs.~(\ref{eq:HYP1})-(\ref{eq:HYP3}), the indices $\mu$, $\lambda$, $\rho$ and $\sigma$ are all different.  The coefficients $d_i$ are defined by
\bea d_1=\frac{2}{3}\alpha_1(1+\alpha_2(1+\alpha_3)), \ \ \ d_2=\frac{4}{3}\alpha_1\alpha_2(1+2\alpha_3), \ \ \ d_3=8\alpha_1\alpha_2\alpha_3,
\eea
where we take in our simulations the standard Hasenfratz \etal\ values $\alpha_1=0.75$, $\alpha_2=0.6$, $\alpha_3=0.3$~\cite{Hasenfratz:2001hp}.

\subsection {Renormalization factor $Z_{B_K}$}

The 4-quark operator renormalization needed for $B_K$ through one-loop can be written in terms of integrals that appear in the renormalization of bilinear operators.  We thus calculate the renormalization factors for the quark bilinear operators ${\cal O} =\overline{q}\Gamma q$.  The  bilinear operator gets renormalized in the $\overline{MS}$ scheme according to 
\bea {\cal O}^{\overline{MS}}_\Gamma (\mu) = (1-w_0^2)^{-1}Z_w^{-1}u_0Z_\Gamma(\mu a){\cal O}^{lat}_\Gamma (1/a),
\eea
where $Z_w$ renormalizes the domain-wall height.  It is convenient to define the quark wavefunction renormalization factor $Z_2$ implicitly via the relation
\bea q^{\overline{MS}} = (1-w_0)^{-1/2}Z^{-1/2}_w (u_0 Z_2)^{1/2}q^{lat}.
\eea
Using the Feynman rules presented in the previous sub-sections, we then have for the vertex correction to the bilinear operator in the $\overline{MS}$, NDR scheme \cite{Aoki:2002iq}
\bea \frac{Z_\Gamma}{Z_2} = 1+\frac{g^2C_F}{16\pi^2}\left[A_\Gamma \ln(\mu a)^2 + A_\Gamma(1-\ln\pi^2)+B_\Gamma -16\pi^2 I_\Gamma \right],
\eea
with 
\bea A_\Gamma = \frac{h_2(\Gamma)}{4}, \ \ \ \ B_\Gamma = -\frac{h_2(\Gamma)}{4}+V_\Gamma^{\overline{MS}},
\eea
where $h_2(\Gamma)=4 (V,A)$, $16 (P,S)$, $0 (T)$; $V_\Gamma^{\overline{MS}}= -1/2 (V,A)$, $2 (P,S)$, $0 (T)$; and $I_\Gamma$ is a finite lattice integral,
\bea\label{eq:Igamma} 
	I_\Gamma  &=&  \frac{1}{4g^2C_F} \int_k\Biggl\{
	\sum_{s,t}
	\textrm{Tr}\left[
	L(s)
	V_{1\mu}(0,k) 
	\langle \psi(k,s) \bar q(-k)\rangle 
	\Gamma 
	\langle q(k) \bar \psi(-k,t)\rangle
	V_{1\nu}(-k,0)
	R(t)\Gamma^\dagger \right]
	\nonumber \\&&{}\times
	h_{\mu\lambda}(k)D_{\lambda\sigma}(k)h_{\nu\sigma}(k) 
	- 4 g^2 C_F A_\Gamma \frac{\theta(\pi^2-k^2)}{(k^2)^2} \Biggr\},
\eea
with the trace over Dirac spin and 
\be
	\int_k \equiv \int \frac{d^4k}{(2\pi)^4}\ .
\ee 
The last term in Eq.~(\ref{eq:Igamma}) subtracts an IR (infrared) divergence from the integral.  By chiral symmetry, the renormalization factors for the vector and axial-vector currents are equal; the renormalization factors for scalar and pseudoscalar currents are also equal by chiral symmetry \cite{Aoki:1998vv}.  The Feynman diagram for the vertex correction is given in Fig.~\ref{fig:lat_pt}.

The renormalization factor matching the lattice calculation of $B_K$ to the $\overline{\textrm{MS}}$ scheme can be written \cite{Aoki:2002iq}
\bea Z_{B_K}(\mu a) = \frac{(1-w_0^2)^{-2}Z_w^{-2}Z_+(\mu a)}{(1-w_0^2)^{-2}Z^{-2}_w Z_A(\mu a)^2} = \frac{Z_+(\mu a)}{Z_A(\mu a)^2},
\eea
where $Z_+$ is the renormalization factor for the operator $O_K^{\Delta S=2}$, and $Z_A$ renormalizes the axial current.  It is useful to define $B_K$ in this way, since the tadpole and self-energy corrections cancel.  The renormalization factor contains the running of the operator from the lattice scale $a^{-1}$ to the continuum scale $\mu$.  In the $\overline{\textrm{MS}}$ scheme with naive dimensional regularization (NDR), we obtain \cite{Aoki:2002iq}
\bea Z_{B_K}^{\bar{\textrm{MS}},\textrm{NDR}}(\mu a) = 1+\frac{\alpha_s}{4\pi}\left[-4 \ln(\mu a) + z_{B_K}^{\bar{MS},NDR}\right],
\eea
where
\bea z_{B_K}^{\bar{\textrm{MS}},\textrm{NDR}}=-\frac{11}{3} + 2\ln \pi^2 +\frac{2}{3}(16\pi^2)(I_{S}-I_V),
\eea
with $I_{S,V}$ defined in Eq.~(\ref{eq:Igamma}).

\section{Matching Scheme and Perturbative Running for $Z_{B_K}$}
\label{app:NPR}

Although the functions used to convert the renormalization factor $Z_{B_K}$ from the RI/MOM scheme to  the $\bar{\textrm{MS}}$ scheme are the same as those shown in the appendices of Ref.~\cite{Aoki:2007xm}, we display them here for completeness.

\subsection{The QCD $\beta$-function in the $\bar{\textrm{MS}}$ scheme}

In this work we calculate the value of the coupling constant $\alpha_{s}^{\bar{\textrm{MS}}}\left(\mu\right)$ at any scale using the four-loop (NNNLO) running formula of Ref.~\cite{vanRitbergen:1997va}:
\begin{eqnarray}\label{eq:alpha_s_run}
	\frac{\partial }{\partial\ln\mu^{2}}
	\left(\frac{\alpha_s}{\pi}\right)
	& = & \beta\left(\alpha_{s}\right)\nonumber \\
 	& = & -\beta_0\left(\frac{\alpha_s}{\pi}\right)^2
	- \beta_1\left(\frac{\alpha_s}{\pi}\right)^3
	- \beta_2\left(\frac{\alpha_s}{\pi}\right)^4
	- \beta_3\left(\frac{\alpha_s}{\pi}\right)^5
	+\CO\left(\alpha_s^6\right)\, ,
\end{eqnarray}
where
\begin{eqnarray}\label{eq:beta-function}
	\beta_0 & = & \frac{1}{4}
	\left(11-\frac{2}{3}n_{f}\right),\nonumber \\
	\beta_1 & = & \frac{1}{16}
	\left(102-\frac{38}{3}n_{f}\right),\nonumber \\
	\beta_2 & = & \frac{1}{64}
	\left(\frac{2857}{2}-\frac{5033}{18}
	n_{f}+\frac{325}{54}n_{f}^{2}\right),\nonumber \\
	\beta_3 & = & \frac{1}{256}\left[\frac{149753}{6}
	+3564\zeta_{3}-\left(\frac{1078361}{162}+\frac{6508}{27}
	\zeta_{3}\right)n_{f}\right.\nonumber \\
 	&& \left.+\left(\frac{50065}{162}+\frac{6472}{81}\zeta_{3}
	\right)n_{f}^{2}+\frac{1093}{729}n_{f}^{3}\right] \ .
\end{eqnarray}
We implement this numerically by starting with the world average of the strong coupling constant at the $Z$-boson mass~\cite{Amsler:2008zzb}, 
\begin{equation}\label{eq:alpha_s-Mz}
	\alpha_{s}^{\left(5\right)}\left(m_{Z}\right)=0.1176 \pm 0.0020\, ,
\end{equation}
where the superscript indicates that this is determined in the region with five active quark flavors.  We then run $\alpha_{s}$ below the bottom and charm quark thresholds imposing the matching conditions
\begin{equation}\label{eq:alpha_s run match}
	\alpha_{s}^{\left(5\right)}\left(m_{b}\right) 
	=\alpha_{s}^{\left(4\right)}\left(m_{b}\right)
	\qquad\textrm{and}\qquad
	\alpha_{s}^{\left(4\right)}\left(m_{c}\right) 
	=\alpha_{s}^{\left(3\right)}\left(m_{c}\right)\, ,
\end{equation}
in order to determine $\alpha_{s}^{\left( 3\right)}\left( \mu \right)$ at any scale in the 3-flavor theory.  

\subsection{Perturbative Running and Scheme Matching for $Z_{B_{K}}$\label{sec:Zbk run factor}}

We convert the renormalization factor $Z_{B_K}$ between the scale-invariant, $\overline{\rm MS}$, and RI/MOM schemes using the one-loop renormalization group running formulae with $n_{f}=3$~\cite{Ciuchini:1997bw}: 
\begin{equation}\label{eq:Z_bk_SI}
	Z_{B_{K}}^{\mathrm{SI}}\left(n_{f}\right)
	= w_{\mathrm{scheme}}^{-1}\left(\mu,n_{f}\right)
	Z_{B_{K}}^{\mathrm{scheme}}
	\left(\mu,n_{f}\right)\, ,
\end{equation}
where
\begin{equation}\label{eq:w-RI-MOM}
	w_{\mathrm{scheme}}^{-1}\left(\mu,n_{f}\right)
	=\alpha_{s}^{\overline{\rm MS}}\left(\mu\right)^{-\gamma_{0}/2\beta_{0}}
	\left[1+ \frac{\alpha_{s}^{\overline{\rm MS}}\left(\mu\right)}{4\pi}
	J_{\mathrm{scheme}}^{\left(n_{f}\right)}\right]
\end{equation}
and
\begin{eqnarray}
	J_{\mathrm{RI/MOM}}^{\left(n_{f}\right)} & = &
	-\frac{17397-2070n_{f}+104n_{f}^{2}}{6\left(33-2n_{f}\right)^{2}}
	+8\ln2\, ,
	\label{eq:J-RI-MOM}\\
	J_{\mathrm{\overline{MS}}}^{\left(n_{f}\right)} & = &
	\frac{13095-1626n_{f}+8n_{f}^{2}}{6\left(33-2n_{f}\right)^{2}}\, .
	\label{eq:J-MSbar}
\end{eqnarray}
%

\section{Non-perturbative Mixing Coefficients}
\label{app:NPRmix}

In this appendix we present results for the mixing coefficients between the operator $\CO_K$ and other lattice operators of different chiralities.  We compute them nonperturbatively using the method of Rome-Southampton as discussed in Sec.~\ref{sec:RomeSH}.

The renormalized operator that contributes to $B_K$ in the continuum, which has a $VV+AA$ chiral structure, receives contributions from several lattice operators:
\begin{equation}\label{eq:ren_LL_oper}
	\CO_K^\textrm{ren} = \sum_i Z_{VV+AA,i}\, \CO^{0}_i\, ,
\end{equation}
where $i \in \{VV+AA,VV-AA,SS-PP,SS+PP,TT\}$.  Because the operator mixings require two flips of chirality, the off-diagonal coefficients are suppressed by $O((a m_{\rm res})^2)$~\cite{Christ:2005xh,Aoki:2007xm}, which is $\sim 10^{-6}$ on the coarse lattice and even smaller on the fine lattice.  We therefore expect the contributions to $B_K$ from wrong-chirality lattice operators to be negligible.

\begin{figure}
\begin{center}
\includegraphics[width=4in]{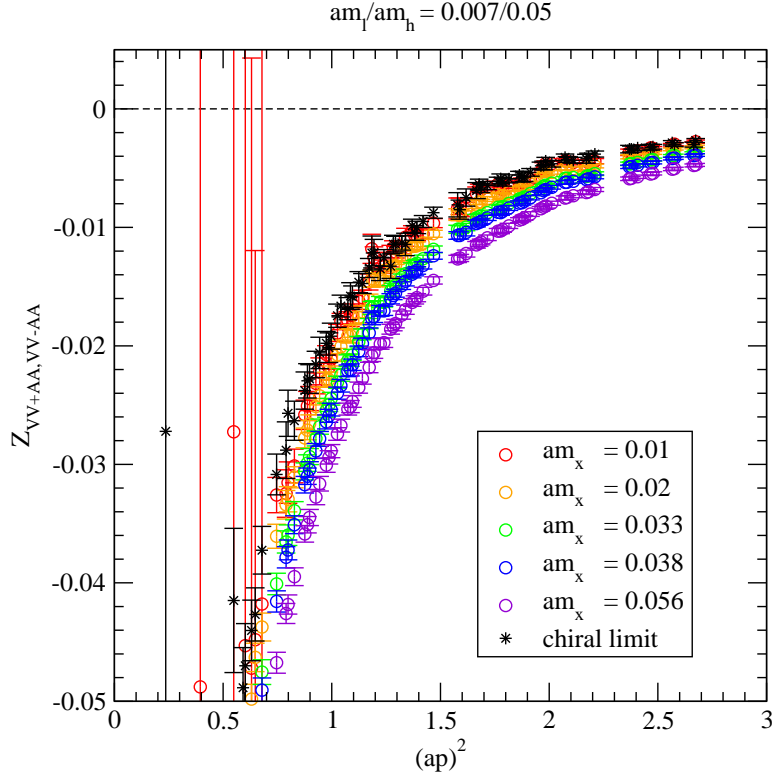}
\caption{Mixing coefficient $Z_{VV+AA,VV-AA}$ versus $(ap)^2$ at several valence quark masses on the $am_l/am_h = 0.007/0.05$ coarse ensemble.  The stars indicate the value of the mixing coefficient in the limit that the valence quark mass and the light sea quark mass go to zero.}
\label{fig:coarse_VVmAAsea007}
\end{center}
\end{figure}

\begin{figure}
\begin{center}
\includegraphics[width=4in]{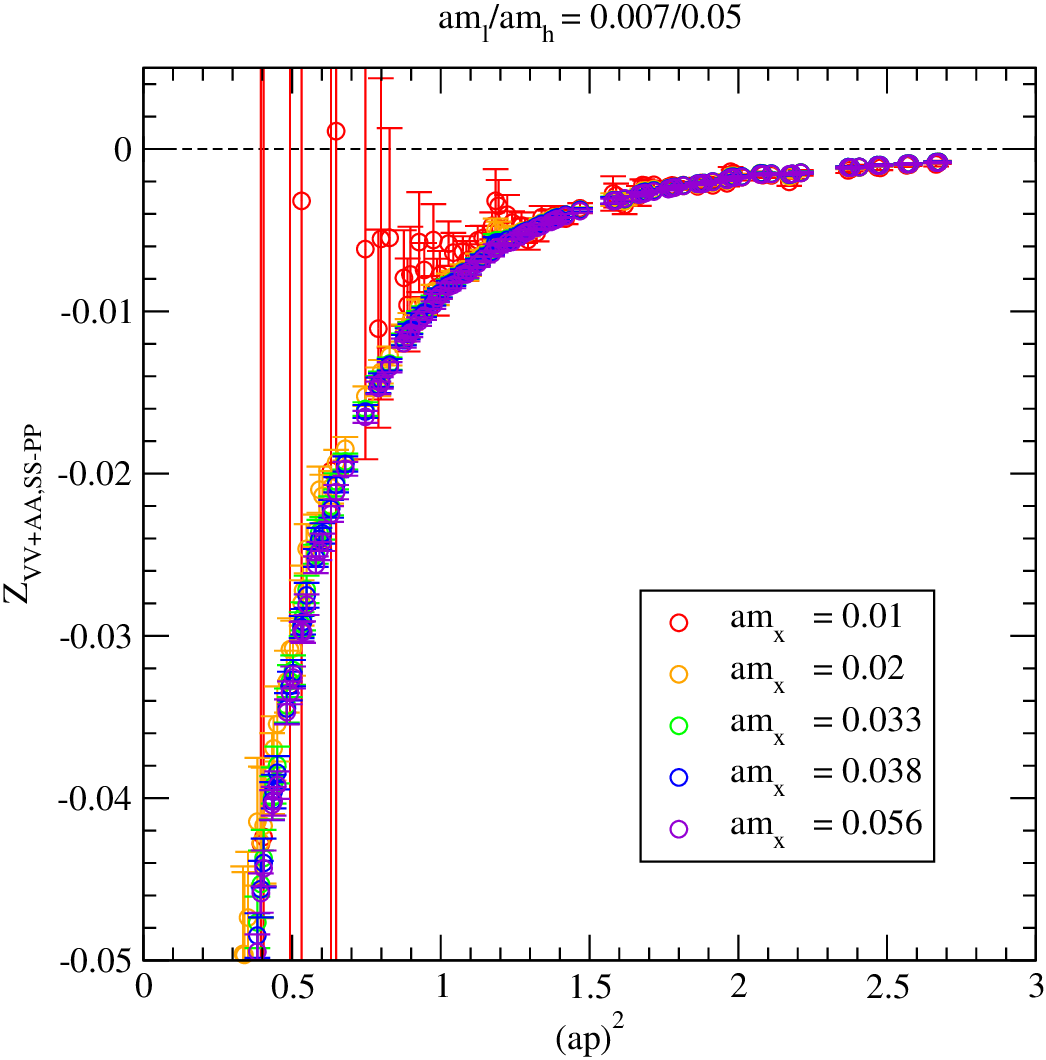}
\caption{Mixing coefficient $Z_{VV+AA,SS-PP}$ versus $(ap)^2$ at several valence quark masses on the $am_l/am_h = 0.007/0.05$ coarse ensemble.}
\label{fig:coarse_SSmPPsea007}
\end{center}
\end{figure}

\begin{figure}
\begin{center}
\includegraphics[width=4in]{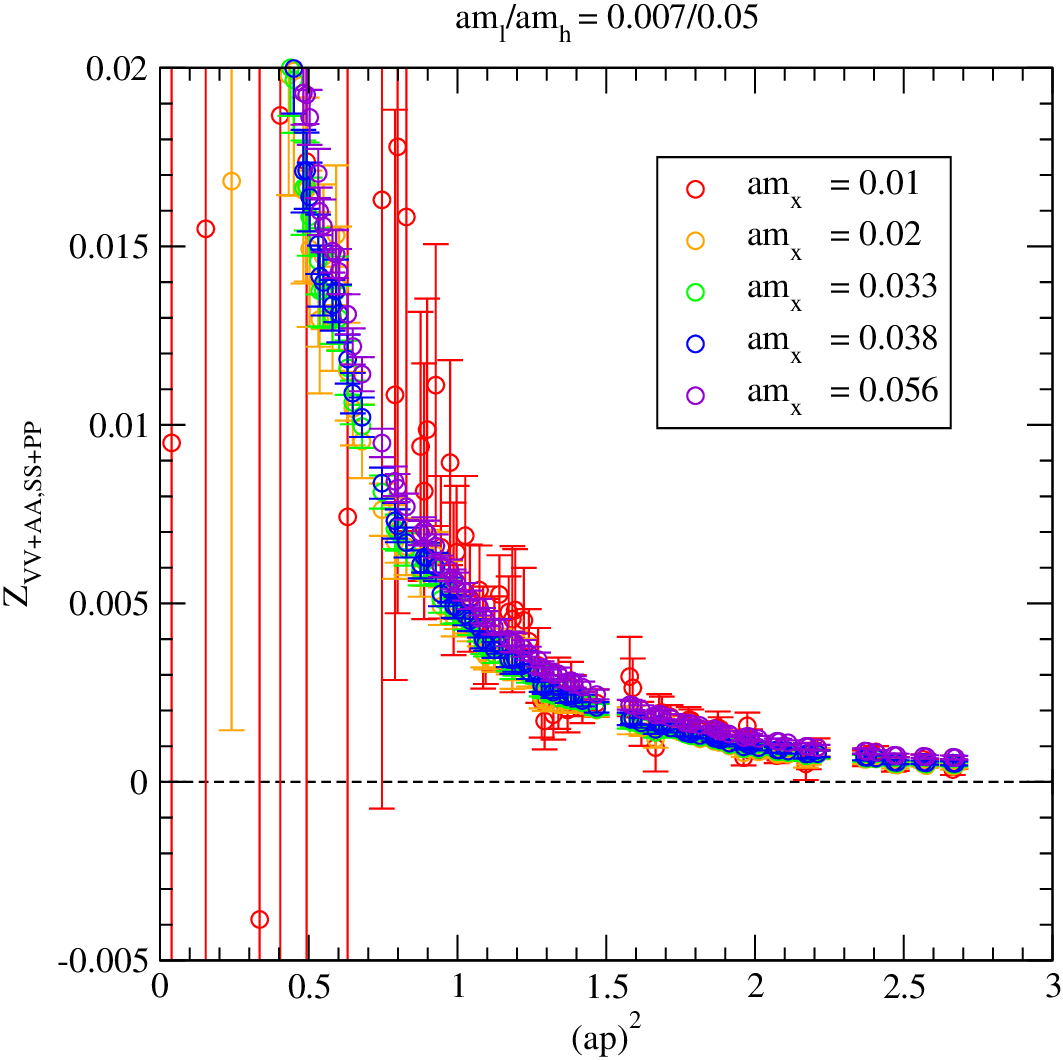}
\caption{Mixing coefficient $Z_{VV+AA,SS+PP}$ versus $(ap)^2$ at several valence quark masses on the $am_l/am_h = 0.007/0.05$ coarse ensemble.}
\label{fig:coarse_SSpPPsea007}
\end{center}
\end{figure}

\begin{figure}
\begin{center}
\includegraphics[width=4in]{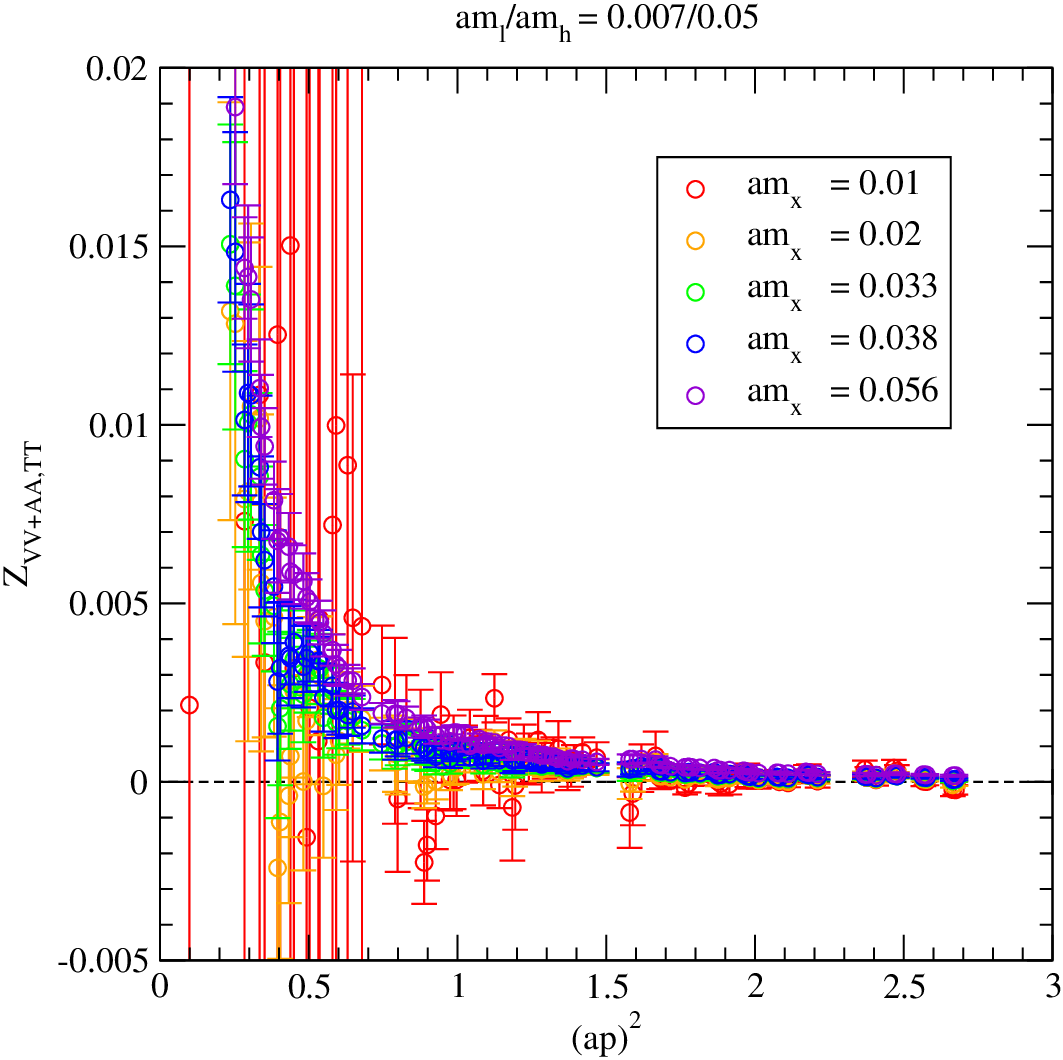}
\caption{Mixing coefficient $Z_{VV+AA,TT}$ versus $(ap)^2$ at several valence quark masses on the $am_l/am_h = 0.007/0.05$ coarse ensemble.}
\label{fig:coarse_TTsea007}
\end{center}
\end{figure}

In practice, however, we find that the mixing coefficients are not of $O((a m_{\rm res})^2)$ when we compute them using exceptional kinematics.  This is because the choice of external momenta in the renormalization factor calculation leads to additional chiral symmetry breaking, as discussed in section~\ref{sec:LA_LV}.  Figures~\ref{fig:coarse_VVmAAsea007}--\ref{fig:coarse_TTsea007} show the mixing coefficients as a function of $(ap)^2$ for five valence quark masses on the $am_l/am_h = 0.007/0.05$ coarse ensemble and in the chiral limit.    At $p \approx {\rm 2 GeV}$, the mixing coefficients are still all quite small compared to $Z_{B_K}$.  The largest is the mixing of $\CO_K$ with the $VV-AA$ operator, which is $\sim 0.01$.  We observe coefficients of approximately the same size on the fine lattice, since this effect is not due to the lattice spacing or residual quark mass.  

\begin{figure}
\begin{center}
\includegraphics[width=4in]{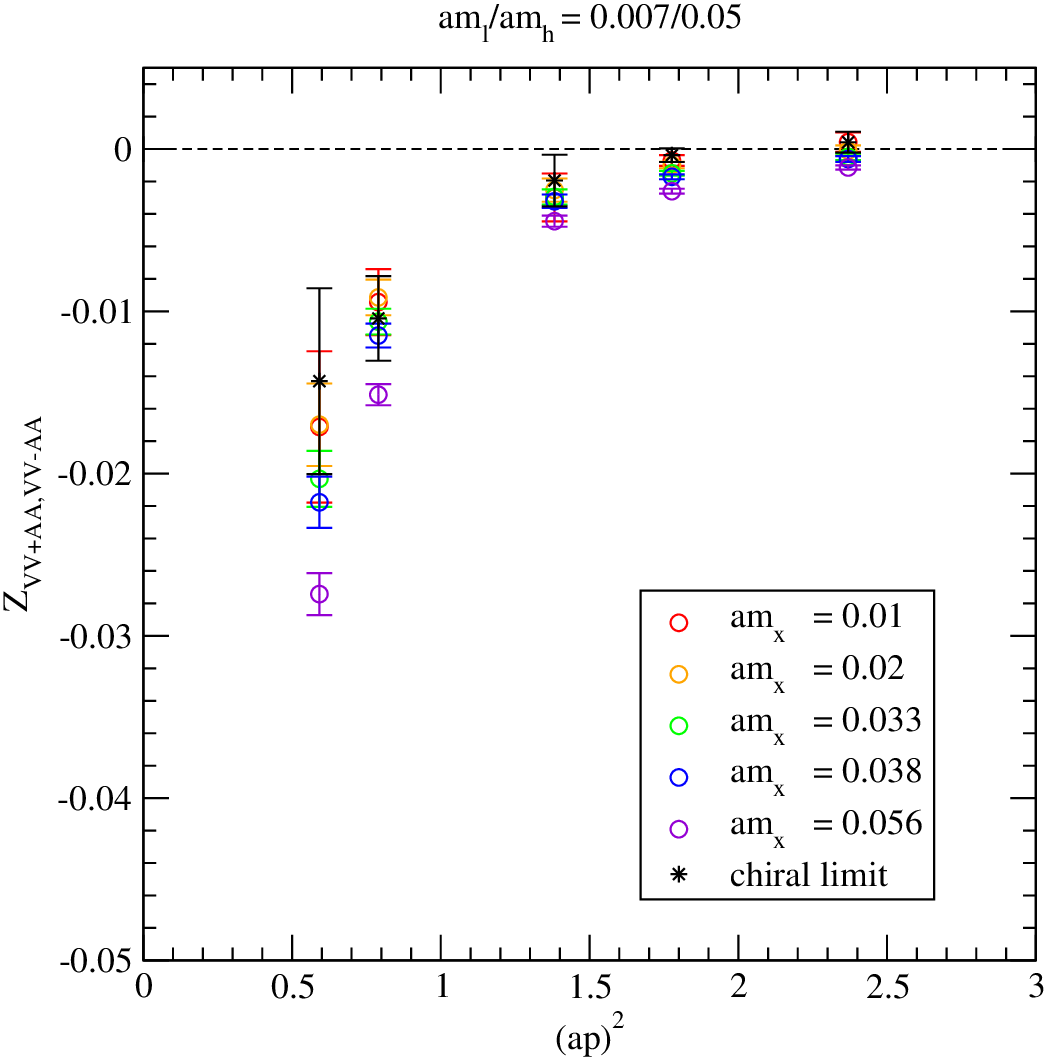}
\caption{Mixing coefficient $Z_{VV+AA,VV-AA}$ versus $(ap)^2$ at several valence quark masses on the $am_l/am_h = 0.007/0.05$ coarse ensemble, using non-exceptional kinematics.  The stars indicate the value of the mixing coefficient in the limit that the valence quark mass and the light sea quark mass go to zero.}
\label{fig:coarse_VVmAAsea007_nonexp}
\end{center}
\end{figure}

\begin{figure}
\begin{center}
\includegraphics[width=4in]{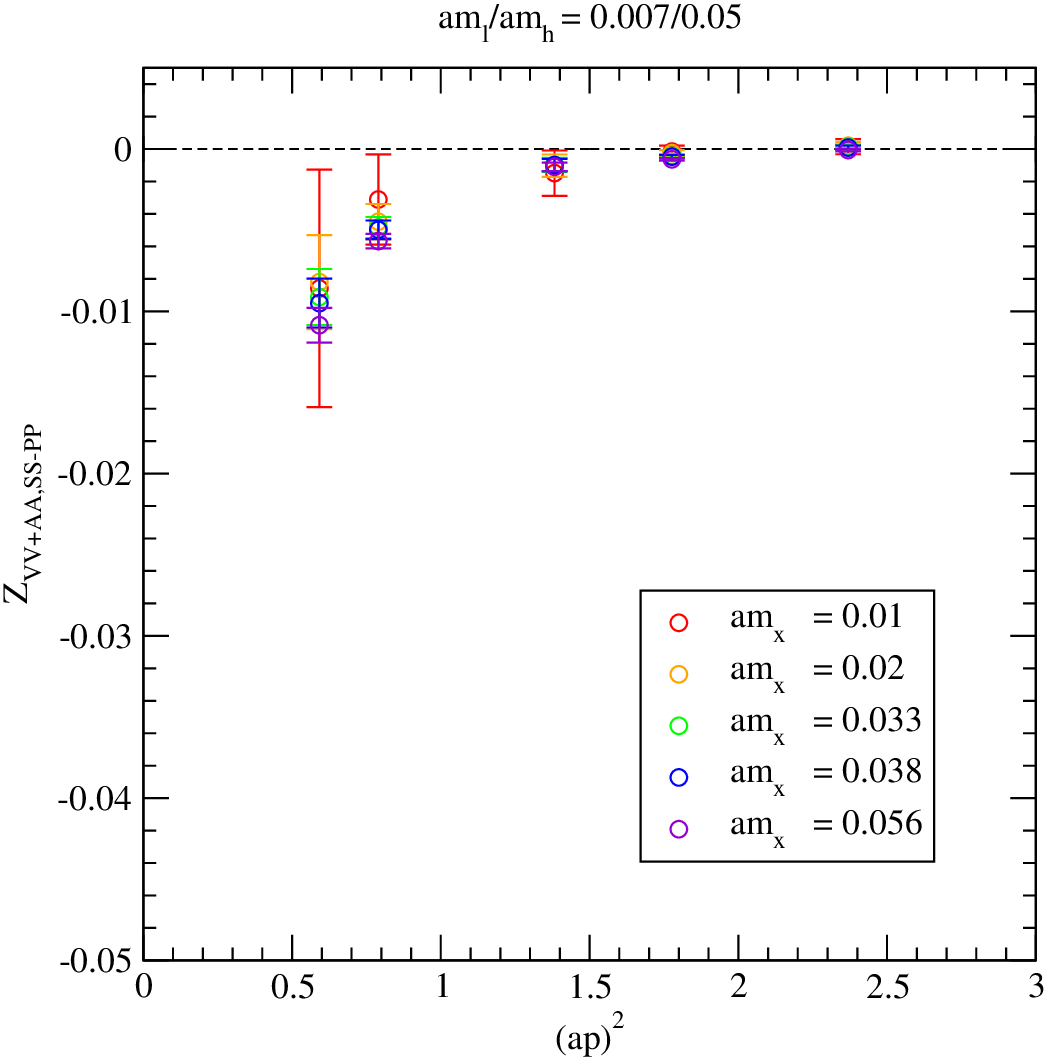}
\caption{Mixing coefficient $Z_{VV+AA,SS-PP}$ versus $(ap)^2$ at several valence quark masses on the $am_l/am_h = 0.007/0.05$ coarse ensemble, using non-exceptional kinematics.}
\label{fig:coarse_SSmPPsea007_nonexp}
\end{center}
\end{figure}

\begin{figure}
\begin{center}
\includegraphics[width=4in]{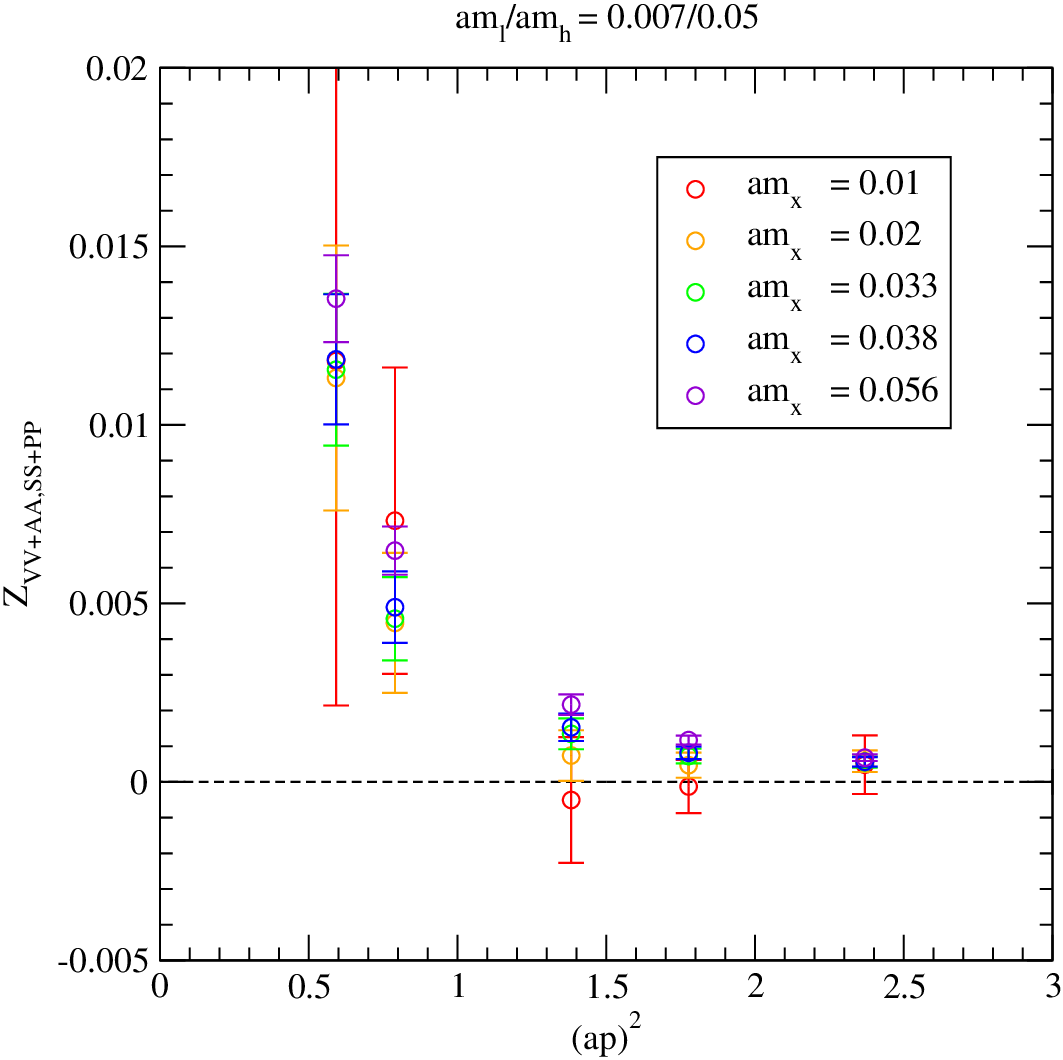}
\caption{Mixing coefficient $Z_{VV+AA,SS+PP}$ versus $(ap)^2$ at several valence quark masses on the $am_l/am_h = 0.007/0.05$ coarse ensemble, using non-exceptional kinematics.}
\label{fig:coarse_SSpPPsea007_nonexp}
\end{center}
\end{figure}

\begin{figure}
\begin{center}
\includegraphics[width=4in]{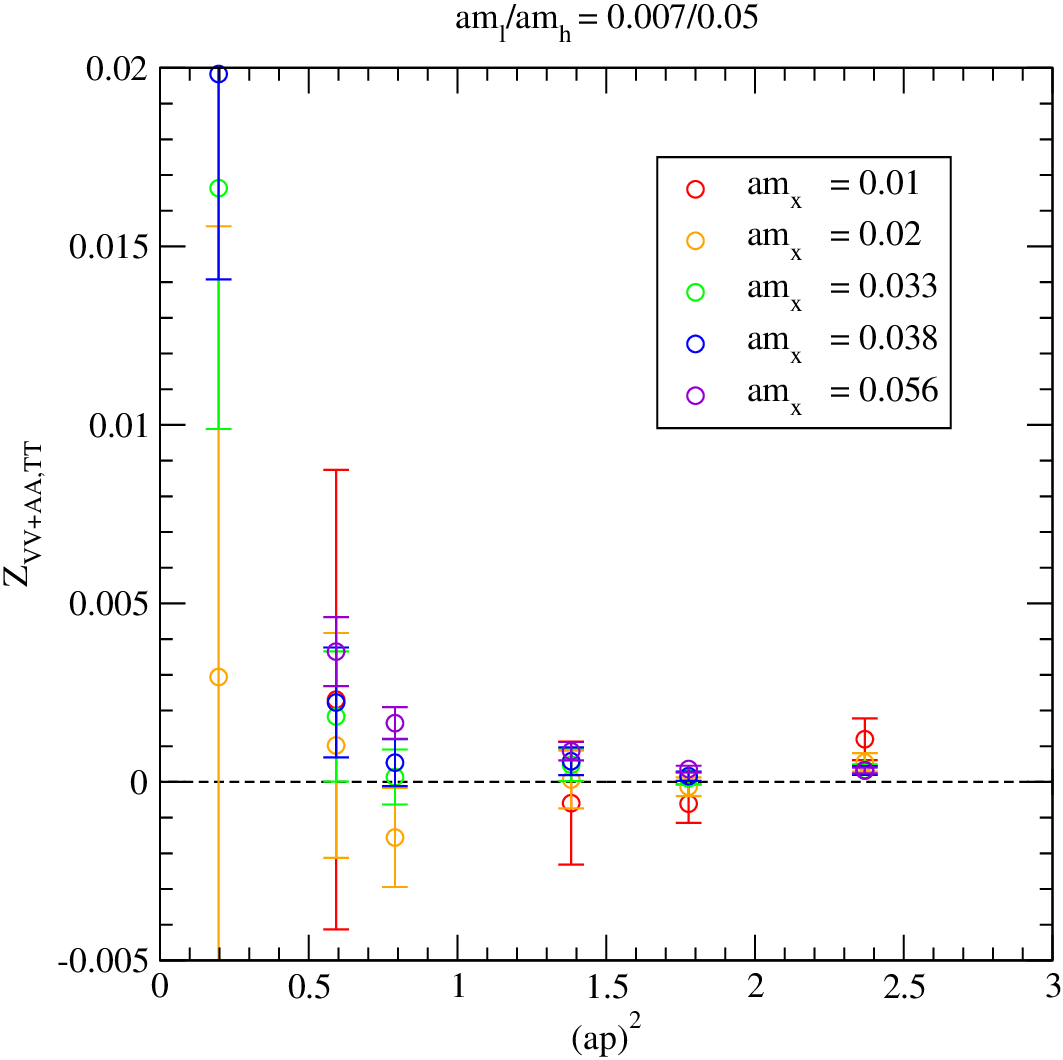}
\caption{Mixing coefficient for $Z_{VV+AA,TT}$ versus $(ap)^2$ at several valence quark masses on the $am_l/am_h = 0.007/0.05$ coarse ensemble, using non-exceptional kinematics.
}
\label{fig:coarse_TTsea007_nonexp}
\end{center}
\end{figure}

\begin{figure}
\begin{center}
\includegraphics[width=4in]{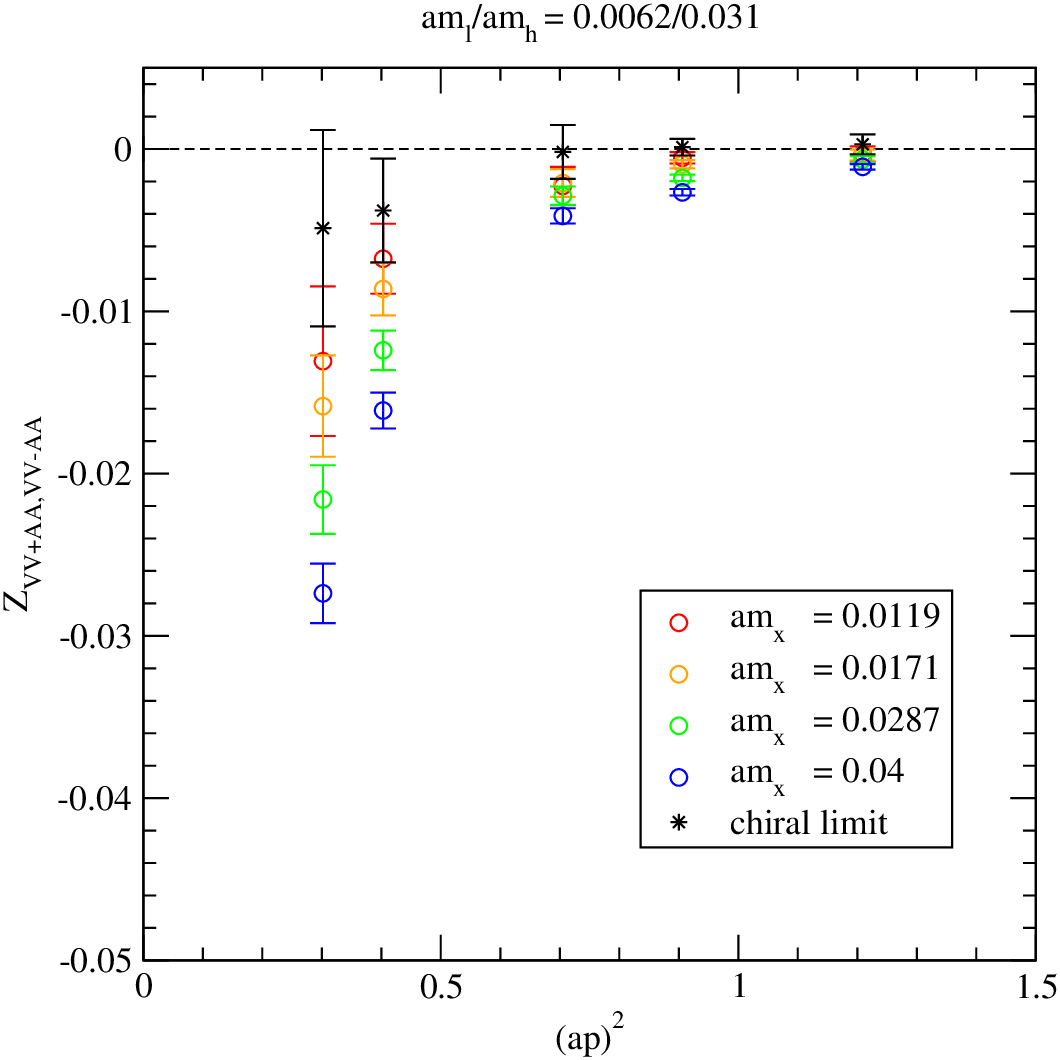}
\caption{Mixing coefficient $Z_{VV+AA,VV-AA}$ versus $(ap)^2$ at several valence quark masses on the $am_l/am_h = 0.0062/0.031$ fine ensemble, using non-exceptional kinematics. The stars indicate the value of the mixing coefficient in the limit that the valence quark mass and the light sea quark mass go to zero.
}
\label{fig:fine_VVmAAsea0062_nonexp}
\end{center}
\end{figure}

\begin{figure}
\begin{center}
\includegraphics[width=4in]{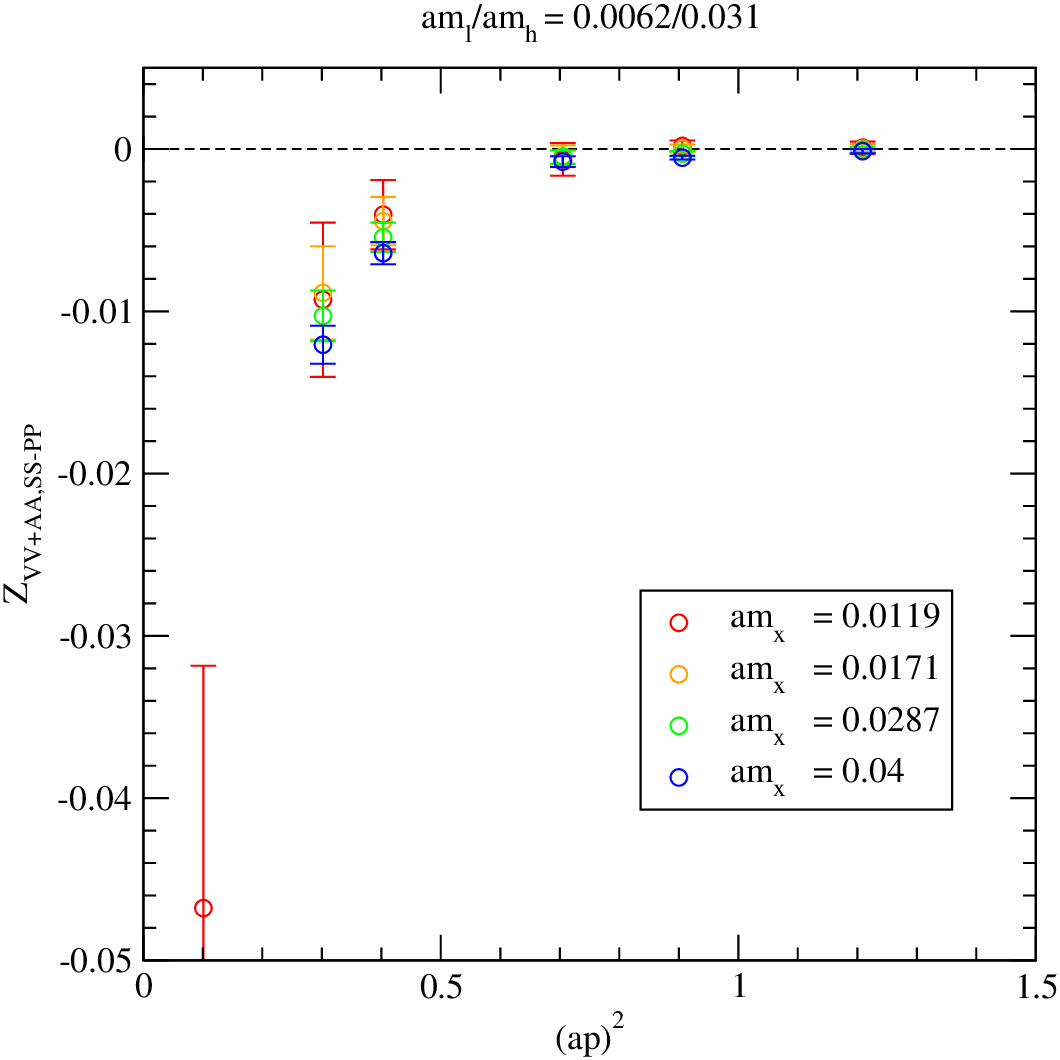}
\caption{Mixing coefficient $Z_{VV+AA,SS-PP}$ versus $(ap)^2$ at several valence quark masses on the $am_l/am_h = 0.0062/0.031$ fine ensemble, using non-exceptional kinematics.}
\label{fig:fine_SSmPPsea0062_nonexp}
\end{center}
\end{figure}

\begin{figure}
\begin{center}
\includegraphics[width=4in]{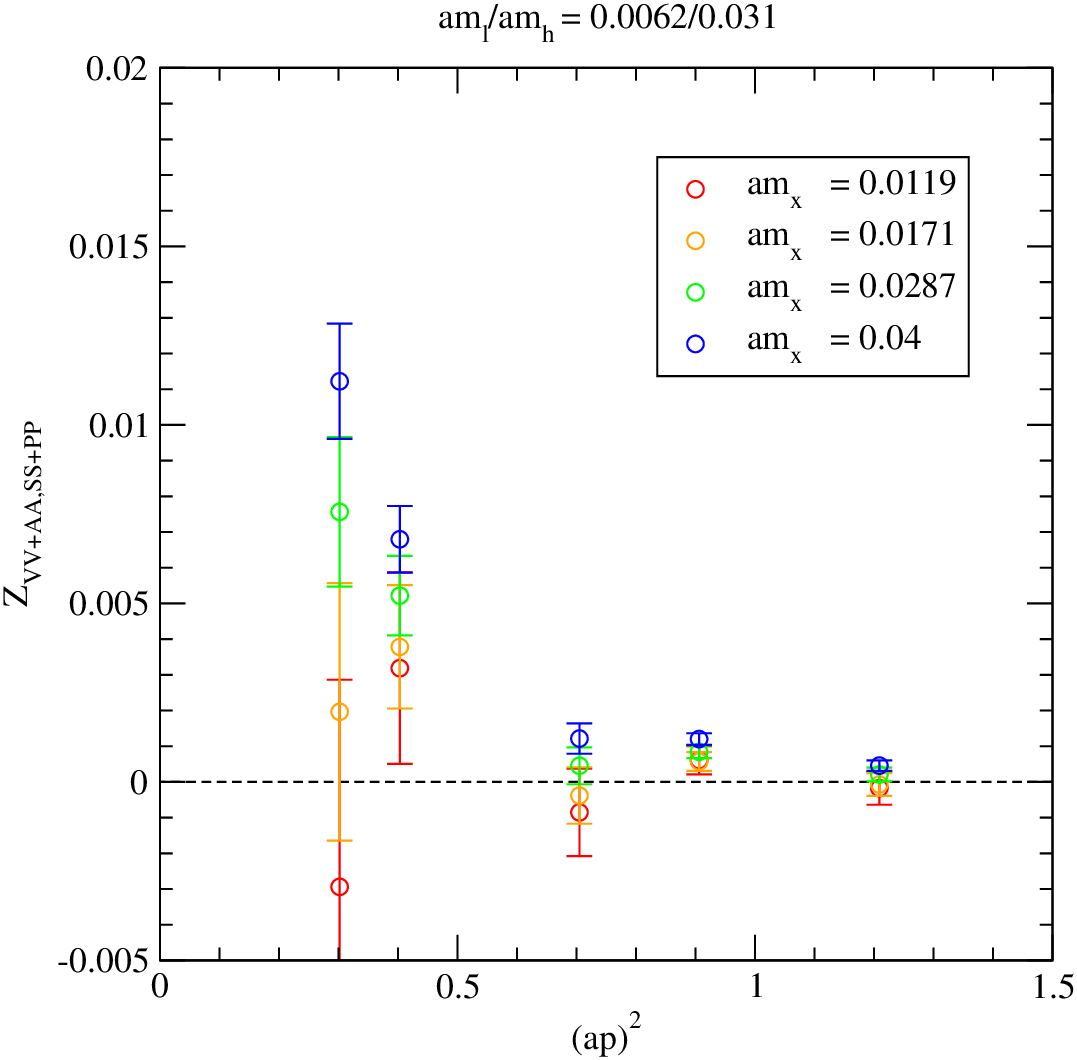}
\caption{Mixing coefficient $Z_{VV+AA,SS+PP}$ versus $(ap)^2$ at several valence quark masses on the $am_l/am_h = 0.0062/0.031$ fine ensemble, using non-exceptional kinematics.}
\label{fig:fine_SSpPPsea0062_nonexp}
\end{center}
\end{figure}

\begin{figure}
\begin{center}
\includegraphics[width=4in]{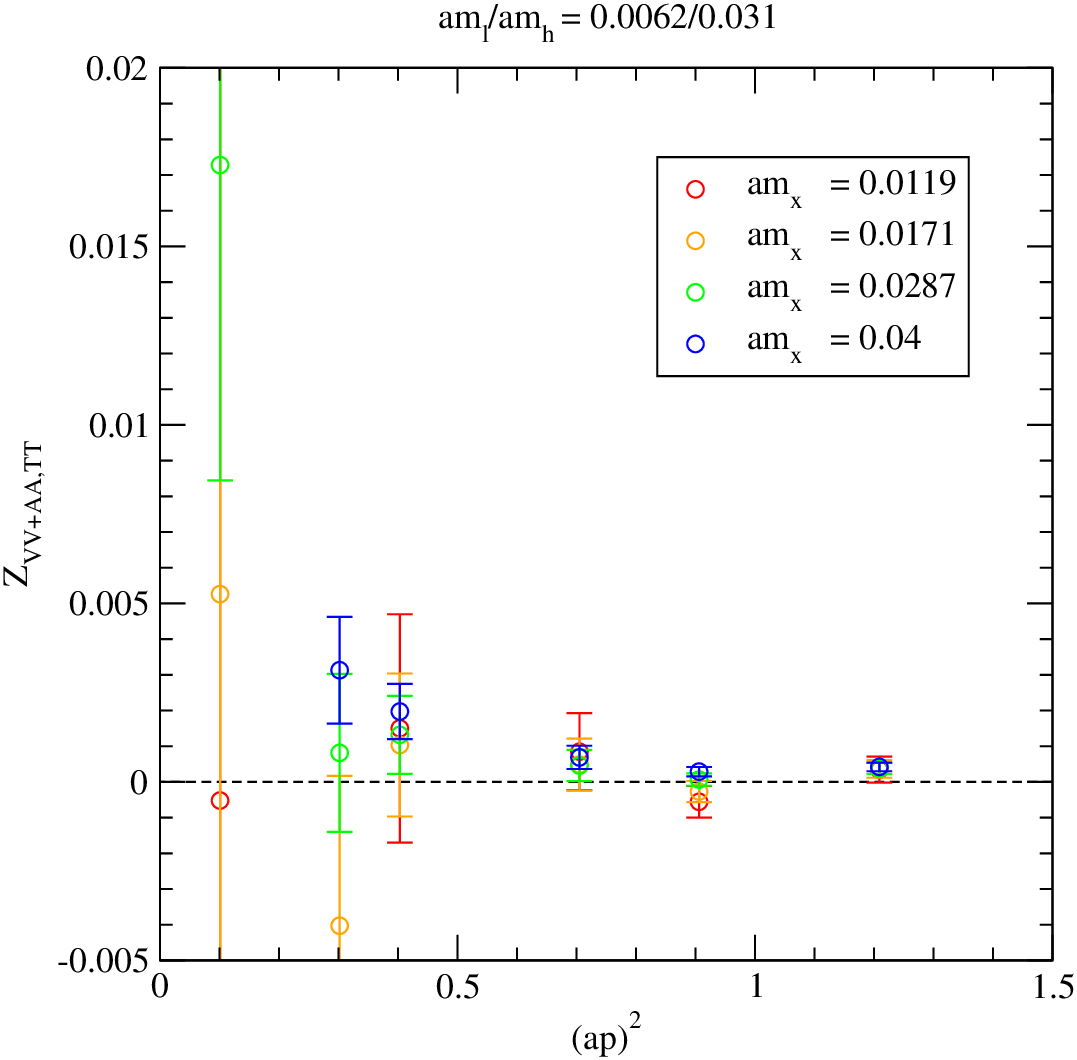}
\caption{Mixing coefficient $Z_{VV+AA,TT}$ versus $(ap)^2$ at several valence quark masses on the $am_l/am_h = 0.0062/0.031$ fine ensemble, using non-exceptional kinematics.}
\label{fig:fine_TTsea0062_nonexp}
\end{center}
\end{figure}

Although the size of the mixing coefficients as computed with exceptional kinematics is not negligible, the results are contaminated by chiral symmetry breaking effects and are potentially unreliable.  We therefore repeat the mixing coefficient calculation using non-exceptional kinematics.  The results are shown for the coarse lattice in Figs.~\ref{fig:coarse_VVmAAsea007_nonexp}--\ref{fig:coarse_TTsea007_nonexp} and for the fine lattice in Figs.~\ref{fig:fine_VVmAAsea0062_nonexp}--\ref{fig:fine_TTsea0062_nonexp}.  Although the mixing coefficients determined with non-exceptional kinematics have larger statistical errors, their values are smaller than when determined with exceptional kinematics.  Furthermore, the new mixing coefficients are consistent with zero in the chiral limit.  This confirms the hypothesis that the source of the large mixing coefficients is simply the choice of non-exceptional kinematics, and that the sizes of the true mixing coefficients are consistent with theoretical estimates.

Although we find that the mixing coefficients are consistent with zero in the RI/MOM scheme using non-exceptional kinematics, we can still use this information to aid in our determination of $Z_{B_K}$ using exceptional kinematics.  This is because, ultimately, irrespective of the lattice scheme used to obtain the mixing coefficients, one must obtain the same mixing coefficients once the results are converted to the $\bar{\textrm{MS}}$ scheme. A vanishing contribution to BK from a particular operator in the RI/MOM scheme with non-exceptional kinematics implies a vanishing contribution in the $\bar{\textrm{MS}}$ scheme, since they are related multiplicatively.  Generically, once an operator's contribution is zero in any scheme, its contribution is zero in all schemes that are multiplicatively related.  Note, however, that once an operator's contribution is nonzero, its particular value is scheme-dependent.


\bibliography{SuperBib}

\end{document}